\documentclass[aoas,preprint]{imsart}

\RequirePackage[OT1]{fontenc}
\RequirePackage{amsthm,amsmath}
\RequirePackage{natbib}
\usepackage{mathrsfs}
\usepackage{appendix}
\usepackage{subfigure}
\usepackage{float}
\usepackage{graphicx}
\usepackage{caption}
\usepackage{amsfonts}
\usepackage{enumerate}
\usepackage{algorithm}
\usepackage{algorithmicx}
\usepackage{rotating}
\usepackage{pdfpages}
\arxiv{arXiv:0000.0000}
\startlocaldefs
\numberwithin{equation}{section}
\theoremstyle{plain}

\newtheorem*{theorem*}{Theorem}

\newtheorem*{thmdef*}{Theorem}

\newtheorem*{defn*}{Definition}
\newtheorem*{prop*}{Proposition}

\endlocaldefs

\newcommand{\calS}{{\cal S}}

\newcommand{\calL}{{\cal L}}
\newcommand{\calT}{{\cal T}}
\newcommand{\calR}{{\cal R}}
\newcommand{\calU}{{\cal U}}
\newcommand{\calV}{{\cal V}}

\newcommand{\bbeta}{ \mbox{\boldmath $ \beta $} }

\newcommand{\btheta}{ \mbox{\boldmath $\theta$}}

\newcommand{\bmu}{ \mbox{\boldmath $\mu$} }

\newcommand{\bzero}{\textbf{0}}

\newcommand{\ba}{\textbf{a}}

\newcommand{\bC}{\textbf{C}}

\newcommand{\bF}{\textbf{F}}

\newcommand{\bK}{\textbf{K}}
\newcommand{\bh}{\textbf{h}}

\newcommand{\bs}{\textbf{s}}

\newcommand{\bV}{\textbf{V}}
\newcommand{\bw}{\textbf{w}}

\newcommand{\bx}{\textbf{x}}
\newcommand{\bX}{\textbf{X}}
\newcommand{\by}{\textbf{y}}

\newcommand{\tildebC}{\tilde{\bC}}

\newcommand{\tildeC}{\tilde{C}}

\newcommand{\given}{\,|\,}

\newcommand{\eps}{\epsilon}
\newcommand{\taus}{\tau^2}
\newcommand{\sigs}{\sigma^2}

\usepackage{xr}
\begin{document}

\begin{frontmatter}
\title{Non-separable Dynamic Nearest-Neighbor Gaussian Process Models for Large spatio-temporal Data With an Application to Particulate Matter Analysis}
\runtitle{Spatio-temporal NNGP}

\begin{aug}
\author{\fnms{Abhirup} \snm{Datta}\ead[label=e1]{datta013@umn.edu}}
\and
\author{\fnms{Sudipto} \snm{Banerjee}
\ead[label=e2]{sudipto@ucla.edu}}
\and
\author{\fnms{Andrew O.} \snm{Finley}
\ead[label=e3]{finleya@msu.edu}}
\and
\author{\fnms{Nicholas A.S.} \snm{Hamm}
\ead[label=e4]{n.hamm@utwente.nl}}
\and
\author{\fnms{Martijn} \snm{Schaap}
\ead[label=e5]{martijn.schaap@tno.nl}}
\runauthor{A. Datta et al.}

\affiliation{University of Minnesota}
\affiliation{University of California, Los Angeles}
\affiliation{Michigan State University}
\affiliation{University of Twente}
\affiliation{TNO}

\address{University of Minnesota\\
Division of Biostatistics\\
Minneapolis, MN 55455\\
USA\\
\printead{e1}}

\address{University of California, Los Angeles\\
Department of Biostatistics\\
Los Angeles, CA 90095\\
USA\\
\printead{e2}}

\address{Michigan State University\\
Departments of Forestry and Geography\\
East Lansing, MI 48824\\
USA\\
\printead{e3}}

\address{University of Twente\\
Faculty of Geo-Information Science\\
 and Earth Observation (ITC)\\
Enschede, 7500 AE\\
The Netherlands\\
\printead{e4}}

\address{TNO\\
Department of Climate, Air and Sustainability\\
 Utrecht, 3508 TA\\
The Netherlands\\
\printead{e5}}

\end{aug}

\begin{abstract}

Particulate matter (PM) is a class of malicious environmental pollutants known to be detrimental to human health. 
Regulatory efforts aimed at curbing PM levels in different countries often require high resolution space-time maps that can identify red-flag regions exceeding statutory concentration limits. 
Continuous spatio-temporal Gaussian Process (GP) models can deliver maps depicting predicted PM levels and quantify predictive uncertainty. However, GP based approaches are usually thwarted by computational challenges posed by large datasets. We construct a novel class of scalable Dynamic Nearest Neighbor Gaussian Process (DNNGP) models 
that can provide a sparse approximation to any spatio-temporal GP (e.g., with non-separable covariance structures). 
The DNNGP we develop here can be used as a sparsity-inducing prior for spatio-temporal random effects in any Bayesian hierarchical model to deliver full posterior inference. Storage and memory requirements for a DNNGP model are linear in the size of the dataset thereby delivering massive scalability without sacrificing inferential richness. Extensive numerical studies reveal that the DNNGP provides substantially superior approximations to the underlying process than low rank approximations. Finally, we use the DNNGP to analyze a massive air quality dataset to substantially improve predictions of PM levels across Europe in conjunction with the LOTOS-EUROS chemistry transport models (CTMs).
\end{abstract}


\end{frontmatter}

\section{Introduction}\label{Sec: Intro}

Recent years have witnessed considerable growth in statistical modeling of large spatio-temporal datasets; see, for example, the recent books by \cite{geldigfuegut}, \cite{creswikle11} and \cite{ban14} and the references therein for a variety of methods and applications. An especially important domain of application for such models is environmental public health, where analysts and researchers seek map projections for ambient air pollutants measured at monitoring stations and understand the temporal variation in such maps. When inference is sought at the same scale as the observed data, one popular approach is to model the measurements as a time series of spatial processes. This approach encompasses standard time series models with spatial covariance structures \citep{pfief80a, pfief80b, stoff86} and dynamic models \citep{stroud01, gelf05} among numerous other alternatives. 

On the other hand, when inference is sought at arbitrary scales, possibly finer than the observed data (e.g., interpolation over the entire spatial and temporal domains), one constructs stochastic process models to capture dependence using spatio-temporal covariance functions \citep[see, e.g.,][]{cres99,kyria99,gnei02,stein05b,all03,ggg07}. In modeling ambient air pollution data, it is now customary to meld observed measurements with physical model outputs, where the latter can operate at much finer scales. Inference, therefore, is increasingly being sought at arbitrary resolutions using spatio-temporal process models \citep[see, e.g.,][]{gnei10} . Henceforth, we focus upon this setting.

While the richness and flexibility of spatio-temporal process models are indisputable, their computational feasibility and implementation pose major challenges for large datasets. Model-based inference usually involves the inverse and determinant of an $n\times n$ spatio-temporal covariance matrix $\bC(\btheta)$, where $n$ is the number of space-time coordinates at which the data have been observed. When $\bC(\btheta)$ has no exploitable structure, matrix computations typically require $\sim n^3$ floating point operations (flops) and storage in the order of $n^2$ which becomes prohibitive if $n$ is large. Approaches for modeling large covariance matrices in purely spatial settings include low rank models \citep[see, e.g.,][]{hig01, kam03, stein07, stein08, ban08,  cres08, cra08, rasm08, fin09, katzfussmultires}, covariance tapering \citep[see, e.g.,][]{fur06,kauf08,du09,shabytaper,bevil15}, approximations using Gaussian Markov Random Fields (GMRF) \citep[see, e.g.,][]{rueheld04}, products of lower dimensional conditional densities \citep{datta14,ve88,ve92,stein04}, and composite likelihoods \citep[e.g.,][]{eidsvik14}. Extensions to spatio-temporal settings include \citet{cres10}, \citet{fbg12} and \citet{katz12} who extend low-rank spatial processes to dynamic spatio-temporal settings
while \citet{gang14} who opts for a GMRF approach. All these methods use dynamic models defined on fixed temporal lags and do not lend themselves easily to continuous spatio-temporal domains. 

Spatio-temporal process models for continuous space-time modeling of large datasets have received relatively scant attention. 
\citet{bai12} and \citet{bevil12} used composite likelihoods for parameter estimation in a continuous space-time setup. Both these approaches, like their spatial analogues, have focused upon constructing computationally attractive likelihood approximations and have restricted inference only to parameter estimation. Uncertainty estimates are usually based on asymptotic results which are usually inappropriate for irregularly observed datasets. Moreover, prediction at arbitrary locations and time points proceeds by imputing estimates into an interpolator derived from a different process model. This remains expensive for large $n$ and may not reflect predictive uncertainty accurately. 

Our current work offers a highly scalable spatio-temporal process for continuous space-time modeling. 
We expand upon the neighbor-based conditioning set approaches outlined in purely spatial contexts by \cite{ve88}, \cite{stein04} and \cite{datta14}. We derive a scalable version of a spatio-temporal process, which we call the Dynamic Nearest-Neighbor Gaussian Process (DNNGP), using information from smaller sets of neighbors over space and time. This approach offers several benefits. The DNNGP is a well-defined spatio-temporal process whose realizations follow Gaussian distributions with sparse precision matrices. Thus, the DNNGP can act as a sparsity-inducing prior for spatio-temporal random effects in any Bayesian hierarchical model and enables full posterior inference considerably enhancing its applicability. Moreover, it can be used with any spatio-temporal covariance function, thereby accommodating non-separability. Being a process, importantly, allows the DNNGP to provide inference at arbitrary resolutions and, in particular, enables predictions at new spatial locations and time points in posterior predictive fashion. The DNNGP also delivers a substantially superior approximation to the underlying process than, for example, by low rank approximations \citep[see, e.g,][ for problems with low-rank approximations]{Stein13}. Finally, storage and memory requirements for a DNNGP model are linear in the number of observations, so it efficiently scales up to massive datasets without sacrificing richness and flexibility in modeling and inference.

The remainder of the article is organized as follows. In Section~\ref{sec:dpm10} we present the details of a massive environmental pollutants dataset and the need for a full Bayesian analysis. Section~\ref{sec:scale_dngpp} elucidates a general framework for building scalable spatio-temporal processes and uses it to construct a sparsity-inducing DNNGP over a spatio-temporal domain. Section~\ref{sec:dnngp} describes efficient schemes for fixed as well as adaptive neighbor selection, which are used in the DNNGP. Section~\ref{sec:dbayes} details a Bayesian hierarchical model with a DNNGP prior and its implementation using Markov Chain Monte Carlo (MCMC) algorithms. Section~\ref{sec:dillu} illustrates the performance of DNNGP using simulated datasets. In Section~\ref{sec:dpm10_analysis} we present a detailed analysis of our environmental pollutants dataset. We conclude the manuscript in Section~\ref{sec:dconc} with a brief review and pointers to future research.

\section{PM$_{10}$ pollution analysis}\label{sec:dpm10}
Exposure to airborne particulate matter (PM) is known to increase human morbidity and mortality \citep{BrunekreefH02a,LoomisEtAl13a,HoekEtAl13a}. In response to these and other health impact studies, regulatory agencies have introduced policies to monitor and regulate PM concentrations. For example, the European Commission's air quality standards limit PM$_{10}$ (PM$<$10 $\mu$m in diameter) concentrations to an average of 50 $\mu$g m$^{-3}$ over 24 hours and of 40 $\mu$g m$^{-3}$ over a year \citep{ECStandards15}.  Measurements made with standard instruments are considered authoritative, but these observations are sparse and maps at finer scales are needed for monitoring progress with mitigation strategies and for monitoring compliance.  Hence, accurately quantifying uncertainty in predicted PM concentrations is critical.

Substantial work has been aimed at developing regional scale chemistry transport models (CTM) for use in generating such maps. CTM's, however, have been shown to systematically underestimate observed PM$_{10}$ concentrations, due to lack of information and understanding about emissions and formation pathways~\citep{SternEtAl08a}. Empirical regression~\citep{BrauerEtAl11a} or geostatistical models~\citep{LloydA04a} are an alternative to CTM's for predicting continuous surfaces of PM$_{10}$. Empirical models may give accurate results, but are restricted to the conditions under which they are developed~\citep{MandersSH09a}. Assimilating monitoring station observations and CTM output, with appropriate bias adjustments, has been shown to provide improvements over using either data source alone ~\citep{vdKassteeleS06a,DenbyEtAl08a,CandianiEtAl13a,hamm2015}. In such settings, the CTM output enters as a model covariate and the measured station observations are the response. In addition to delivering more informed and realistic maps, analyses conducted using the models detailed in Section~\ref{sec:dbayes} can provide estimates of spatial and temporal dependence not accounted for by the CTM and hence provide insights useful for improving the transport models. 

We focus on the development and illustration of continuous space-time process models capable of delivering predictive maps and forecasts for PM$_{10}$ and similar pollutants using sparse monitoring networks and CTM output. We coupled observed PM$_{10}$ measurements across central Europe with corresponding output from the LOTOS-EUROS~\citep{SchaapEtAl08a} CTM. Inferential objectives included $i$) delivering continuous maps of PM$_{10}$ with associated uncertainty, $ii$) producing statistically valid forecast maps given CTM projections, and $iii$) developing inference on space and time residual structure, i.e., space and time lags, that can help identify lurking processes missing in the CTM.  The study area and dataset are the same as those used by \cite{hamm2015} and the reader is referred to that paper for more background information. Note that the current paper works with a 2-year time series, whereas \cite{hamm2015} focused on daily analysis of a limited number of pollution events.

\begin{figure}
\centering
\subfigure[April 3, 2009]{\includegraphics[trim={0 0.5cm 0 1cm},clip,width=2.25in]{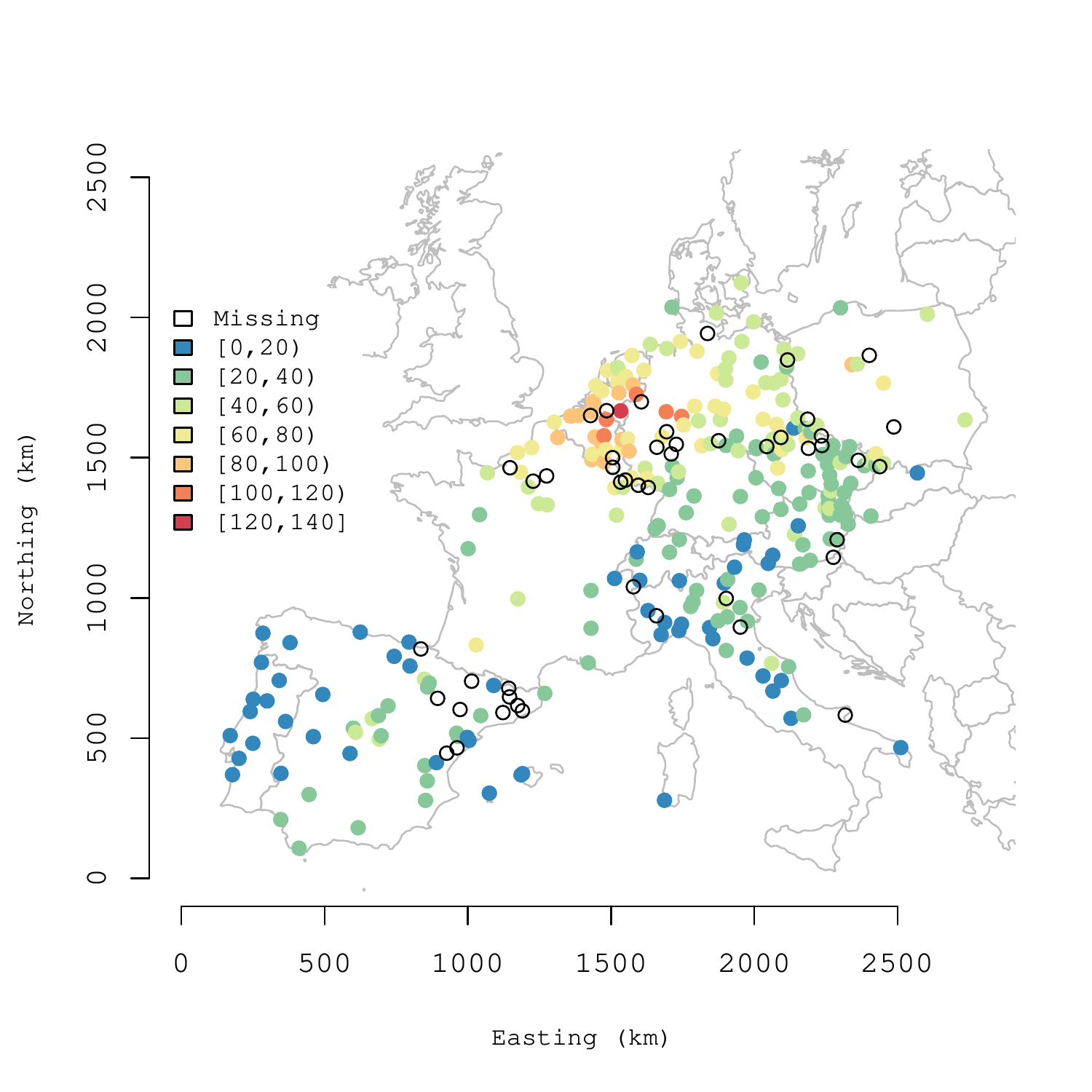}\label{dec-01-09-y-obs}} 
\hfil
\subfigure[April 5, 2009]{\includegraphics[trim={0 0.5cm 0 1cm},clip,width=2.25in]{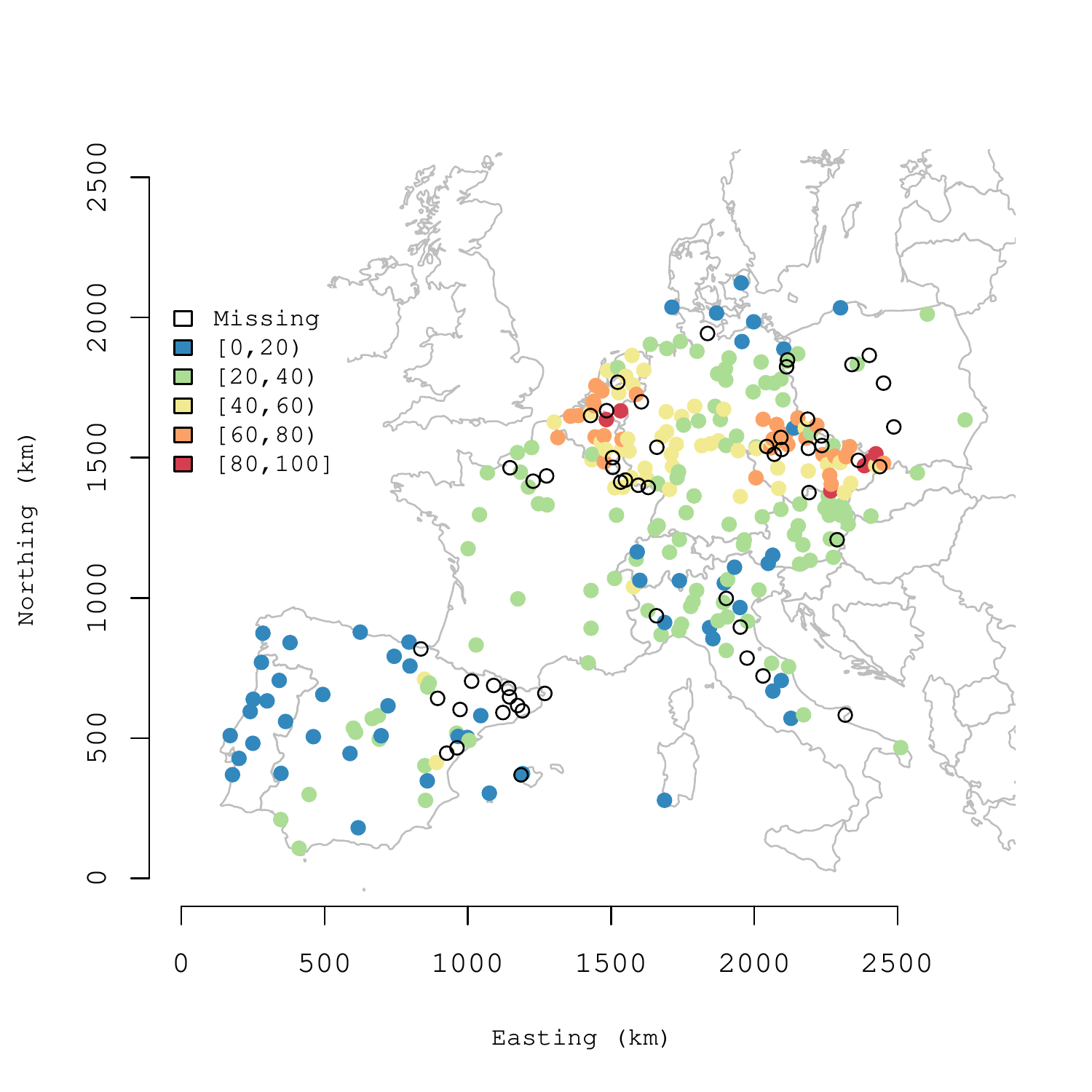}\label{dec-30-09-y-obs}} 
\caption{Observed PM$_{10}$ $\mu$g m$^{-3}$ for two example dates.}\label{y-obs}
\end{figure}

\subsection{Study area}\label{sec:s.site}
The study domain comprises mainland European countries with a substantial number of available PM$_{10}$ observations. The countries included were Portugal, Spain, Italy, France, Switzerland, Belgium, The Netherlands, Germany, Denmark, Austria, Poland, The Czech Republic, Slovakia and Slovenia. All data were projected to the European Terrestrial Reference System 1989 (ETRS) Lambert Azimuthal Equal-Area (LAEA) projection which gives a coordinate reference system for the whole of Europe. 

\subsection{Observed measurements}\label{sec:s.obs}
Air quality observations for the study area were drawn from the Airbase (\emph{Air} quality data\emph{base})\footnote{http://acm.eionet.europa.eu/databases/airbase (accessed 26 September 2014)}. Daily PM$_{10}$ concentrations were extracted for January 1 2008 through December 30 2009 resulting in a maximum of $M$=730 observations at each of $N=308$ monitoring stations. Airbase daily values are averaged over the within-day hourly values when at least 18 hourly measurements are available, otherwise no data are provided. Airbase monitors are classified by type of area (rural, urban, suburban) and by type (background, industrial, traffic or unknown). Only rural background monitors were used in our study. This is common for comparing measured observations to coarse resolution CTM simulations~\citep{DenbyEtAl08a}. Monitoring stations above 800 m altitude were also excluded. These tend to be located in areas of variable topography and the accuracy of the CTM for locations that shift from inside to outside the atmospheric mixing layer is known to be poor. No further quality control was performed on the data. The locations of the 308 stations used in the subsequent analysis are shown in Figure~\ref{y-obs} with associated observed and missing PM$_{10}$ for two example dates. Of the 224,840 ($M\times N$) potential observations across 730 day time series and 308 stations, 41,761 observations were missing due to sensor failure or removal, and post-processing removal by Airbase. These missing values were predicted using the proposed models.

\subsection{LOTOS-EUROS CTM data}\label{sec:s.LE}
LOTOS-EUROS (v1.8) is a 3D CTM that simulates air pollution in the lower troposphere. The simulator's geographic projection is longitude-latitude at a resolution of $0.50^{\circ}$ longitude $\times 0.25^{\circ}$ latitude (approximately 25 km $\times$ 25 km). LOTOS-EUROS simulates the evolution of the components of particulate matter separately. Hence, this CTM incorporates the dispersion, formation and removal of sulfate, nitrate, ammonium, sea salt, dust, primary organic and elemental carbon and non-specified primary material, although it does not incorporate secondary organic aerosol. \cite{HendriksEtAl13a} provide a detailed description of LOTOS-EUROS.

The hour-by-hour calculations of European air quality in 2008-2009 were driven by the European Centre for Medium Range Weather Forecasting (ECMWF). Emissions were taken from the MACC (Monitoring Atmospheric Composition and Climate) emissions database \citep{PouliotEtAl12a}. Boundary conditions were taken from the global MACC service~\citep{FlemmingEtAl09a}. The LOTOS-EUROS hourly model output was averaged to daily mean PM$_{10}$ concentrations. LOTOS-EUROS grid cells that were spatially coincident with the Airbase observations were extracted and used as the covariate in the subsequent model.

CTM grid cell values nearest to station locations were used for subsequent model development. No attempt was made to match the spatial support (resolution) of the CTM simulations and station observations. The support of the CTM is 25 km, but the support of the observations is vague. Rural background observations were deliberately chosen because they are distant from urban areas and pollution sources.  They are, therefore considered representative of background, ambient pollution conditions and appropriate for matching with moderate resolution CTM-output \citep{DenbyEtAl08a,hamm2015}.  This assumption is further backed up by empirical studies indicating that PM$_{10}$ concentrations are dominated by rural background values even in urban areas \citep{EeftensEtAl12a}.

\section{Scalable Dynamic Nearest-Neighbor Gaussian Processes}\label{sec:scale_dngpp} 
Let $\{w(\ell) : \ell \in \calL\}$ be a zero-centered continuous spatio-temporal process \citep[see, e.g.,][for details]{gnei10}, where $\calL = \calS\times \calT$ with $\calS \subset \Re^d$ (usually $d=2$ or $3$) is the spatial region, $\calT \subset [0,\infty)$ is the time domain and $\ell = (\bs,t)$ is a space-time coordinate with spatial location $\bs \in \calS$ and time point $t\in \calT$. Such processes are specified with a spatio-temporal \emph{covariance function} $\mbox{Cov}\{w(\ell_i), w(\ell_j)\} = C(\ell_i, \ell_j\given \btheta)$. For any finite collection $\calU=\{ \ell_1, \ell_2, \ldots, \ell_n \}$ in $\calL$, let $\bw_\calU = (w(\ell_1)), w(\ell_2), \ldots, w(\ell_n))'$ be the realizations of the process over $\calU$. Also, for two finite sets $\calU$ and $\calV$ containing $n$ and $m$ points in $\calL$, respectively, we define the $n\times m$ matrix $\bC_{\calU,\calV}(\btheta)=\mbox{Cov}(\bw_\calU,\bw_\calV \given \btheta)$, where the covariances are evaluated using $C(\cdot, \cdot\given \btheta)$. When $\calU$ or $\calV$ contains a single point, $\bC_{\calU,\calV}$ is a row or column vector. A valid spatio-temporal covariance function ensures that $\bC_{\calU,\calU}(\btheta)$ is positive definite for any finite set $\calU$. In particular, for spatio-temporal Gaussian processes, $\bw_{\calU}$ has a multivariate normal distribution $N(\bzero, \bC_{\calU,\calU}(\btheta))$ and the $(i,j)$th element of $\bC_{\calU,\calU}(\btheta)$ is $C(\ell_i,\ell_j\given \btheta)$.  

Storage and computations involving $\bC_{\calU,\calU}(\btheta)$ can become impractical when $n$ is large relative to the resources available. For full Bayesian inference on a continuous domain, we seek a scalable (in terms of flops and storage) spatio-temporal Gaussian process that will provide an excellent approximation to a full spatio-temporal process with any specified covariance function. We outline a general framework that first uses a set of points in $\calL$ to construct a computationally efficient approximation for the random field and extends the finite dimensional distribution over this set to a process. To ease the notation, we will suppress the explicit dependence of matrices and vectors on $\btheta$ whenever the context is clear.   

Let $\calR = \{\ell_1^*, \ell_2^*,\ldots,\ell_r^*\}$ be a fixed finite set of $r$ points in $\calL$. We refer to $\calR$ as a \emph{reference set}. We construct a spatio-temporal process $w(\ell)$ on $\calL$ by first specifying $\bw_{\calR}= (w(\ell_1^*), w(\ell_2^*),\ldots,w(\ell_r^*))' \sim N(\bzero, \bK(\btheta))$, where $\bK(\btheta)$ is any $r\times r$ positive definite matrix and then defining
\begin{equation}\label{eq:sgp}
w(\ell) = \sum_{i=1}^r a_{i}(\ell) w(\ell_i^*)  + \eta(\ell) \mbox{ for any } \ell \notin \calR\; ,
\end{equation}
where 
$\eta(\ell)$ is a zero-centered Gaussian process independent of $\bw_\calR$ and such that $\mbox{Cov}\{\eta(\ell_i), \eta(\ell_j)\}=0$ for any two distinct points in $\calL$.

Observe that $w(\ell)$ in (\ref{eq:sgp}) is a well defined spatio-temporal Gaussian process on $\calL$ for \emph{any} choice of $a_{i}(\ell)$'s, as long as $\bK(\btheta)$ is positive definite. 
For example, $w(\ell)$ is a Gaussian process with covariance function $C(\cdot,\cdot\given\btheta)$ if we set $\bK(\btheta)=\bC_{\calR,\calR}(\btheta)$, $\ba(\ell) = \bC_{\calR,\calR}^{-1}\bC_{\calR,\ell}$ where $\ba(\ell)$ is $r\times 1$ with elements $a_{i}(\ell)$, and $\eta(\ell) \stackrel{ind}{\sim} N\left(0,  C(\ell,\ell \given \btheta)- \bC_{\ell,\calR}\bC_{\calR,\calR}^{-1}\bC_{\calR,\ell}\right)$. Now (\ref{eq:sgp}) represents the `kriging' equation for a location $\ell$ based on observations over $\calR$ \citep{creswikle11}. Dimension reduction can be achieved with suitable choices for $\bK(\btheta)$ and $\ba(\ell)$. Low-rank spatio-temporal processes emerge when we choose $\calR$ to be a smaller set of `knots' (or `centers'). Additionally, specifying $\eta(\ell)$ to be a diagonal or sparse residual process yields $w(\ell)$ to be a non-degenerate (or bias-adjusted) low rank Gaussian Process \citep{ban08,fin09,sang12}.

Because of demonstrably impaired inferential performance of low-rank models in purely spatial contexts at scales similar to ours \citep[see, e.g.,][]{Stein13,datta14}, we use the framework in (\ref{eq:sgp}) to construct a class of sparse spatio-temporal process models. To be specific, let the reference set $\calR$ be an enumeration of $r=MN$ points in $\calL$, so that each $\ell^*_i$ in $\calR$ corresponds to some $(\bs_j,t_k)$ for $j=1,2,\ldots,N$ and $k=1,2,\ldots,M$. For any $\ell^*_{i} = (\bs_j,t_k)$ in $\calR$ we define a \emph{history set} $H(\ell_{i}^*)$ as the collection of all locations observed at times before $t_k$ and of all points at time $t_k$ with spatial locations in $\{\bs_1, \bs_2, \ldots, \bs_{j-1}\}$. Thus, $H(\ell_{i}^*) =\{(\bs_p, t_q) \given  p=1,2,\ldots,N,\, q= 1,2,\ldots,(k-1) \} \cup \{ (\bs_p,t_k) \given p=1,2,\ldots,(j-1)\}$. For any location $\ell_{i}^*$ in $\calR$, let $N(\ell_{i}^*)$ be a subset of the history set $H(\ell_i^*)$. Also, for any location $\ell\notin\calR$, let $N(\ell)$ denote any finite subset of $\calR$. We refer to the sets $N(\ell)$ as a `neighbor set' for the location $\ell$ and describe their construction later. 

We now turn to our choices for $\bK(\btheta)$ and $\ba(\ell)$ in (\ref{eq:sgp}). Let $w(\ell)\sim GP(0, C(\cdot,\cdot\given\btheta))$. We choose $\bK(\btheta)$ to effectuate a sparse approximation for the joint density of the realizations of $w(\ell)$ over $\calR$, i.e., $N(\bw_{\calR}\given \bzero, \bC_{\calR,\calR}(\btheta))$. Adapting the ideas underlying likelihood approximations in \cite{ve88} and \cite{datta14}, we specify $\bK(\btheta)$ to be the $r \times r$ matrix such that 
\begin{align}\label{eq:nngpref}
N(\bw_{\calR}\given \bzero, \bC_{\calR,\calR}(\btheta)) &= \prod_{i=1}^r p(w(\ell_{i}^*) \given \bw_{H(\ell_{i}^*)}) \nonumber\\
&\approx \prod_{i=1}^r p(w(\ell_{i}^*) \given \bw_{N(\ell_{i}^*)} ) = N(\bw_\calR \given \bzero, \bK(\btheta))\; .
\end{align}
Here, $H(\ell_1^*)$ is the empty set (hence, so is $N(\ell_{1}^*)$) and $p(w(\ell_{1}^*) \given \bw_{H(\ell_{1}^*)}) = p(w(\ell_{1}^*) \given \bw_{N(\ell_{1}^*)} ) = p(w(\ell_{1}^*))$. The underlying idea behind the approximation in Equation~\ref{eq:nngpref} is to compress the conditioning sets from $H(\ell_i^*)$ to $N(\ell_i^*)$ so that the resulting approximation is a multivariate normal distribution with a sparse precision matrix $\bK^{-1}$. This implies 
\begin{equation}\label{eq:condexp}
{E}[w(\ell_{i}^*)\given \bw_{H_{\ell_i^*}}] = \mbox{E}[w(\ell_{i}^*)\given \bw_{N(\ell_{i}^*)}] = \ba_{N(\ell_i^*)}'\bw_{N(\ell_i^*)}
\end{equation} 
where $\ba_{N(\ell_i^*)} = \bC_{N(\ell_i^*),N(\ell^*_i)}^{-1}\bC_{N(\ell^*_i),\ell^*_i}$. Also, $\bK$ is determined by $\bC_{\calR,\calR}$ because $\bK^{-1} = \bV'\bF^{-1}\bV$, where $\bF$ is a diagonal matrix with diagonal entries $f_{\ell_i^*} = \mbox{Var}(w(\ell_i^*)$  $\given \bw_{N(\ell_i^*)})= C(\ell_i^*,\ell_i^* \given \btheta) - \bC_{\ell^*_i,N(\ell^*_i)}\bC_{N(\ell_i^*),N(\ell^*_i)}^{-1}\bC_{N(\ell^*_i),\ell^*_i}$ and $\bV$ is the $r\times r$ matrix with entries $v_{i,j}$ such that $v_{i,i}=1$ and $v_{i,j}=0$ whenever $\ell_i^*\notin N(\ell_j^*)$. The remaining entries in column $j$ of $\bV$ are specified by setting the subvector $\bV_{c(\ell_j^*),j} = - \ba_{N(\ell_j^*)}$, where $c(\ell_j^*) = \{i\given \ell_i^*\in N(\ell_j^*)\}$. 
If $m (<< r)$ denotes the limiting size of the neighbor sets $N(\ell)$, then the columns of $\bV$ are sparse with at most $m+1$ non-zero elements. Consequently, $\bK^{-1}$ has at most $O(rm^2)$ non-zero elements \citep[this is the spatial-temporal analogue of the result in ][]{datta14}. Hence, the approximation in (\ref{eq:nngpref}) produces a sparsity-inducing proper prior distribution for the spatio-temporal random effects over $\calR$ that closely approximates the realizations from a $GP(0, C(\cdot,\cdot\given\btheta))$. 

Turning to the vector of coefficients $\ba(\ell)$ in (\ref{eq:sgp}), 
we extend the idea in (\ref{eq:condexp}) to any point $\ell\notin \calR$ by requiring that $\mbox{E}[w(\ell)\given \bw_{\calR}] = \mbox{E}[w(\ell)\given \bw_{N(\ell)}]$. This is achieved by setting $a_{i}(\ell) = 0$ in (\ref{eq:sgp}) whenever $\ell^*_i\notin N(\ell)$ for any point $\ell\notin \calR$. Hence, if $N(\ell)$ contains $m$ points, then at most $m$ of the elements in the $r\times 1$ vector $\ba(\ell)$ can be nonzero. These nonzero entries are determined from the above conditional expectation given $N(\ell)$. To be precise, 
if $\ba_{N(\ell)}$ is the $m\times 1$ vector of these $m$ entries, then we solve $\bC_{N(\ell),N(\ell)}\ba_{N(\ell)} = \bC_{N(\ell),\ell}$ for $\ba_{N(\ell)}$.  Also note that $\ba'(\ell)\bw_{\calR} = \ba_{N(\ell)}'\bw_{N(\ell)}$. Finally, to complete the process specifications in (\ref{eq:sgp}), we specify $\eta(\ell)\stackrel{ind}{\sim}N(0,f_{\ell})$, where $f_{\ell} = \mbox{Var}(w(\ell)\given \bw_{N(\ell)}) = C(\ell,\ell \given \btheta) - \bC_{\ell,N(\ell)}\bC_{N(\ell),N(\ell)}^{-1}\bC_{N(\ell),\ell}$. The covariance function $\tildeC(\cdot, \cdot \given \btheta)$ of the resulting Gaussian Process is given by:
\begin{align}\label{eq:nngpcov}
\tildeC(\ell_i,\ell_j \given \btheta) =  \left\{
\begin{array}{ll}
K_{p,q} \mbox{ if } \ell_i=\ell_p^* \mbox{ and } \ell_j=\ell_q^* \mbox{ are both in } \calR \\
\ba'(\ell_i)\bK_{*q} \mbox{ if } \ell_i \notin \calR \mbox{ and } \ell_j=\ell_q^* \in \calR \\
\ba'(\ell_i)\bK\ba(\ell_j) + I(\ell_i=\ell_j)f_{\ell_i} \mbox{ if } \ell_i \notin \calR \mbox{ and } \ell_j \notin \calR\; ,
\end{array}
 \right.
\end{align}
where $K_{p,q}$ is element $(p,q)$ and $\bK_{*q}$ is column $q$ in $\bK$.
	
 
Owing to the sparsity of $\bK^{-1}$, the likelihood $N(\bw_\calR \given \bzero, \bK)$ can be evaluated using $O(rm^3)$ flops for any given $\btheta$. Substantial computational savings accrue because $m$ is usually very small (also see later sections).  
Furthermore as $\eta(\ell)$ yields a diagonal covariance matrix and $\ba(\ell)$ has at most $m$ non-zero elements, for any finite set $\calV$ outside $\calR$, the flop count for computing the density $p(\bw_\calV \given \bw_\calR, \btheta)$ will be linear in the size of $\calV$. We have now constructed a scalable spatio-temporal Gaussian Process from a \emph{parent} spatio-temporal $GP(0, C(\cdot, \cdot\given \btheta))$ using small neighbor sets $N(\ell)$. We denote this \emph{Dynamic Nearest Neighbor Gaussian Process} (DNNGP) as $DNNGP(0, \tildeC(\cdot, \cdot \given \btheta))$, where $\tildeC(\cdot, \cdot \given \btheta))$ denotes the covariance function of this new GP.

\section{Constructing Neighbor-Sets}\label{sec:dnngp}
\subsection{Simple Neighbor Selection}\label{sec:simple}
Spatial correlation functions usually decay with increasing inter-site distance, so the set of nearest neighbors based on the inter-site distances represents locations exhibiting highest correlation with the given location. This has motivated use of nearest neighbors to construct these small neighbor sets \citep{ve88, datta14}. On the other hand, spatio-temporal covariances between two points typically depend on the spatial as well as the temporal lag between the points. To be specific, non-separable isotropic spatio-temporal covariance functions can be written as $C((\bs_1,t_1),(\bs_2,t_2) \given \btheta) = C(h,u \given \btheta)$ where $h=|| \bs_1-\bs_2 || $ and $u=|t_1-t_2|$. This often precludes defining any universal distance function $d: (\calS \times \calT)^2 \rightarrow \mathbb{R}^+$ such that $C((\bs_1,t_1),(\bs_2, t_2) \given \btheta)$ will be monotonic with respect to $d((\bs_1,t_1),(\bs_2,t_2))$ for all choices of $\btheta$.

In the light of the above discussion, we define ``nearest neighbors'' in a spatio-temporal domain using the spatio-temporal covariance function itself as a proxy for distance. To elucidate, for any three points $(\bs_1,t_1)$, $(\bs_2,t_2)$ and $(\bs_3,t_3)$, we say that $(\bs_1,t_1)$ is nearer to $(\bs_2,t_2)$ than to $(\bs_3,t_3)$ if $C((\bs_1,t_1),(\bs_2,t_2) \given \btheta) \; > C((\bs_1,t_1),(\bs_3,t_3) \given \btheta)$. Subsequently, this definition of ``distance'' is used to find $m$ nearest neighbors for any location.

\begin{figure}[!t]
\begin{center}
\subfigure[$N_{\theta=1}(s_i,t_j)$]{\includegraphics[width=4cm, trim={5cm 0cm 4cm 0cm},clip]{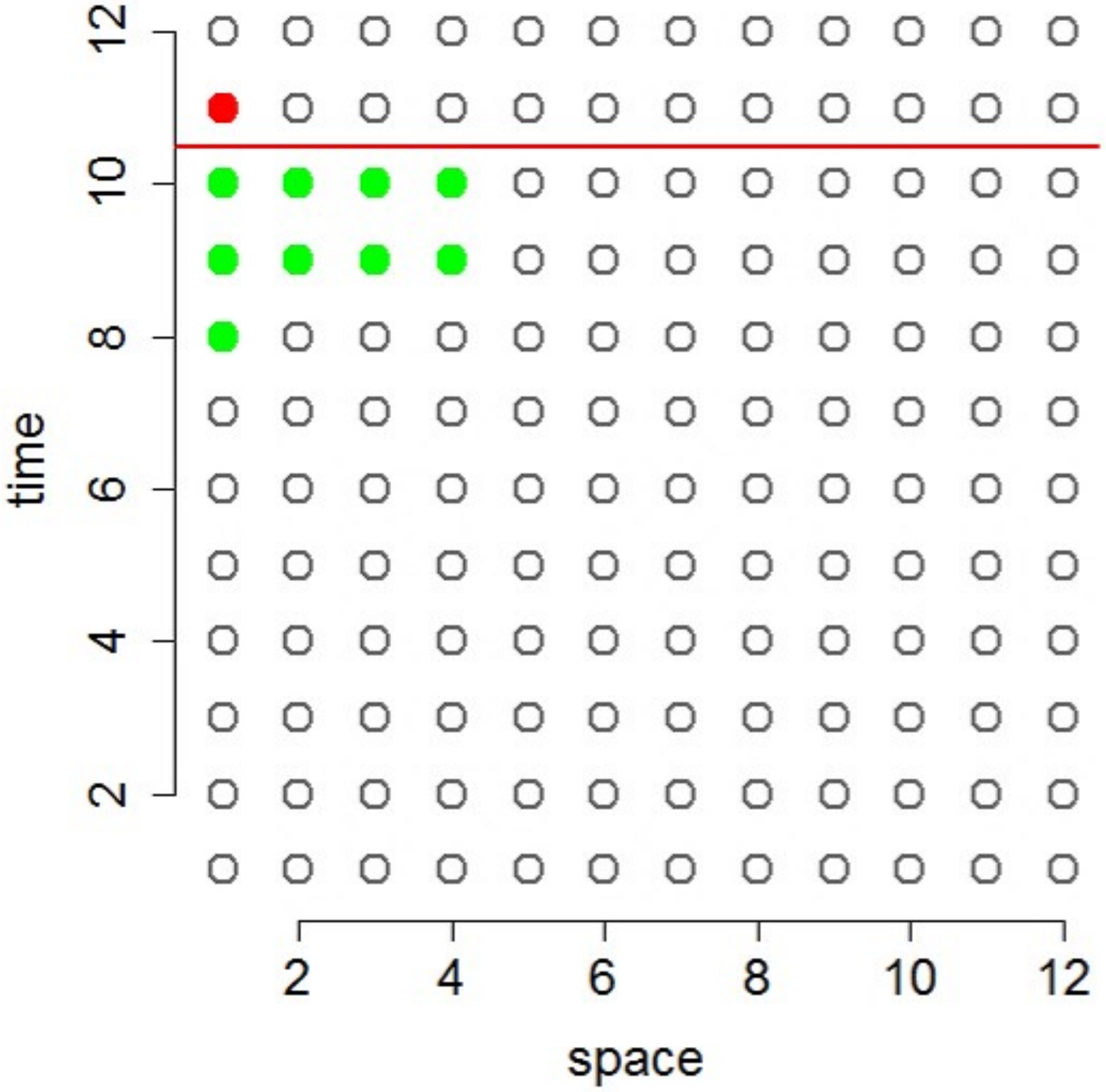}\label{fig:nei1}}
\subfigure[$N_{\theta=2}(s_i,t_j)$]{\includegraphics[width=4cm, trim={5cm 0cm 4cm 0cm},clip]{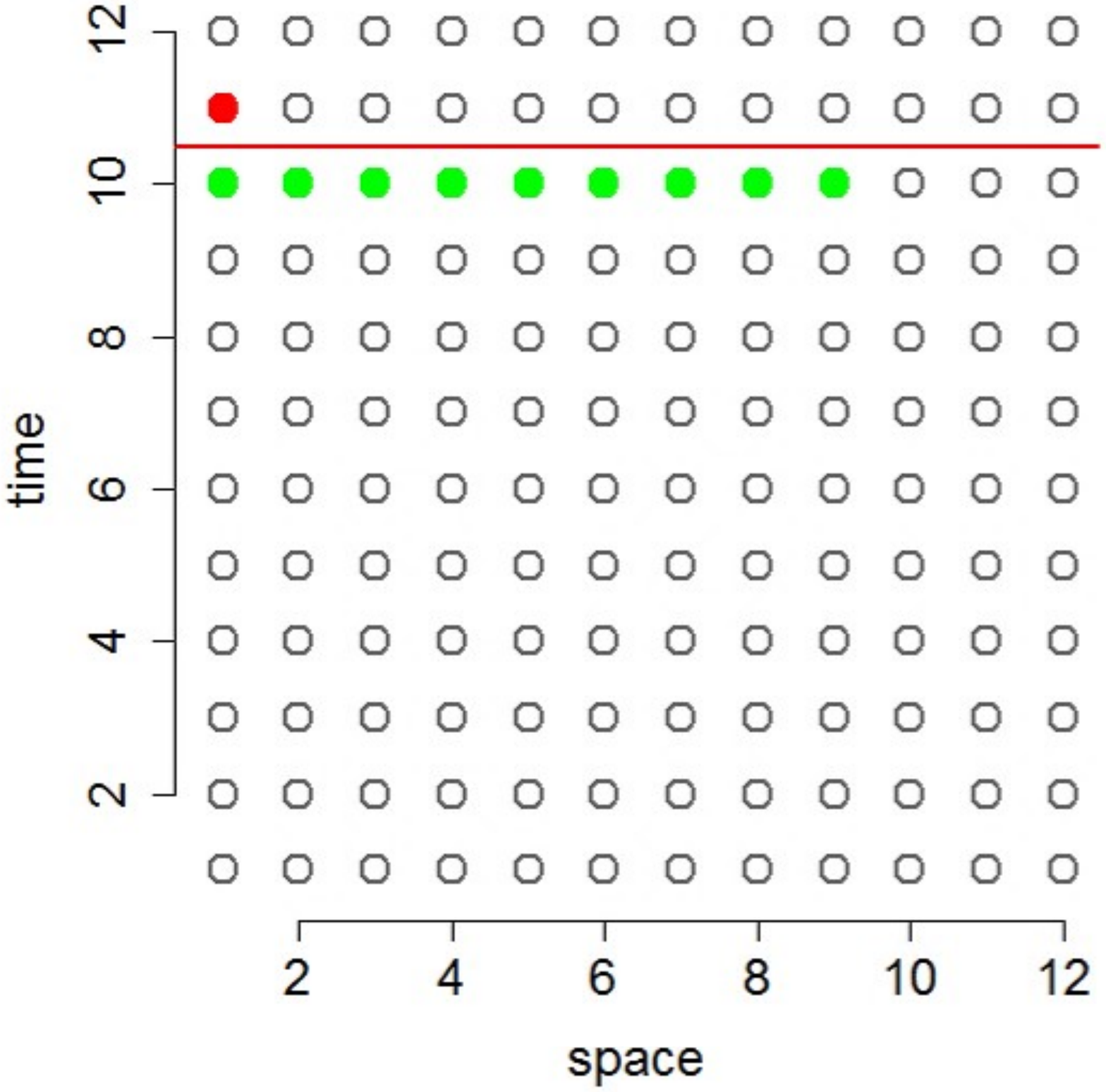}\label{fig:nei2}}
\subfigure[Simple neighbor sets]{\includegraphics[width=4cm, trim={5cm 0cm 4cm 0cm},clip]{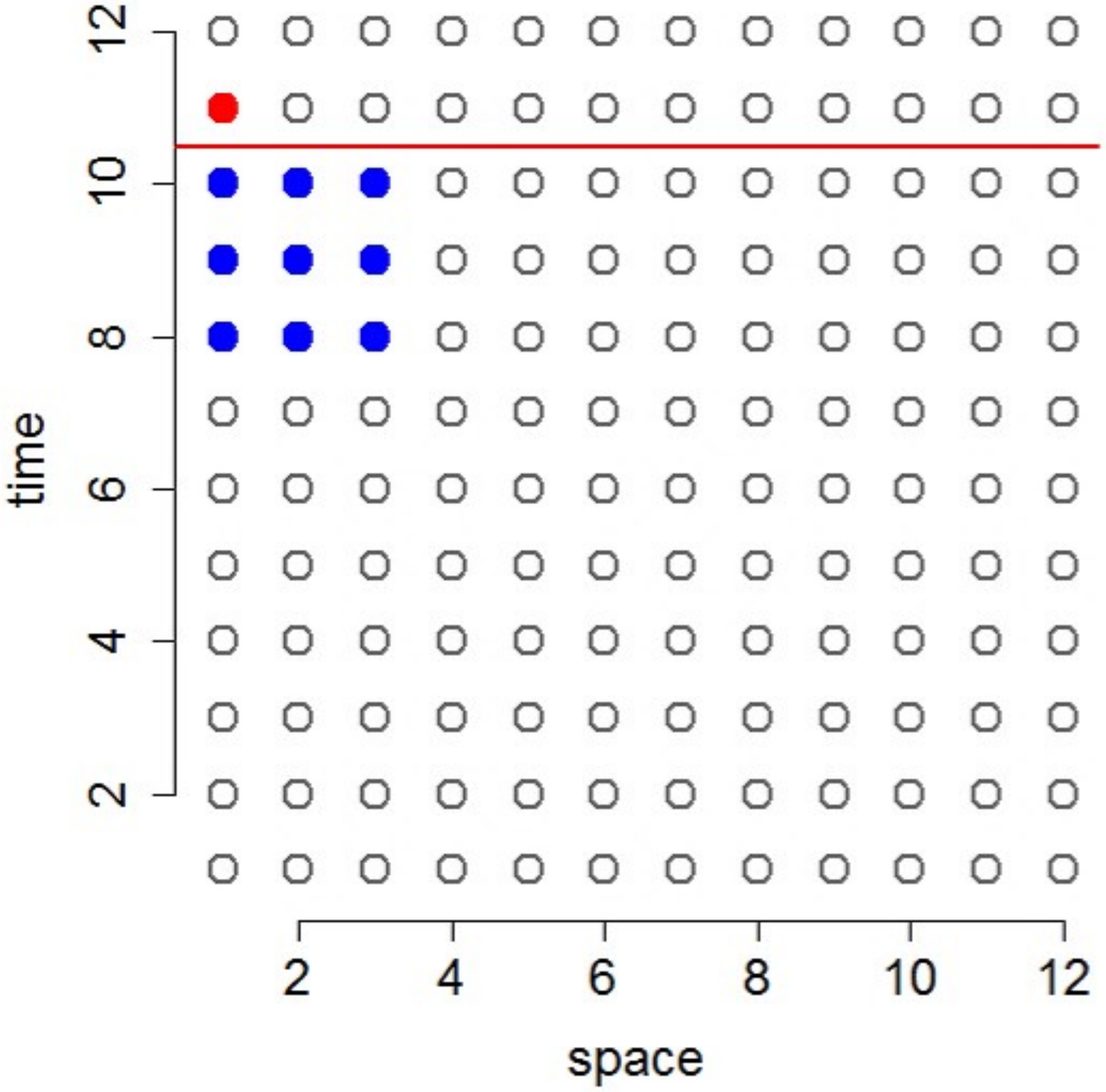}\label{fig:simple}}
\caption{True and simple neighbor sets for a $ 12 \times 12$ spatio-temporal dataset with one-dimensional spatial domain and covariance function $C((s_1,t_1),(s_2,t_2) \given \theta) = \exp (- |s_1-s_2|^2 - \theta |t_1-t_2|^2)$. All points below the red horizontal line constitute the history set for the red point $(s_i,t_j)$. Green points denote $N_\theta(s_i,t_j)$ -- the sets of  $m (=9)$ true nearest neighbors with $\theta=1$ (figure (a)) and $\theta=2$ (figure (b)). The blue points in figure (c) denotes the simple neighbor set.}\label{fig:nei}
\end{center}
\end{figure}

Of course, this choice of nearest neighbors depends on the choice of the covariance function $C$ and $\btheta$. Since the purpose of the DNNGP is to provide a scalable approximation of the parent GP, we always choose $C(\cdot, \cdot \given \btheta)$ to be same as the covariance function of the parent GP. However, for every location $(\bs_i,t_j)$, its neighbor set, denoted by $N_\theta(\bs_i,t_j)$, still depends on $\btheta$. This is illustrated in Figures~\ref{fig:nei1} and \ref{fig:nei2} which shows how neighbor sets can differ drastically based on the choice of $\btheta$. 

In most applications, $\btheta$ is unknown precluding the use of these newly defined neighbor sets $N_\theta(\bs_i,t_j)$ to construct the DNNGP. We propose a simple intuitive method to construct neighbor sets. We choose $m$ to be a perfect square and construct a simple neighbor set of size $m$ using $\sqrt m$ spatial nearest neighbors and $\sqrt m$ temporal nearest neighbors. Figure \ref{fig:simple} illustrates the simple neighbor set of size $m=9$ for the red point. 
In order to formally define the simple neighbor sets, we denote $S=\{\bs_1,\bs_2,\ldots,\bs_N\}$, $S_i=\{\bs_1,\bs_2,\ldots,\bs_{i-1}\}$ and $T=\{t_1,t_2,\ldots,t_M\}$. Furthermore, for any finite set of spatial locations $V \subseteq S$, let $A(\bs,V,m)$ denote the set of $m$ nearest neighbors in $V$ for the location $\bs$. For any point $(\bs_i,t_j) \in \calR$ we define the simple neighbor sets
\begin{equation}\label{eq:simple}
N(\bs_i,t_j)=\bigcup^{\sqrt m-1}_{k=1} \{(\bs,t_{j-k}) \given \bs \in A (\bs_i,S,\sqrt m)\} \bigcup  \{(\bs,t_j) \given \bs \in A (\bs_i,S_i,\sqrt m) \}
\end{equation}
The above construction implies that the neighbor set for any point in $\calR$ consists of $\sqrt m$ spatial nearest neighbors of the preceding $\sqrt m$ time points. For arbitrary $(\bs,t)\notin \calR$, $N(\bs,t)$ is simply defined as the Cartesian product of $\sqrt m$ nearest neighbors for $\bs$ in $\calS$ with $\sqrt m$ nearest neighbors of $t$ in $\calT$. 

In many applications, one desirable property of the spatio-temporal covariance functions is {\em natural monotonicity}, i.e. $C(h,u)$ is decreasing in $h$ for fixed $u$ and decreasing in $u$ for fixed $h$. All Mat\`ern-based space-time separable covariances and many non-separable classes of covariance functions possess this property \citep{stein2013,omidi15}. If $C(\cdot,\cdot \given \btheta)$ possesses natural monotonicity, then $N(\bs_i,t_j)$ defined in Equation~\ref{eq:simple} is guaranteed to contain at least $\sqrt m-1$ nearest neighbors of $(\bs_i,t_j)$ in $H(\bs_i,t_j)$. Thus, the neighbor sets defined above do not depend on any parameter and, for any value of $\btheta$, will contain a few nearest neighbors. 

\subsection{Adaptive Neighbor Selection}\label{sec:adapt}
The simple neighbor selection scheme described in Section~\ref{sec:simple} does not depend on $\btheta$ and is undoubtedly useful for fast implementation of the DNNGP. However, for some  values of $\btheta$, the neighbor sets may often consist of very few nearest neighbors. This issue is illustrated in Figure~\ref{fig:nei} where the simple neighbor set (blue points) contained $7$ out of $9$ true nearest neighbors (green points) for $\theta=1$ but only $3$ out of $9$ true nearest neighbors for $\theta=2$. 
We see that for different choices of the covariance parameters the simple neighbor sets contain different proportions of the true nearest neighbors. The problem is exacerbated in extreme cases with variation only along the spatial or temporal direction. In such cases, the neighbor sets defined in (\ref{eq:simple}) will contain only about $\sqrt m$ nearest neighbors and $m-\sqrt m$ uncorrelated points.

Ideally, if $\btheta$ was known, one could have simply evaluated the pairwise correlations between any point $(\bs_i,t_j)$ in $\calR$ and all points in its history set $H(\bs_i,t_j)$ to obtain $N_\theta(\bs_i,t_j)$ --- the set of $m$ true nearest neighbors. In practice, however, we encounter a computational roadblock because $\btheta$ is unknown and for every new value of $\btheta$ in an iterative optimizer or Markov Chain Monte Carlo sampler, we need to redo the search for the neighbor sets within the history sets. As the history sets are typically large this is computationally challenging. For example, in Figure \ref{fig:nei}, the history set for the red point is composed of all points below the red horizontal line. So, evaluating the pairwise correlations required for updating neighbor sets of all points in $\calR$ and $n$ datapoints outside $\calR$, will use $O(r^2+nr)$ flops at each iteration. The reference set $\calR$ is typically chosen to match the scale of the observed dataset to achieve a reasonable approximation of the parent GP by DNNGP. Hence, for large datasets this updating becomes a deterrent. In fact, \citet{ve88} and \citet{stein04} admit that this challenge has inhibited the use of correlation based neighbor sets in a spatial setting. \citet{jones97} permitted locations within a small prefixed temporal lag of a given location to be eligible for neighbors. However, this assumption will fail to capture any long term temporal dependence present in the datasets.

We now provide an algorithm that efficiently updates the neighbor sets after every update of $\btheta$. The underlying idea is to restrict the search for the neighbor sets to carefully constructed small subsets of the history sets. These small {\em eligible sets} $E(\bs_i,t_j)$ are constructed in such a manner that, despite being much smaller than the history sets, they are guaranteed to contain the true nearest neighbor sets $N_\theta(\bs_i,t_j)$ for all choices of the parameter $\btheta$. So, for each $\btheta$ we can evaluate the pairwise correlations between $(\bs_i,t_j)$ and only the points in $E(\bs_i,t_j)$ and still find the true set of $m$-nearest neighbors.  

\begin{figure}[!t]
\begin{center}
\subfigure[Ineligible point]{\includegraphics[width=4cm, trim={5cm 0cm 4cm 0cm},clip]{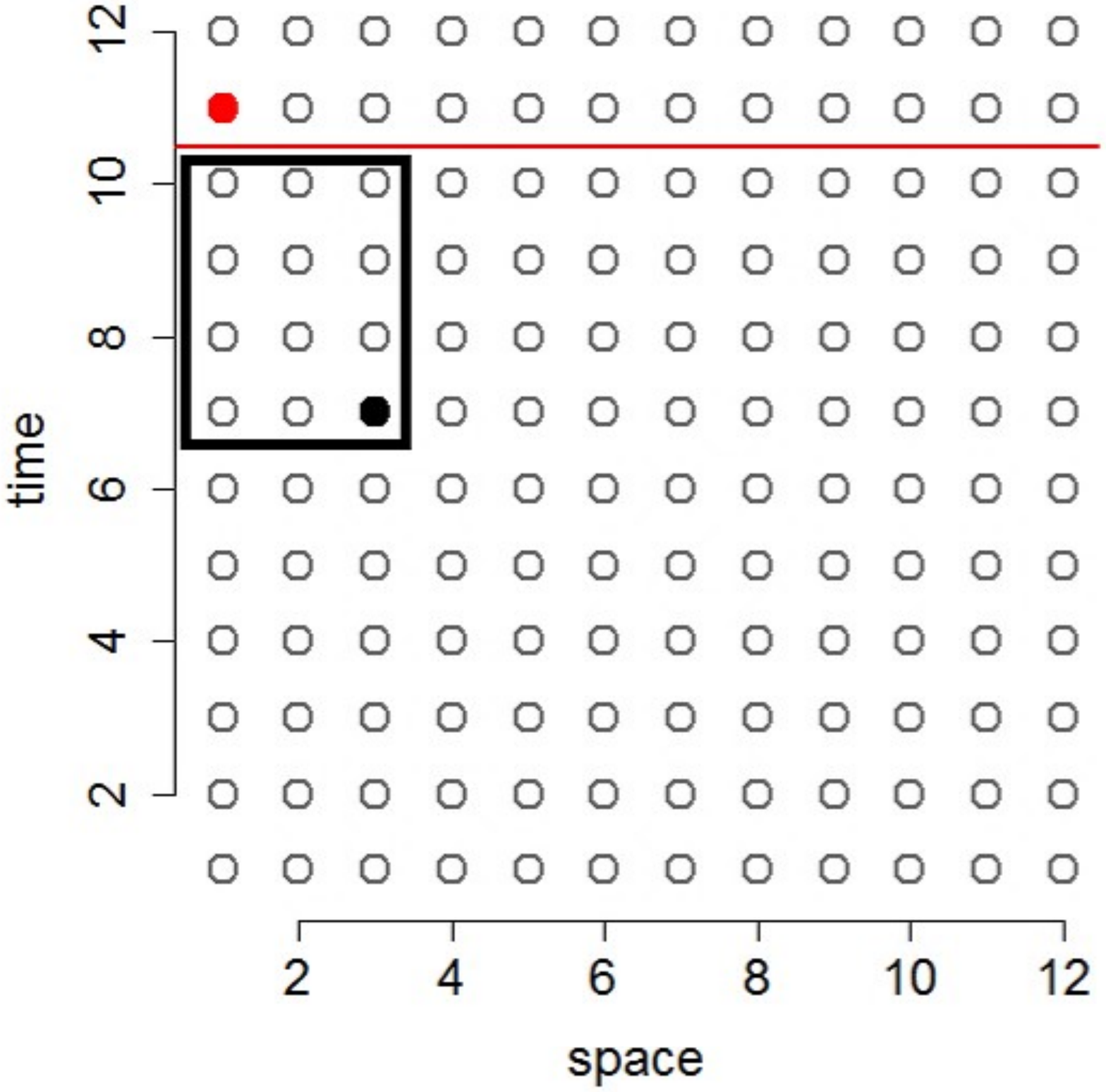}\label{fig:const1}}
\subfigure[Eligible point]{\includegraphics[width=4cm, trim={5cm 0cm 4cm 0cm},clip]{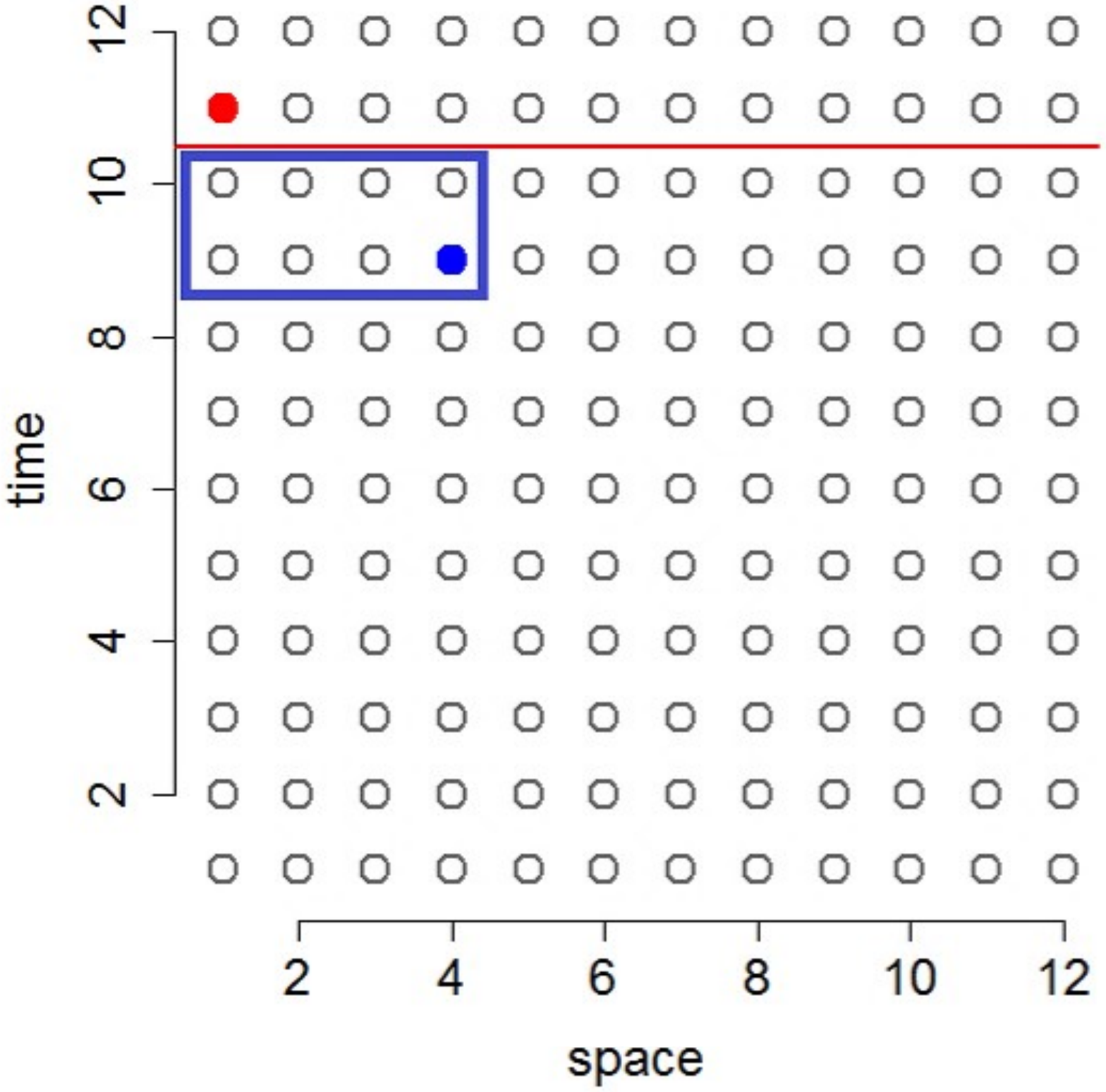}\label{fig:const2}}
\subfigure[Full eligible set]{\includegraphics[width=4cm, trim={5cm 0cm 4cm 0cm},clip]{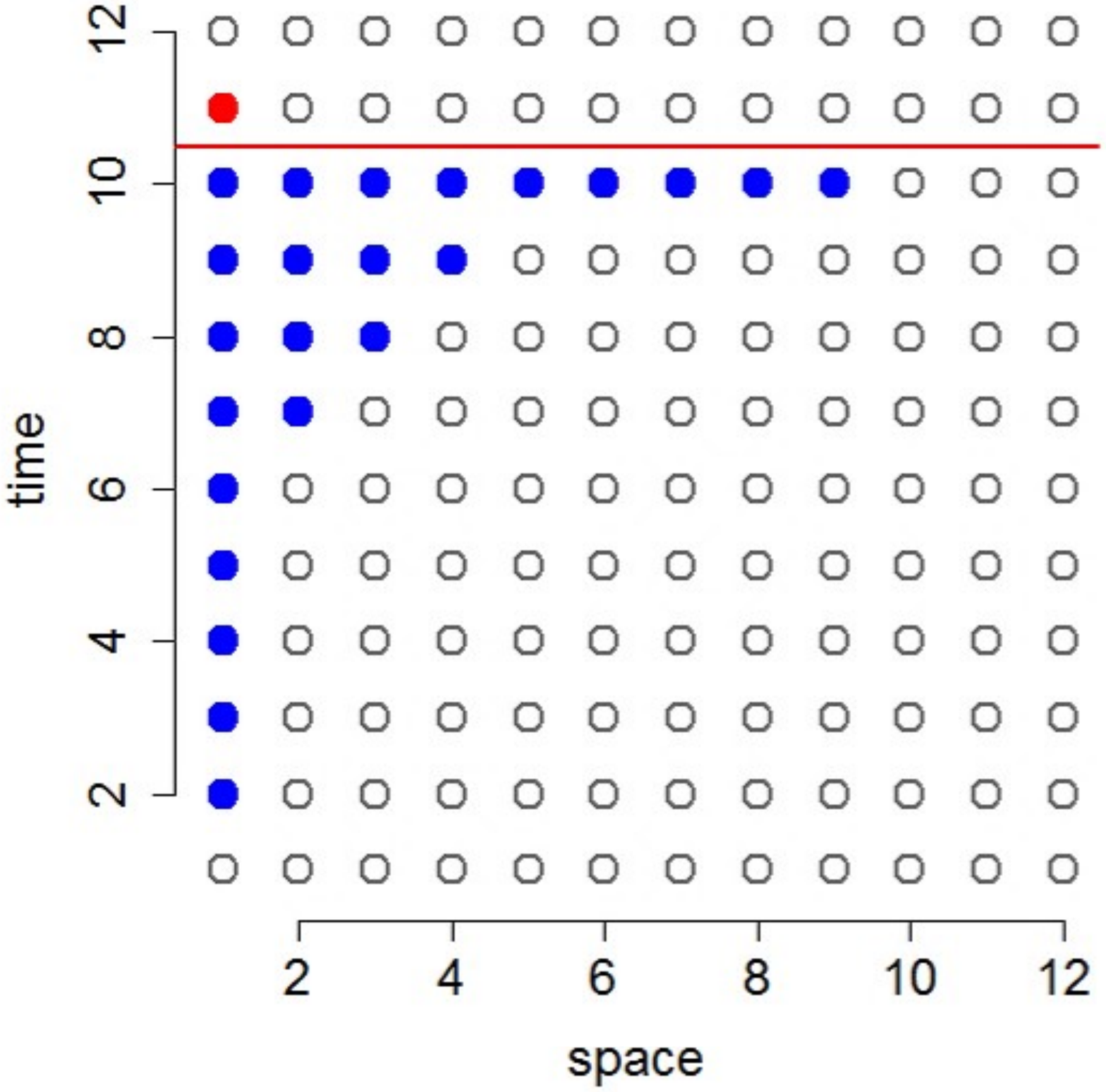}\label{fig:eli}}
\caption{Construction of eligible sets for finding nearest neighbor sets of size $m=9$: In figure (a) the black point is ineligible bacause the black rectangle contains more than $m=9$ points. In figure (b) the blue point will belong to $E(\bs_i,t_j)$ as the blue rectangle contains less than $m=9$ points. Figure (c) shows the final eligible set obtained by repeating this algorithm for all points in the history set (below the red line).}\label{fig:adapt}
\end{center}
\end{figure}

Figures \ref{fig:const1} and \ref{fig:const2} illustrate how to  determine which points belong to $E(\bs_i,t_j)$. Let $h$ and $u$ denote the spatial and temporal lags with the black point and the red point in Figure \ref{fig:const1}. All other points in the black rectangle have spatial lag $\leq h$ and temporal lag $\leq u$ with the red point. So if the covariance function $C(h,u \given \btheta)$ possess natural monotonicity, the black point has lowest correlation with the red point among all the points in the black rectangle. For the black point to be in the the set of $m$ nearest neighbors $N_\theta(\bs_i,t_j)$ for any $\btheta$, all other points in the black rectangle should also be included. Since, this is not possible as the black rectangle contains $12$ points and $m=9$, the black point becomes ineligible. By a similar logic, the blue rectangle in Figure \ref{fig:const2} contains $8 (< m)$ points and is included in $E(\bs_i,t_j)$. Proceeding like this, we can easily determine the entire eligible set (Figure \ref{fig:eli}) without any knowledge of the parameter $\btheta$.

A formal construction of eligible sets is provided in Section S1 of the supplemental article \cite{dnngpsupp}. Proposition S1 proves that eligible sets are guaranteed to contain the neighbor sets for all  choices of $\btheta$. This result has substantial consequences because the size of the eligible sets is approximately equal to $4m$. The eligible sets needs to be calculated only once before the MCMC as they are free of any parameter choices. Subsequently, for every new update of $\btheta$ in a MCMC sampler or an iterative solver, one can search for a new set of $m$-nearest neighbors $N_\theta(\bs_i,t_j)$ only within the eligible sets and use $N_\theta(\bs_i,t_j)$ as the conditioning sets to construct the DNNGP. We summarize the MCMC steps of the DNNGP with adaptive neighbor selection in Algorithm~\ref{algo:dyn}. 

\begin{algorithm}[!h]
\caption{Algorithm for adaptive neighbor selection in dynamic NNGP}
\label{algo:dyn}
\begin{algorithmic}[1]
\\ Compute the eligible sets $E(\bs_i,t_j)$ for all $(\bs_i,t_j)$ in $\calR$ from Eqn. S1
\\ At the $l^{th}$ iteration of the MCMC:
 \begin{enumerate}[(a)]
\item Calculate $C((\bs,t),(\bs_i,t_j) \given \btheta^{(l)})$ for all $(\bs,t)$ in $E(\bs_i,t_j)$
\item Define $N{_\theta}(\bs_i,t_j)^{(l)}$ as the set of $m$ locations in $E(\bs_i,t_j)$ which maximizes $C((\bs,t),(\bs_i,t_j) \given \btheta^{(l)})$
\item Repeat steps (a) and (b) for all $(\bs_i,t_j)$ in $\calR$
\item Update $\btheta^{(l+1)}$ based on the new set of neighbor sets computed in step (c) using Metropolis step specified in (\ref{Eq: fulltheta})
\end{enumerate}
\\ Repeat Step 2 for $N$ MCMC iterations
\end{algorithmic}
\end{algorithm}

As the size of the sets are approximately $4m$, for every $(\bs_i,t_j)$ we need to evaluate only $4m$ pairwise correlations. So the total computational complexity of the search is now reduced to $O(4m(n+r))$ from $O(nr+r^2)$. This is at par with the scale of implementing the remainder of the algorithm. With this adaptive neighbor selection scheme we gain the advantage of selecting the set of $m$-nearest neighbors at every update while retaining the scalability of the DNNGP. Parallel computing resources, if available, can be greatly utilized to further reduce computations as the search for eligible sets for each point (Algorithm~\ref{algo:dyn}: Step (c)) can proceed independent of one another.

%

\section{Bayesian DNNGP model}\label{sec:dbayes}
We consider a spatio-temporal dataset observed at locations $\bs_1,\bs_2, \ldots, \bs_N$ and at time points $t_1,t_2,\ldots, t_M$. Note that there may not be data for all locations at all time points i.e. we allow missing data. Let $\{\ell_1,\ell_2,\ldots,\ell_{n}\}$ be an enumeration of $n=MN$ points in $\calL$, where each $\ell_i$ is an ordered pair $(\bs_j,t_k)$. Let $y(\ell_i)$ be a univariate response corresponding to $\ell_i$ and let $\bx(\ell_{i})$ be a corresponding $p\times 1$ vector of spatio-temporally referenced predictors. A spatio-temporal regression model relates the response and the predictors as
\begin{equation}\label{eq:model}
y(\ell_{i}) = \bx'(\ell_{i})\bbeta + w(\ell_{i}) + \eps(\ell_{i})\;,\;\quad i=1,2,\ldots,MN\; ,
\end{equation}
where $\bbeta$ denotes the coefficient vector for the predictors, $w(\ell_{i})$ is the spatio-temporally varying intercept and $\eps(\ell_i)$ is the random noise customarily assumed to be independent and identically distributed copies from $N(0,\taus)$. 

Usually $w(\ell_{i})$'s are modeled as realizations of a spatio-temporal GP. To ensure scalability, we will construct a DNNGP from a parent GP with a non-separable spatio-temporal isotropic covariance function $C((\bs+\bh,t+u),(\bs,t) \given \btheta)$, introduced by \cite{gnei02},
\begin{align}\label{eq:cov}
\frac \sigs {2^{\nu-1}\Gamma(\nu)(a|u|^{2\alpha}+1)^{\delta+\kappa}}\times \left(\frac {c\|\bh\|}{(a|u|^{2\alpha}+1)^{\kappa/2}}\right)^\nu \times K_\nu\left(\frac {c\|\bh\|}{(a|u|^{2\alpha}+1)^{\kappa/2}}\right)\;,
\end{align}
where $\bh$ and $u$ refers to the spatial and temporal lags between $(\bs+\bh,t+u)$ and $(\bs,t)$ and $\btheta= \{\sigs, \alpha, \kappa, \delta, \nu, a, c\}$. The spatial covariance function at temporal lag zero corresponds to the Whittle-Matern class with marginal variance $\sigs$, smoothness parameter $\nu$ and decay parameter $c$. The parameters $\alpha$ and $a$ control smoothness and decay, respectively, for the temporal process, while $\kappa$ captures non-separability between space and time. 

A straightforward choice of the reference set $\calR$ is the set $\{\ell_1,\ell_2,\ldots,\ell_{n}\}$. While this set will typically be large, its size does not adversely affect the computations. This choice has been shown to yield excellent approximations to the parent random field \citep[][]{ve88,stein04}. Also, while several alternate choices of reference sets (like choosing the points over a regular grid) are possible, it is unlikely they will provide any additional computational or inferential benefits; this has been demonstrated in purely spatial contexts by \cite{datta14}. Hence, we choose $\calR = \{\ell_1,\ell_2,\ldots,\ell_{n}\}$, i.e., $\ell^*_i=\ell_i$ for $i=1,2,\ldots,n$. 

A full hierarchical model with a DNNGP prior on $w(\ell)$ is given by 
\begin{align}\label{eq:hierlike}
& p(\btheta) \times IG(\tau^2\given a_{\tau},b_{\tau})\times N(\bbeta\given \bmu_{\beta}, \bV_{\beta}) \times N(\bw_ \calR \given \bzero, \tildebC_ {\calR,\calR})\nonumber \\
 &\qquad \qquad  \times \prod_{i=1}^{n} N(\by(\ell_{i})\given \bx(\ell_{i})'\bbeta + \bw(\ell_{i}), \taus)\; ,
\end{align}
where $p(\btheta)$ is the prior on $\btheta$, and $IG(\tau^2\given a_{\tau}, b_{\tau})$ denotes the Inverse-Gamma density with shape $a_{\tau}$ and rate $b_{\tau}$. Below we describe an efficient MCMC algorithm using Gibbs and Metropolis steps only to carry out full inference from the posterior in Equation~\ref{eq:hierlike}. 

\subsection{Gibbs' sampler steps}\label{sec:dgibbs}
Let $S_o$ be the points in $\calR$ where the $y(\ell_i)$'s is observed and $I(\ell_i)$ denote the binary indicator for presence or absence of data at $\ell_i$. Let $\by$ be the $n_o \times 1$ vector formed by stacking the responses observed and $\bX$ be the corresponding $n_o \times p$ design matrix. 
The full conditional distribution of $\bbeta$ is $N(\bV^*_{\beta}\bmu^*_{\beta},\bV^*_{\bbeta})$ where $\bV^*_{\beta} = (\bV_{\beta}^{-1}+\bX'\bX/\taus)^{-1}$ and $\bmu^*_ \beta = (\bV_{\beta}^{-1}\bmu_{\beta}+\bX'(\by-\bw_{S_o})/\taus)$. The full conditional distribution of $\taus$ follows $IG\left(a_{\tau}+\frac {n_o}2,b_{\tau}+\frac {1}{2}(\by-\bX\bbeta-\bw_{S_o})'(\by-\bX\bbeta-\bw_{S_o})\right)$.

We update the elements of $\bw_\calR$ sequentially. For any two locations $\ell_1$ and $\ell_2$ in $\calL$, if $\ell_1 \in N(\ell_2)$ and is the $j$-th member of $N(\ell_2)$, then we define $b_{\ell_2,\ell_1}$ as the $j$-th entry of $\ba_{N(\ell_2)}$. Let $U(\ell_1)=\{\ell_2 \in \calR \given \ell_1 \in N(\ell_2) \}$ and
for every $\ell_2 \in U(\ell_1)$, define, $a_{\ell_2,\ell_1}=w(\ell_2)-\sum_{\ell \in N(\ell_2), \ell \neq \ell_1} w(\ell) b_{\ell_2,\ell}$. Then, for $i=1,2,\ldots,n$ the full conditional distribution for $\bw(\ell_{i})$ is $N\left(v(\ell_{i}) \mu(\ell_{i}),v(\ell_{i}) \right)$, where
\begin{align}\label{Eq: fullws1}
v(\ell_{i}) &= \left( \frac{I(\ell_i)}{\taus} + \frac{1}{f_{\ell_{i}}} + \frac{\sum_{\ell \in U(\ell_{i})} b^2_{\ell,\ell_{i}}}{f_{\ell}} \right)^{-1}\; \mbox{ and} \nonumber\\
\mu(\ell_{i}) &= \frac{y(\ell_{i})-\bx(\ell_{i})'\bbeta}{\taus}I(\ell_i) + \frac{\ba_{N(\ell_{i})}'\bw_{N(\ell_{i})}}{f_{\ell_{i}}} + \sum_{\ell \in U(\ell_{i})} \frac{b_{\ell,\ell_{i}} a_{\ell,\ell_i}}{f_{\ell}}\; .
\end{align}
If $U(\ell_{i})$ is empty for some $\ell_{i}$, then all instances of $\sum_{\ell \in U(\ell_{i})}$ in (\ref{Eq: fullws1}) disappear for that $\bw(\ell_{i})$. 

\subsection{Metropolis step}\label{sec:dmetrop} We update $\btheta$ using a random walk Metropolis step.  The full-conditional for $\btheta$ is proportional to
\begin{equation}\label{Eq: fulltheta}
p(\btheta)p(\bw_\calR \given \btheta) 
\propto p(\btheta) \times \prod_{i=1}^{n} N\left(\bw(\ell_{i}) \given \ba_{N(\ell_{i})}'\bw_ {N(\ell_{i})}, f_ {\ell_{i}}\right)\; . 
\end{equation}
Since none of the above updates involve expensive matrix decompositions, the likelihood can be evaluated very efficiently. The algorithm for updating the parameters of a hierarchical DNNGP model is analogous to the corresponding updates for a purely spatial NNGP model (see \citet{datta14}). The only additional computational burden stems from updating the neighbor sets in the adaptive neighbor selection scheme, but even this can be handled efficiently using eligible sets (Algorithm \ref{algo:dyn}). Hence, the number of floating point operations per update is linear in the number of points in $\calL$.

\subsection{Prediction} \label{Sec:dpred}
Once we have computed the posterior samples of the model parameters and the spatio-temporal random effects over $\calR$, we can execute, cheaply and efficiently, full posterior predictive inference at unobserved locations and time points. The Gibbs' sampler in Section~\ref{sec:dgibbs} generates full posterior distributions of the $\bw$'s at all locations in $\calR$. Let $\ell_{i}^*$ denote a point in $\calR$ where the response is unobserved i.e. $I(\ell_i^*)=0$. We already have posterior distributions of $\bw(\ell_{i}^*)$ and the parameters. We can now generate posterior samples of $ \by(\ell_{i}^*)$ from $N(\bx(\ell_{i}^*)'\bbeta+\bw(\ell_{i}^*),\taus)$. Turning to prediction at a location $\ell$ outside $\calR$, we construct $N(\ell)$ from $E(\ell)$ described in Equation S2 for every posterior sample of $\btheta$. We generate posterior samples of $\bw(\ell)$ from $N(\ba_{N(\ell)}'\bw_{N(\ell)}, f_\ell)$ and, subsequently, draw posterior samples of $y(\ell)$ from $N(\bx(\ell)'\bbeta+\bw(\ell),\taus)$.

\section{Synthetic data analyses}\label{sec:dillu}
\begin{figure}[!t]
\begin{center}
\subfigure[Dataset 1]{\includegraphics[width=4cm, trim={4cm 4cm 4cm 4cm},clip]{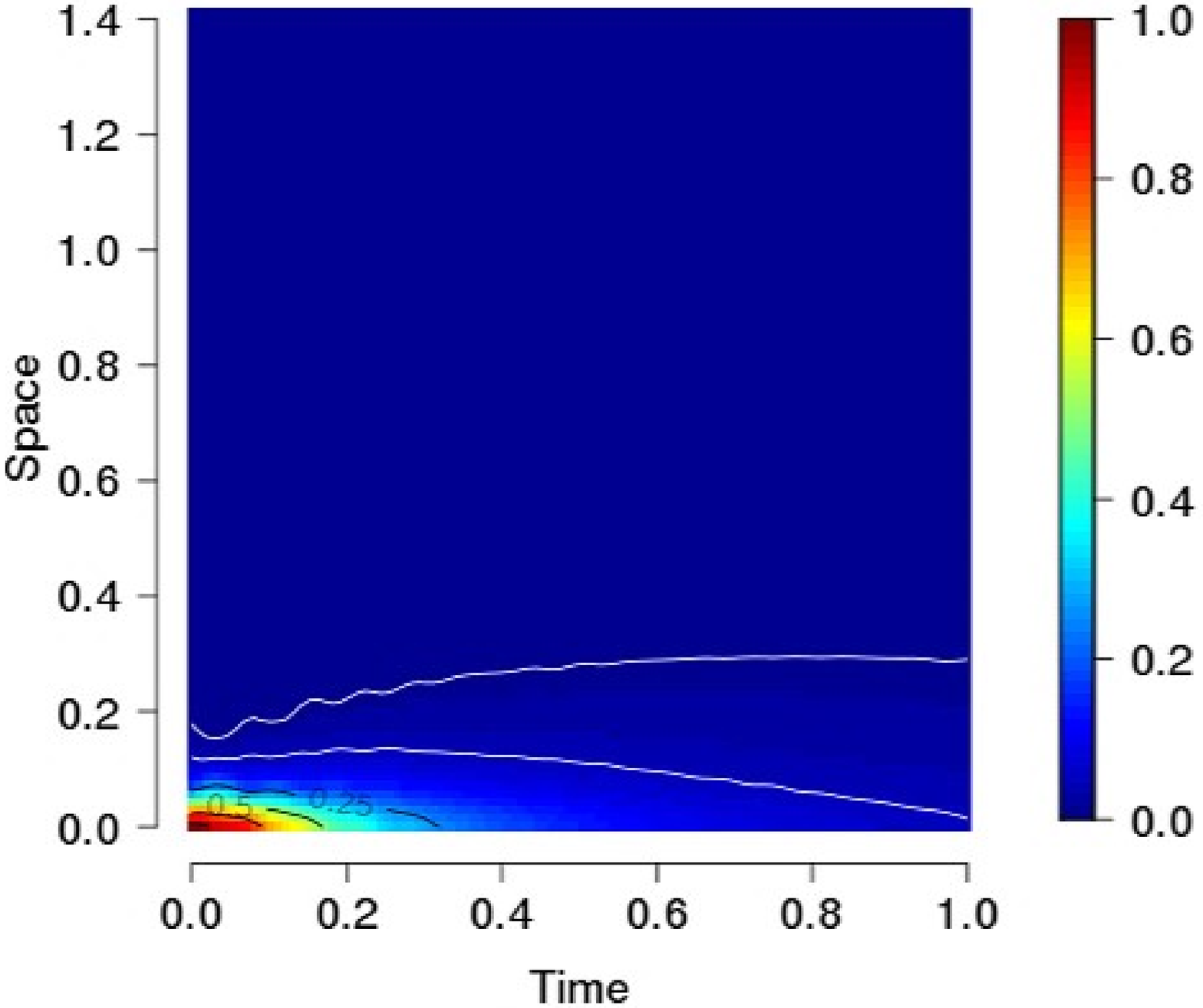}}
\subfigure[Dataset 2]{\includegraphics[width=4cm, trim={4cm 4cm 4cm 4cm},clip]{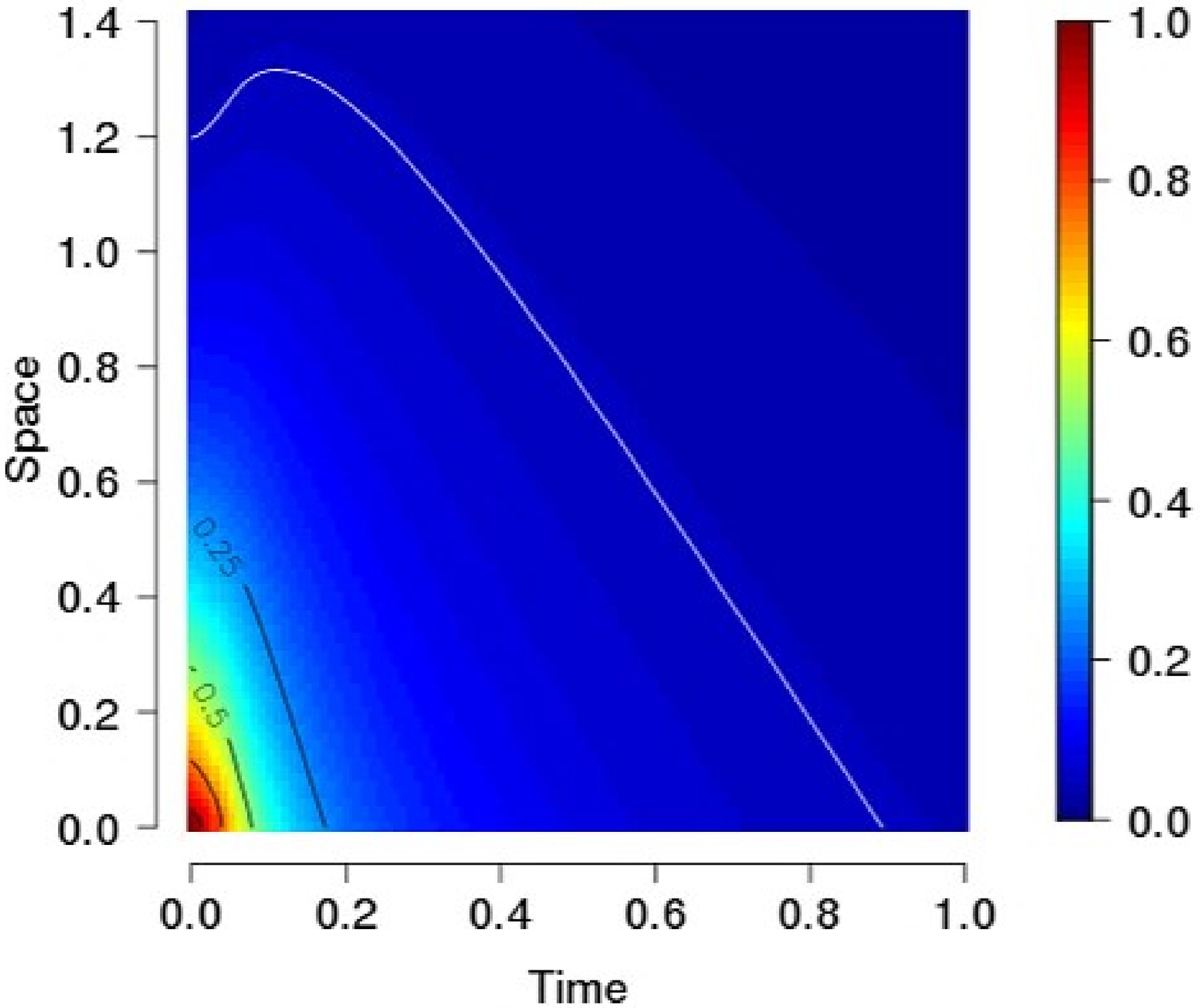}}
\subfigure[Dataset 3]{\includegraphics[width=4cm, trim={4cm 4cm 4cm 4cm},clip]{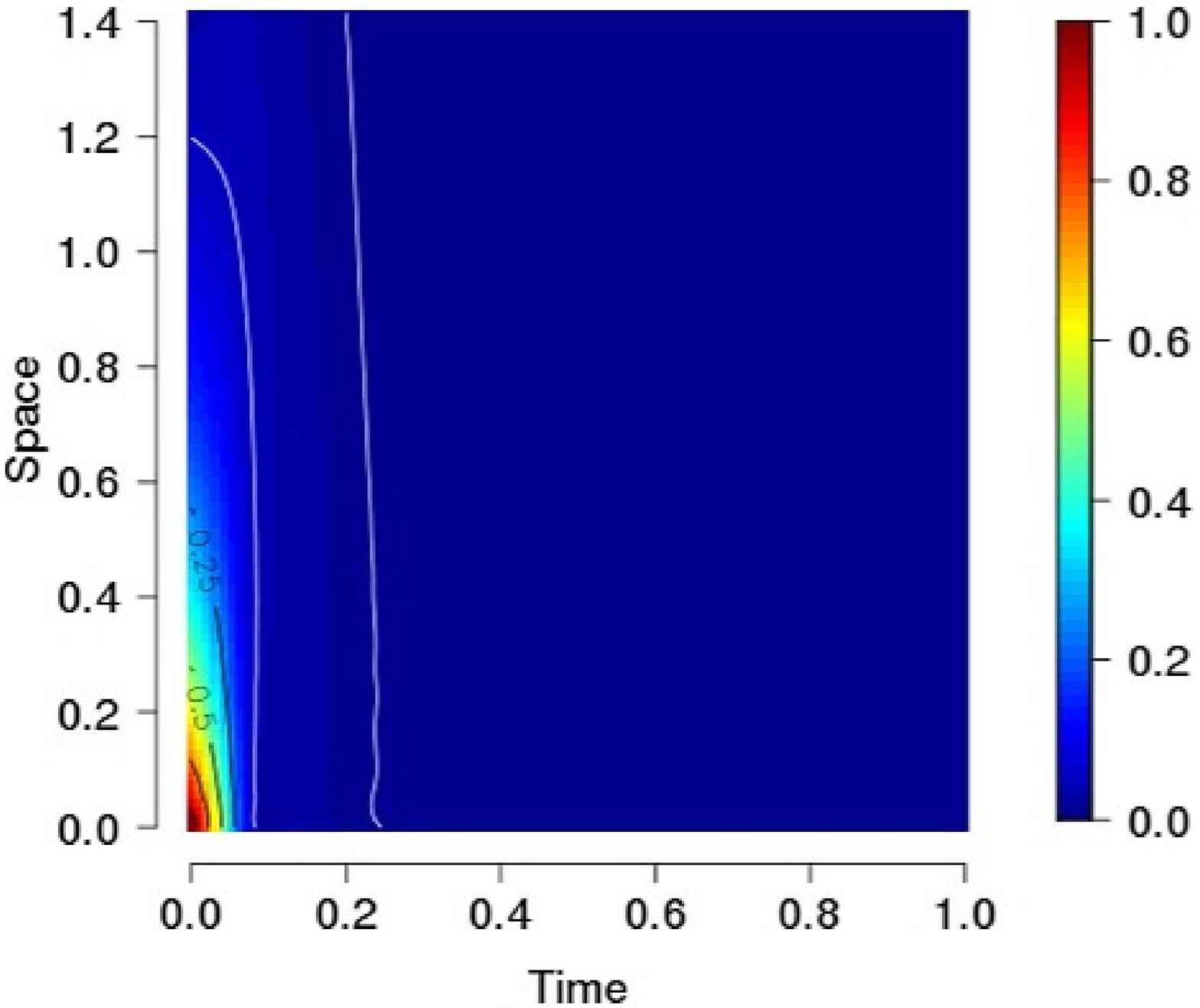}}
\caption{Space-time correlation surface realizations given \emph{true} parameter values in Table~\ref{tab:sim}. Correlation contours are provided, with the two outer white lines corresponding to 0.05 and 0.01.}\label{fig:simcorsurfs}
\end{center}
\end{figure}

In this section we compare the DNNGP, the full rank GP and low rank Gaussian Predictive Process \citep{ban08}. Additional simulation experiments comparing the predictive performance of DNNGP with Local Approximation GP \citep{gram14} are provided in Section S2 of the supplemental article \cite{dnngpsupp}. We generated observations over a $n=15\times 15\times 15=3375$ grid within a unit cube domain. An additional 500 observations used for out-of-sample prediction validation were also located within the domain. All data were generated using model~\ref{eq:model} with $\bx(\ell)$ comprising an intercept and covariate drawn from $N(0,1)$. The spatial covariance matrix $\bC(\btheta)$ was constructed using an exponential form of the non-separable spatio-temporal covariance function (\ref{eq:cov}), viz.,. 
\begin{equation}\label{eq:expcov}
\frac \sigs {(a|u|^{2}+1)^{\kappa}}\text{exp}\left(\frac {-c\|h\|}{(a|u|^{2}+1)^{\kappa/2}}\right),
\end{equation}
where $u = |t_i-t_j|$ and $h = ||\bs_i-\bs_j||$ are the time and space Euclidean norms, respectively. By specifying different values of the decay and interaction parameters in $\btheta=(\sigma^2, \kappa, a, c)$, we generated three datasets that exhibited different covariance structures. The first column in Table~\ref{tab:sim} provides the three specifications for $\btheta$ and Figure~\ref{fig:simcorsurfs} shows the corresponding space-time correlation surface realizations. As illustrated in Figure~\ref{fig:simcorsurfs}, the three datasets exhibit: 1) short spatial range and long temporal range, 2) long spatial and temporal range, and 3) long spatial range and short temporal range. 

For each dataset, model parameters were estimated using: $i$) full Gaussian Process (GP), $ii$) DNNGP with simple neighbor set selection (Simple DNNGP) described in Section~\ref{sec:simple}, $iii$) DNNGP with adaptive neighbor set selection (Adaptive DNNGP) described in Section~\ref{sec:adapt}, and; $iv$) bias-corrected Gaussian Predictive Process (GPP) detailed in \cite{ban08} and \cite{fin09}. DNNGP models were fit using $m=\{16, 25, 36\}$ and the Gaussian Predictive Process model used a regularly spaced grid of $8\times 8 \times 8=512$ knots within the domain.

For all models, the intercept $\beta_0$ and and slope regression parameters, $\beta_1$, were assigned \emph{flat} prior distributions. The variance parameters were assumed to follow inverse-Gamma prior distributions with $\sigma^2\sim IG(2,1)$ and $\tau^2\sim IG(2,0.1)$. The time and space decay parameters received uniform priors that were dataset specific: 1) $a\sim U(1,100)$, $c\sim U(0,50)$; 2) $a\sim U(300,700)$, $c\sim U(0,10)$, and; 3) $a\sim U(1000,3000)$, $c\sim U(0,10)$. The prior for the interaction term matched its theoretical support with $\kappa\sim U(0,1)$. 

\begin{sidewaystable}
\centering
{\tiny
\begin{center}
\caption{Synthetic data analysis parameter estimates and computing time for the candidate models. Parameter posterior summary 50 (2.5, 97.5) percentiles. Bold indicates estimates with 95\% credible intervals that do not include the \emph{true} parameter value.}\label{tab:sim}
\begin{tabular}
[c]{lccccc}%

\hline
&&		GP&	GPP knots=512	&	Adaptive DNNGP m=25&			Simple DNNGP m=25\\
\hline
\multicolumn{6}{c}{Dataset 1}\\
\hline	
$\beta_0$& 1&	0.99 (0.80, 1.12)&1.02 (0.89, 1.16)&	0.97 (0.86, 1.11)&	0.97 (0.86, 1.11)\\
$\beta_1$& 5& 	4.99 (4.97, 5.01)&4.98 (4.94, 5.02)&	4.99 (4.97, 5.01)&	4.99 (4.97, 5.01)\\
$a$& 50&   	46.46 (38.02, 67.46)&\textbf{16.93 (11.91, 29.17)}&	48.37 (37.85, 77.93)&	53.18 (35.93, 83.78)\\
$c$& 25&	25.69 (22.00, 29.49)&22.73 (13.53, 34.20)&	24.77 (20.55, 28.71)&	25.16 (21.91, 29.52)\\
$k$& 0.75&	0.83 (0.61, 0.94)&0.78 (0.39, 0.91)&	0.80 (0.57, 0.99)&	0.75 (0.53, 0.98)\\
$\sigma^2$&1&	\textbf{1.13 (1.03, 1.24)}&\textbf{0.70 (0.56, 0.92)}&	\textbf{1.13 (1.03, 1.24)}&	\textbf{1.14 (1.04, 1.25)}\\
$\tau^2$&0.1&	0.09 (0.07, 0.11)&\textbf{0.95 (0.89, 1.02)}&	0.09 (0.06, 0.11)&	0.09 (0.06, 0.11)\\
\hline
pD&&		2214.57&225.66&		2236.81&		2258.39\\
DIC&&		3700.68&9644.76&		3650.55&		3567.45\\
G&&		104.60&3008.41	&		100.06&			94.50\\
P&&		512.29&3436.52	&		504.66&			493.63\\
D=G+P&&		616.90&6444.93	&		604.72&			588.13\\
RMSPE&&	0.84&	0.95&		0.84&			0.84\\
95\% CI cover \%&&	95.6&94.6&			95.6&			96\\
\hline
\hline
\multicolumn{6}{c}{Dataset 2}\\
\hline					
$\beta_0$& 1&    0.81 (0.48, 1.26)&0.79 (0.26, 1.16)&		1.01 (0.57, 1.27)&		0.92 (0.58, 1.26)\\
$\beta_1$& 5&    4.98 (4.96, 5.00)&4.99 (4.97, 5.02)&		4.98 (4.96, 5.00)&		4.98 (4.96, 5.00)\\
$a$& 500&  	352.82 (301.69, 521.64)&583.59 (391.79, 661.36)&	480.28 (331.63, 662.12)&	410.84 (317.29, 602.21)	\\	
$c$& 2.5&	2.52 (1.93, 3.13)&\textbf{1.67 (1.03, 2.31)}&		2.91 (2.49, 3.37)&	2.59 (1.91, 3.37)\\
$k$& 0.5&	0.56 (0.44, 0.67)&0.39 (0.26, 0.53)&		0.46 (0.36, 0.62)&		0.53 (0.42, 0.67)\\
$\sigma^2$& 1&	1.01 (0.85, 1.31)&1.14 (0.83, 1.77)&		0.94 (0.81, 1.10)&		1.03 (0.83, 1.32)\\
$\tau^2$& 0.1&	0.11 (0.09, 0.13)&\textbf{0.44 (0.41, 0.47)}&		0.10 (0.08, 0.12)&		0.10 (0.08, 0.12)\\
\hline
pD&&		1913.96&312.76&			1999.98		&	2014.78\\
DIC&&		3988.36&7091.84&			3866.38&		3871.09\\
G&&		157.52&1336.94	&			139.28&			137.33\\
P&&		576.02&1609.99	&			550.28&			550.36\\
D=G+P&&		733.53&2946.94	&			689.56&			687.68\\
RMSPE&&		0.53&	0.71&			0.53&			0.53\\
95\% CI cover \%&&	96.4&	93&			94&			93.8\\
\hline
\hline
\multicolumn{6}{c}{Dataset 3}\\
\hline
$\beta_0$& 1&	0.94 (0.66, 1.14)&\textbf{0.55 (0.32, 0.84)}&		0.93 (0.74, 1.17)&		0.87 (0.68, 1.12)\\					
$\beta_1$& 5& 	4.98 (4.96, 5.00)&4.98 (4.95, 5.02)&		4.98 (4.96, 5.00)&		4.98 (4.96, 5.00)\\
$a$& 2000&	1214.02 (1008.23, 2141.16)&1590.77 (1151.78, 2118.63)&	1635.46 (1046.76, 2922.59)&	1495.94 (1019.16, 2751.17)\\		
$c$& 2.5&	2.38 (1.79, 2.95)&\textbf{1.36 (0.73, 2.16)}&		2.25 (1.62, 2.81)&		2.27 (1.60, 2.97)\\
$k$& 0.95&	0.91 (0.72, 0.98)&\textbf{0.68 (0.40, 0.90)}&		0.71 (0.46, 0.98)&		0.81 (0.58, 0.97)\\
$\sigma^2$& 1&	1.03 (0.86, 1.35)&0.91 (0.67, 1.83)&		1.09 (0.89, 1.44)&		1.07 (0.87, 1.42)\\
$\tau^2$& 0.1&	0.11 (0.09, 0.13)&\textbf{0.68 (0.62, 0.74)}&		0.11 (0.09, 0.14)&		0.11 (0.09, 0.14)\\
\hline
pD&&		1990.41&210.11&		1982.28&		1994.77\\
DIC&&		4210.71&8463.33&		4214.68&		4223.88\\
G&&		155.87&2137.55	&		157.24&			155.61\\
P&&		610.01&2424.66	&		611.89&			612.98\\
D=G+P&&		765.89&4562.21	&		769.13&			768.59\\
RMSPE&&		0.78&0.92&		        0.77&			0.77\\
95\% CI cover \%&&	92.8&91.4&			95.6&			95.6\\
\hline
CPU (min)&&7646.96 & 856.54 & 496.12 & 430.88\\
\hline
\end{tabular}
\end{center}
}
\end{sidewaystable}

Candidate model comparison was based on parameter estimates, fit to the observed data, out-of-sample prediction accuracy, and posterior predictive distribution coverage. Goodness-of-fit was assessed using DIC \citep{spieg02} and posterior predictive loss \citep{gelf98}. The DIC is reported along with an estimate of model complexity, pD, while the posterior predictive loss is computed as D=G+P, where G is a goodness-of-fit measure and P measures the number of model parameters. Predictive accuracy for the 500 holdout locations was measured using root mean squared prediction error \citep{rmspe02}. The percent of holdout locations that fell within the candidate models' posterior predictive distribution's 95\% credible interval (CI) was also computed. Inference was based on 15,000 MCMC samples comprising post burn-in samples from three chains of 25,000 iterations (i.e., 5,000 samples from each chain). 

Table~\ref{tab:sim} presents parameter estimation and model assessment metrics. With the exception of $\tau^2$ for Dataset 1, the full GP model recovered the parameter values used to generate the datasets, i.e., the 95\% CIs cover the \emph{true} parameter values. For the DNNGP models, there was negligible difference among parameter estimates for the 15, 25, and 36 neighbor sets. Hence, we report only the $m=25$ cases. There was very little difference between the estimates produced by the Adaptive and Simple DNNGP models and, like the full GP model, they captured the \emph{true} mean and process parameters, with the exception of $\tau^2$ for Dataset 1. Given the extremes in the space and time decay in Datasets 1 and 3, we anticipated the Simple DNNGP model---with at most 5 neighbors in any given time point---would not be able to estimate the covariance parameters. Extensive analysis of simulated data, some of which is reported in Table~\ref{tab:sim}, suggested the Simple DNNGP model performed as well as the Adaptive DNNGP and full GP models. Goodness-of-fit and out-of-sample prediction validation metrics in Table~\ref{tab:sim} also show the full GP and DNNGP models provided comparable results. In contrast the GPP model did not capture many of the process parameters and provided worse fit and prediction than the GP and DNNGP models. The quality of the GPP results would improve with additional knots. However, computing time would also increase. The last row in Table~\ref{tab:sim} provides the CPU time required for each candidate model to generate 25,000 MCMC samples for the $n=3375$ observations. Even with the substantial dimension reduction, the GPP model required about twice the CPU time as the DNNGP models. Compared to the full GP model, the DNNGP models provided substantial computational advantages while delivering comparable results.

\section{Analysis of Airbase and LOTOS-EUROS CTM data}\label{sec:dpm10_analysis}
We consider the model in Equation~\ref{eq:hierlike}, where $y(\ell_{i})$ is a square-root transformed measurement of PM$_{10}$ at space-time coordinate $\ell_i$, $x(\ell_{i})$ is the coinciding square-root transformed output from the LOTOS-EUROS CTM. Given the large dimension of the dataset, $n=N\times M =308 \times 730=\text{224,840}$, the spatio-temporal random effects were modeled as a DNNGP prior derived from a zero-centered GP with the non-separable spatio-temporal covariance function (\ref{eq:expcov}). Exploratory analysis---consisting of semivariogram plots and autocorrelation function plots for simple ordinary least square model residuals---helped guide choice of prior and hyper-parameters for the variance and decay parameters. Specifically, $\sigma^2\sim IG(2, 1)$, $\tau^2\sim IG(2, 0.1)$, $a\sim U(0.1, 5)$, and $c\sim U(0.01, 0.5)$, with $\kappa$ fixed at 0.5.

Candidate models included the $i$) LOTOS-EUROS CTM, $ii$) simple linear regression model with no spatio-temporal effects, i.e., $w(\ell)=0$, and $iii$) Adaptive and Simple DNNGP with $m=\{16, 25, 36\}$. Following Section~\ref{sec:dillu}, candidate model goodness-of-fit to the observed data was assessed using DIC and GPD, whereas predictive performance was assessed using RMSPE and 95\% posterior predictive CI coverage rate for out-of-sample prediction. The holdout set comprised blocks of five days per station---five days of continuous observations were withheld at random from each station's 730 day time series. 

Additionally, prediction using the Adaptive and Simple DNNGP models for a 25\% holdout set selected from April 1-14, 2009 was compared with results from \citet{hamm2015} who considered time invariant spatial regression models for the same two-week period and comparable prediction validation approach.

\begin{sidewaystable}
\centering
{\small
\caption{PM$_{10}$ analysis parameter posterior 50 (2.5, 97.5) percentiles, model fit and prediction metrics, and run time for 25,000 MCMC samples.} \label{tab:realParamsDyn}
\begin{tabular}{cccccc}
\hline 
&&\multicolumn{3}{c}{Adaptive}&Simple\\
\cline{3-5}
Parameter&Non space-time&m=16& m=25& m=36& m=36\\ \hline
$\beta_0$&1.66 (1.64, 1.68)&2.56 (2.53, 2.59)&2.62 (2.59, 2.65)&2.61 (2.58, 2.64)&2.64 (2.61, 2.68)\\
$\beta_1$&0.76 (0.75, 0.76)&0.47 (0.46, 0.47)&0.45 (0.44, 0.46)&0.45 (0.44, 0.46)&0.44 (0.43, 0.45)\\
$a$&--&0.57 (0.57, 0.57)&0.44 (0.44, 0.44)&0.46 (0.46, 0.46)&0.37 (0.37, 0.39)\\
$c$&--&0.08 (0.08, 0.08)&0.07 (0.07, 0.07)&0.07 (0.07, 0.07)&0.05 (0.05, 0.05)\\ 
$\sigma^2$&--&1.49 (1.48, 1.51)&1.64 (1.62, 1.66)&1.56 (1.54, 1.58)&2.06 (2.01, 2.11)\\
$\tau^2$&1.48 (1.47, 1.48)&0.12 (0.12, 0.12)  &0.14 (0.14, 0.14)&0.14 (0.14, 0.14)&0.15 (0.15, 0.16)\\
\hline
P$_D$&2.75&110266.2&122466.2&111190.6&103038.3\\
DIC&586,135.8&279077.3&265720.6&277383.9&286,922.9\\ 
G&432811.9&11538.63&8707.79&11249.11&13521.63\\
P&268036.7&40994.19&36711.28&40532.25&43728.23\\
D&700848.6&52532.82&45419.07&51781.37&57249.86\\\hline
RMSPE&12.75&8.28&8.24&8.2&8.11\\
95\% CI cover \%&93.4&93.33&93.06&93.15&92.86\\\hline
CPU (min)&--&6182.89&15681.8&27660.5&25819\\ 
\hline
\end{tabular}
}
\end{sidewaystable}

\begin{figure}[h!] 
\centerline{
\subfigure[Adaptive m=36]{\includegraphics[trim={0 1cm 0 2cm},clip,width=2.5in]{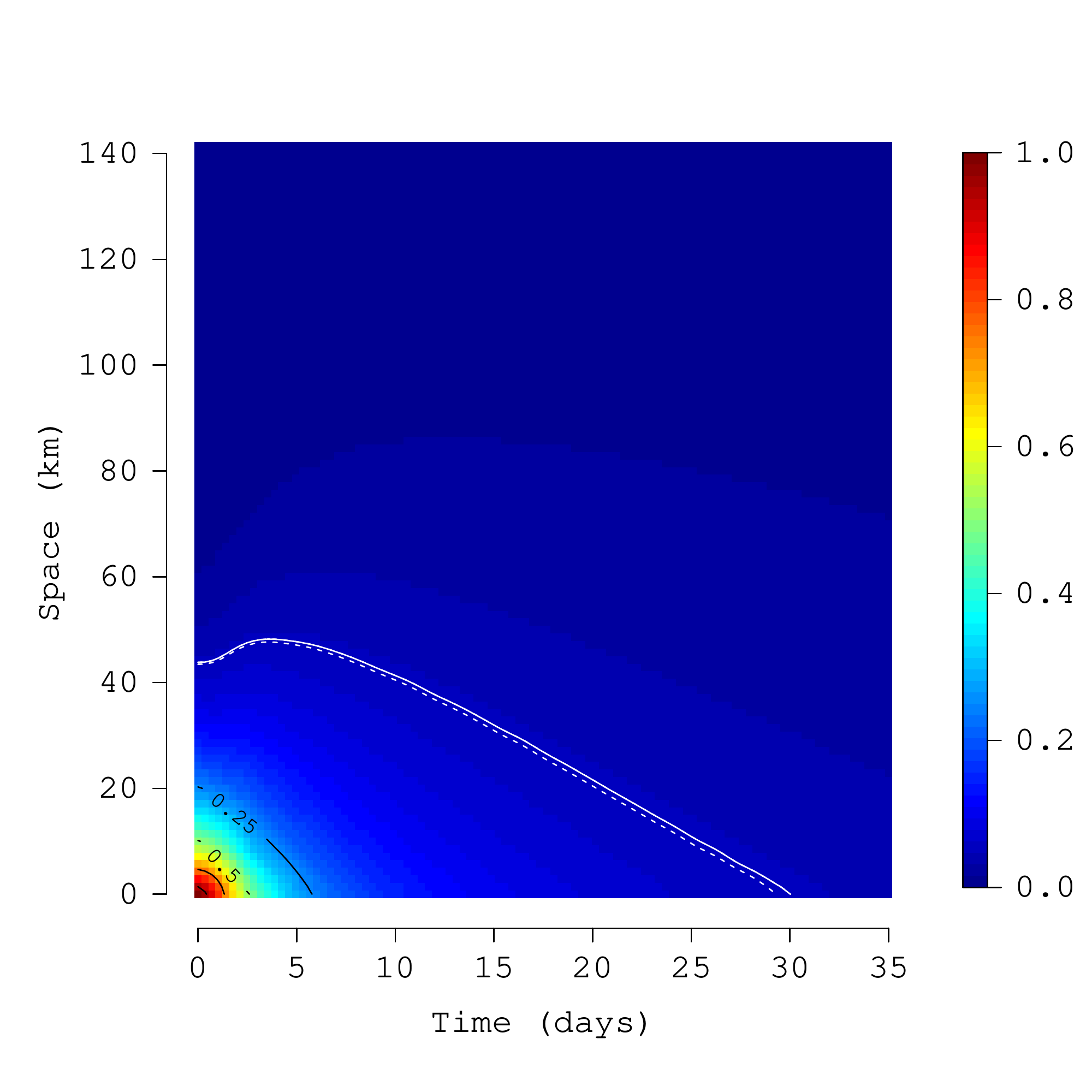}\label{sptime-cor-m35}} 
\subfigure[Simple m=36]{\includegraphics[trim={0 1cm 0 2cm},clip,width=2.5in]{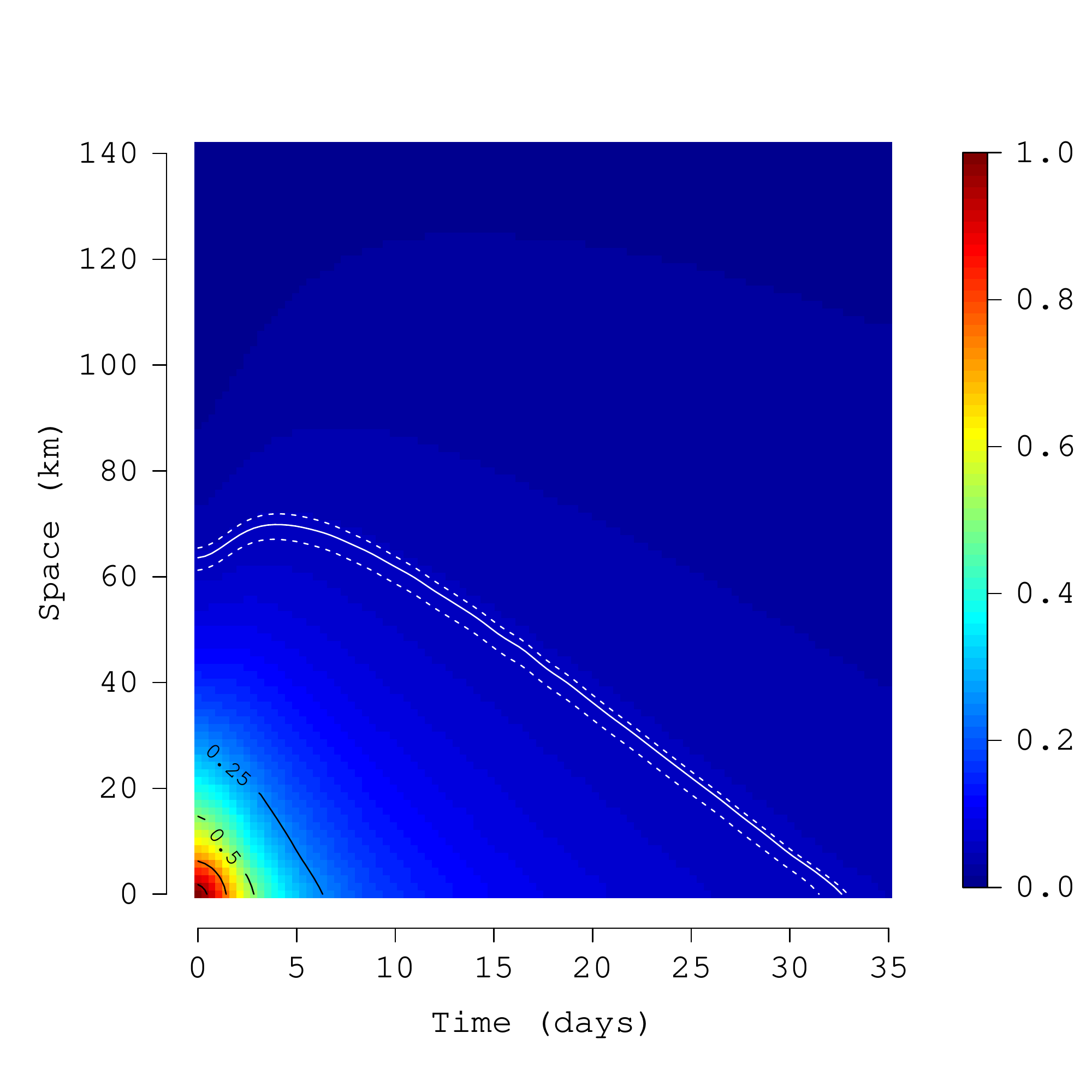}\label{sptime-cor-m35}} 
}
\caption{Space-time correlation posterior distribution median surfaces. Median (white lines) and associated 95\% credible intervals (dotted white lines) for correlation contours of 0.05.}\label{sptime-eff-rng}
\end{figure}

A subset of analysis results are given in Table~\ref{tab:realParamsDyn}. Parameter estimates for the model intercept and regression slope coefficient associated with the CTM output are consistent across the candidate models.  For an accurate CTM it would be expected that $\beta_0 \approx 0$ and $\beta_1 \approx 1$.  The finding that $\beta_0 > 0$ and $0 < \beta_1 < 1$ corroborate previous findings that showed the CTM consistently underestimates PM$_{10}$ \citep{SternEtAl08a,hamm2015}. The spatial and temporal decay parameters differed between the Adaptive and Simple DNNGP models. Figure~\ref{sptime-eff-rng} provides correlation surfaces generated using posterior median values of $a$ and $c$ from the $m=36$ Adaptive and Simple DNNGP models (using values given in Table~\ref{tab:realParamsDyn}). The 0.05 correlation contour on these surfaces suggest the Simple model estimates a moderately longer spatial and temporal range, i.e., $\sim$60 km and $\sim$33 days, versus $\sim$45 km and $\sim$30 days for the Adaptive model. Within a given DNNGP neighbor selection algorithm there is only marginal difference between the covariance parameters estimates when comparing $m$ of 25 and 36. Neighbor sets of less than 25 provided consistently larger temporal decay parameter estimates, i.e., shorter temporal correlation estimates, although even with such few neighbors the models seemed to produce consistent estimates of the spatial decay.

The spatial range of $45$ to $60$ km is an order of magnitude less than that observed by \cite{hamm2015}, who estimated median spatial ranges of 500 to 1500 km.  This is attributed to the inclusion of temporal correlation in the model, which itself accounts for a large amount of the residual spatial structure.  The temporal range is physically reasonable considering the life-time of PM$_{10}$  is in the order of days and its variability is driven by alternating synoptic meteorological conditions, with certain conditions usually lasting for several days to weeks. 

Across all candidate models the Adaptive with $m$=25 provided the lowest values of DIC and D suggesting improved fit to the observed data. This improved fit did not correspond to increased out-of-sample prediction accuracy. Rather, RMSPE consistently decreased with increasing number of neighbors within the Adaptive and Simple model sets. The smallest RMSPE was achieved using the simple neighbor selection with $m$=36. All models achieved reasonable coverage rates. 

\begin{figure}[]
\subfigure[Station 14]{\includegraphics[trim={0 1cm 0 2cm},clip,width=\textwidth]{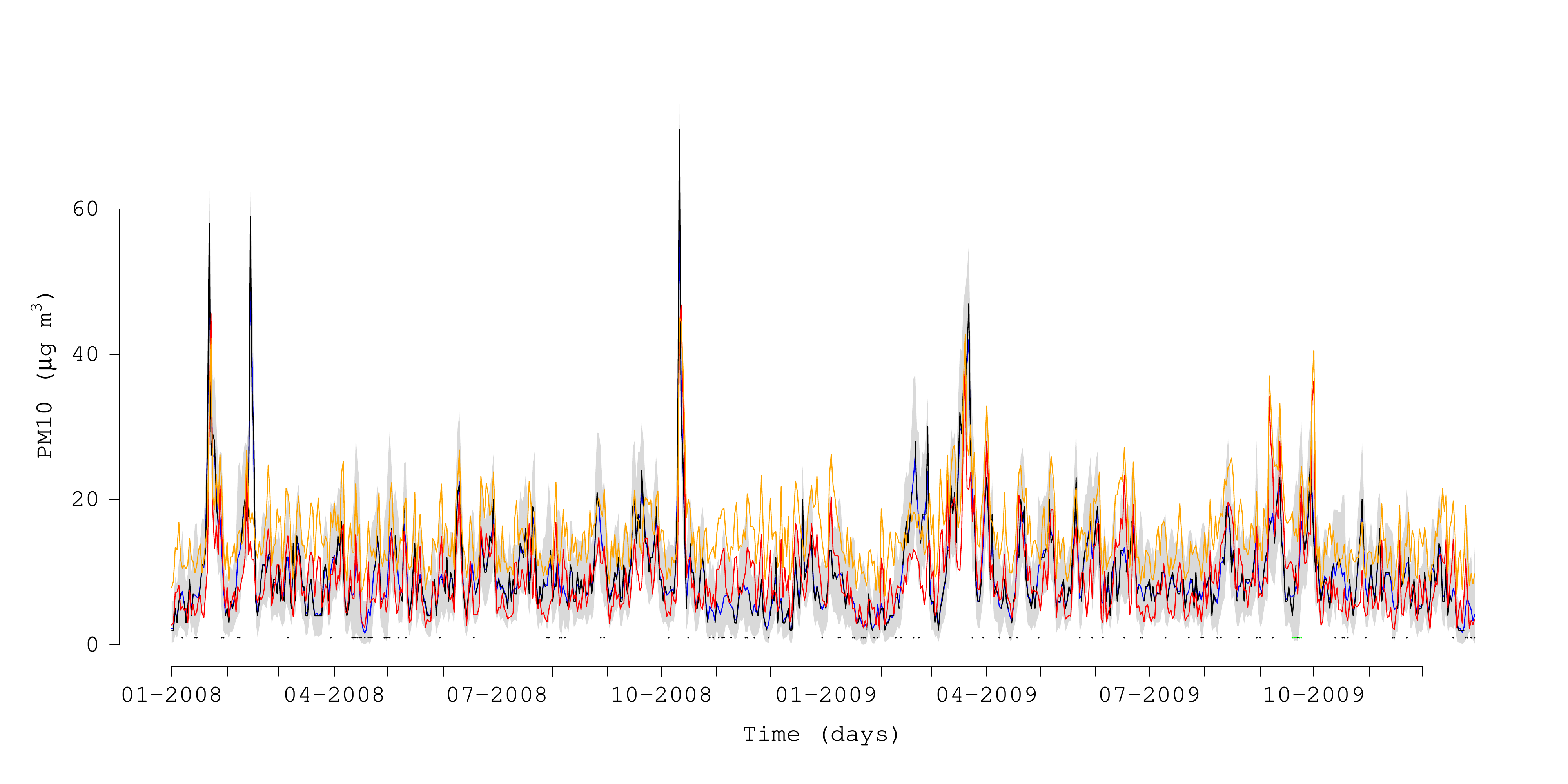}\label{fitted-s14}} \\
\subfigure[Station 47]{\includegraphics[trim={0 1cm 0 2cm},clip,width=\textwidth]{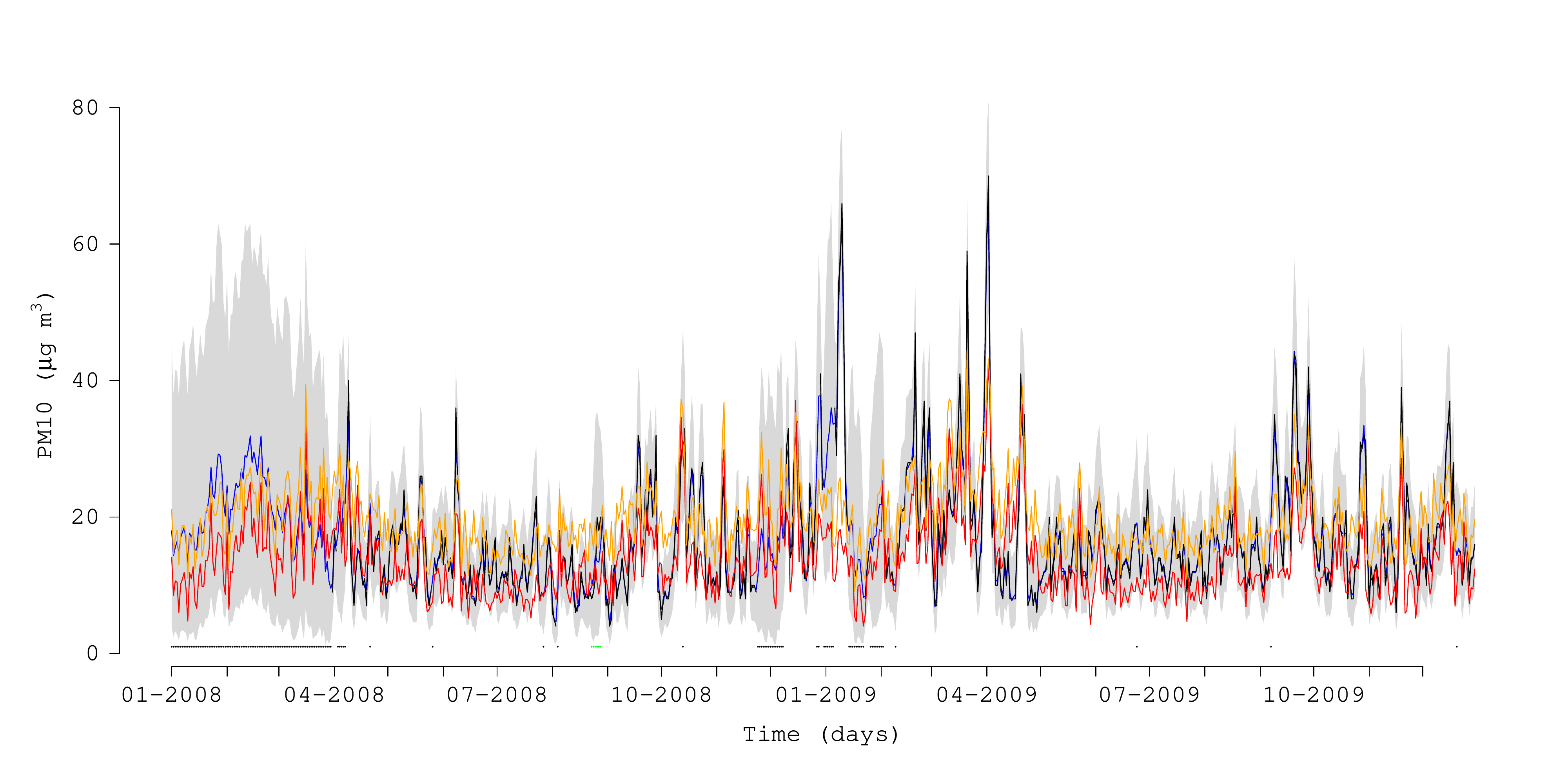}\label{fitted-s47}} \\
\subfigure[Station 151]{\includegraphics[trim={0 1cm 0 2cm},clip,width=\textwidth]{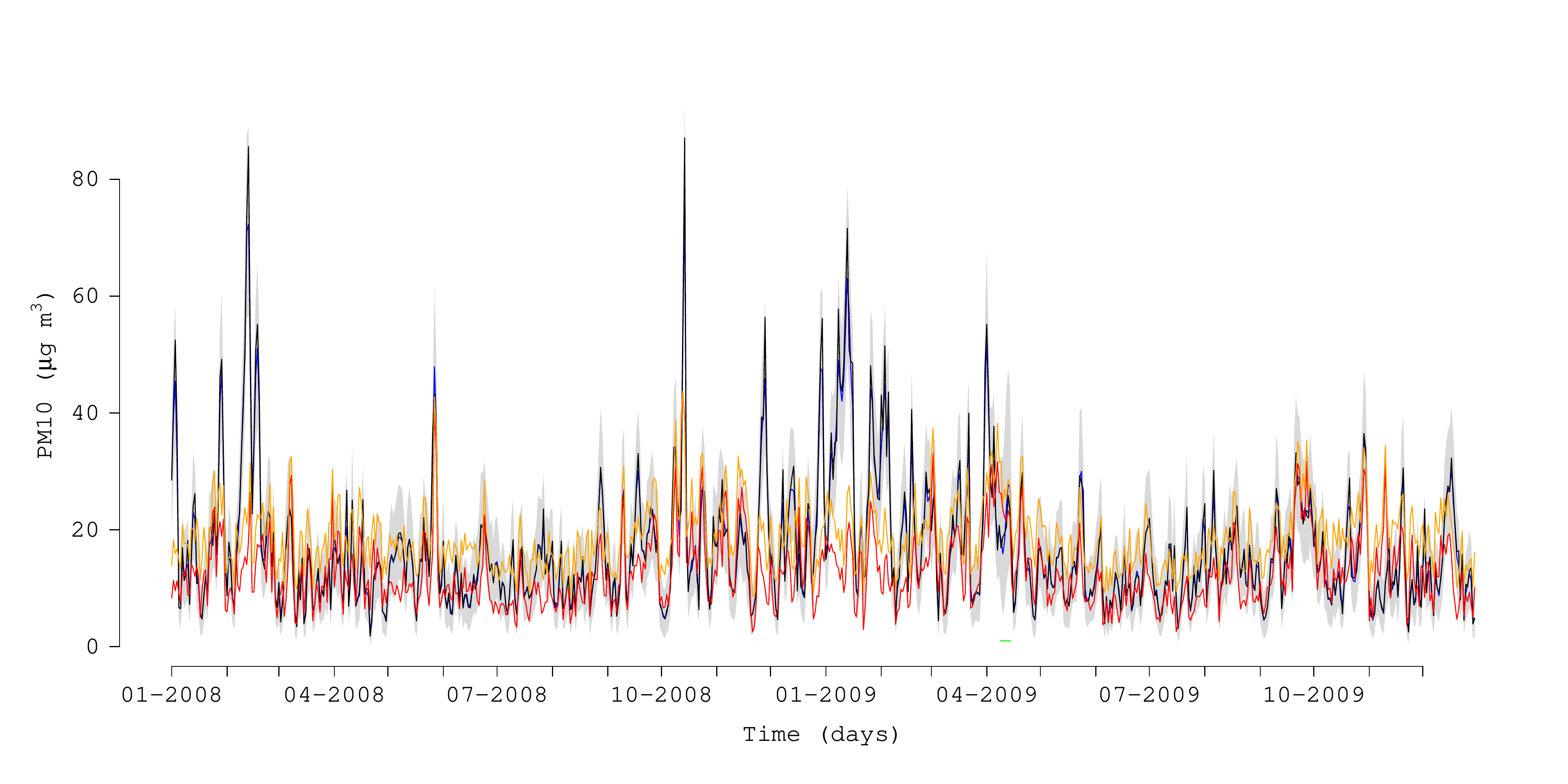}\label{fitted-s151}}\\ 
\caption{Fitted and observed PM$_{10}$ for several example stations. Lines correspond to PM$_{10}$ observed (black), CTM output (red), non space-time, regression (orange), and $m=36$ Adaptive DNNGP (blue) with associated 95\% CI band (gray). Prediction assessment holdout and actual missing observations are indicated with green and black points respectively.}\label{fitted-stations}
\end{figure}

Figure~\ref{fitted-stations} illustrates the observed and candidate model fitted/predicted PM$_{10}$ for three stations. These figures are representative of other stations and show $i$) the downward bias in CTM output, $ii$) improved fit and prediction with the addition of spatio-temporal random effects over non-spatial regression, and $iii$) appropriate widening of CIs for missing station observations. 

Table~\ref{tab:aprilSummary} provides out-of-sample prediction validation metrics for the non space-time and DNNGP Adaptive and Simple models that can be compared with April 1-14, 2009 holdout validation metrics presented in \citet[Table 1]{hamm2015}. Compared to the time invariant (day specific) space-varying intercept (SVI) and space-varying coefficients (SVC) models considered in \cite{hamm2015}, the DNNGP models' RMSPE and bias are lower (more accurate, less biased) while the $R^2$ values are comparable. We also added results for the simple linear regression (SLR) model in the first column of Table~\ref{tab:aprilSummary}. The simple linear regression model does not consider spatio-temporal effects nor does it consider a time varying intercept (unlike the day specific results presented in \citet{hamm2015}) which may explain the poor predictive performance---it is more meaningful to compare the DNNGP model prediction metrics to the days specific metrics presented in \cite{hamm2015}.



\begin{table}[t]
\centering
{\small
\caption{April 1-14, 2009 25\% holdout set prediction summary for comparison with time invariant spatial regression models presented in \cite[Table 1]{hamm2015}.} \label{tab:aprilSummary}
\begin{tabular}{cccccc}
\hline 
&&\multicolumn{2}{c}{Adaptive}&\multicolumn{2}{c}{Simple}\\
       &SLR &m=25& m=36& m=25& m=36\\ \hline
RMSPE&8.48&4.97&5.05&5.06&5.04\\
Bias&0.71&0.20&0.20&0.23&0.22\\
$R^2$&0.14&0.69&0.68&0.68&0.68\\ 
\hline
\end{tabular}
}
\end{table}

\begin{figure}[!h]
\centering
\subfigure[April 3, 2009 PM$_{10}$]{\includegraphics[trim={0 0.5cm 0 1cm},clip,width=2.in]{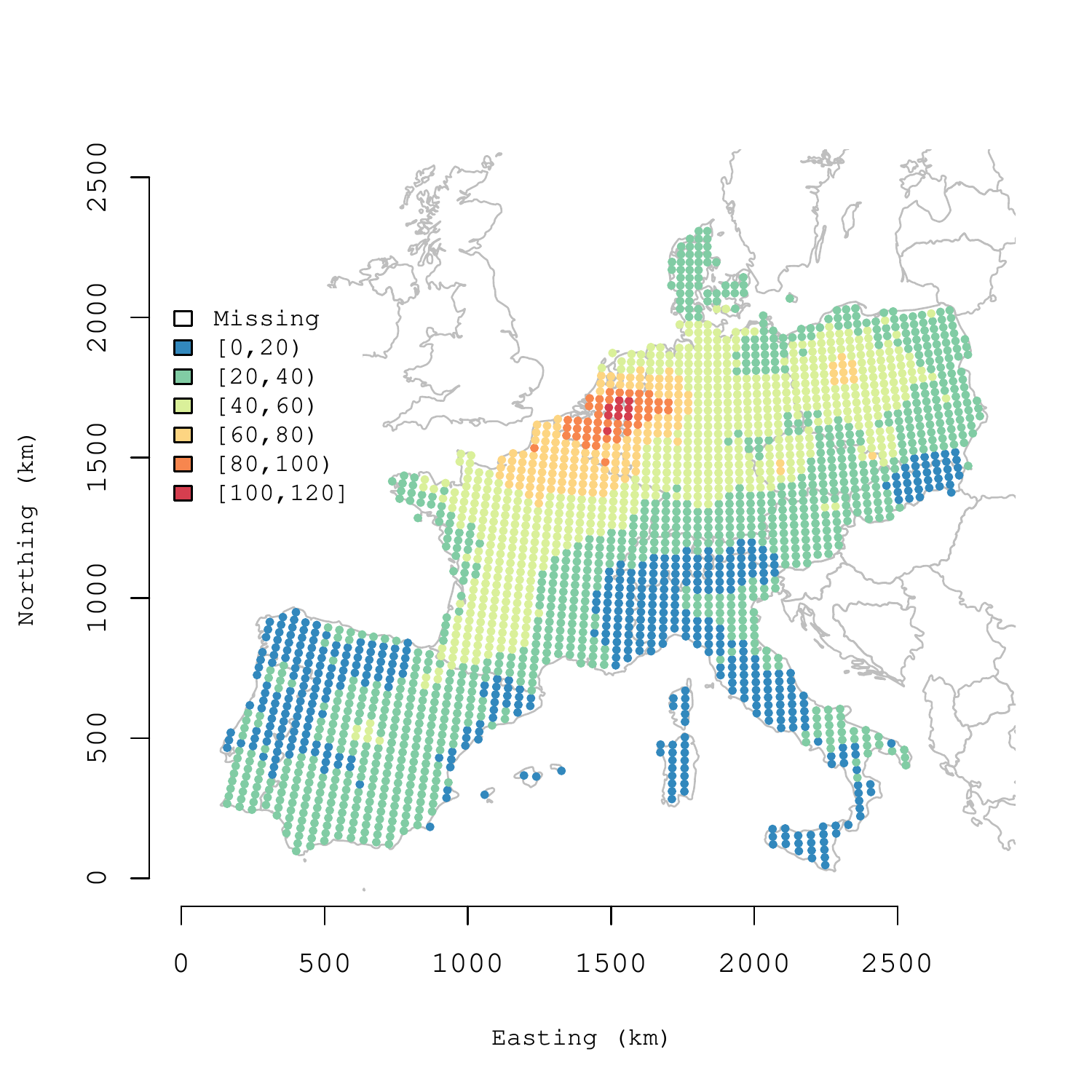}\label{april-3-y-pred}} 
\hfil
\subfigure[April 3, 2009 PM$_{10}$ $>$ 50 $\mu g$ $m^{-3}$]{\includegraphics[trim={0 0.5cm 0 1cm},clip,width=2.in]{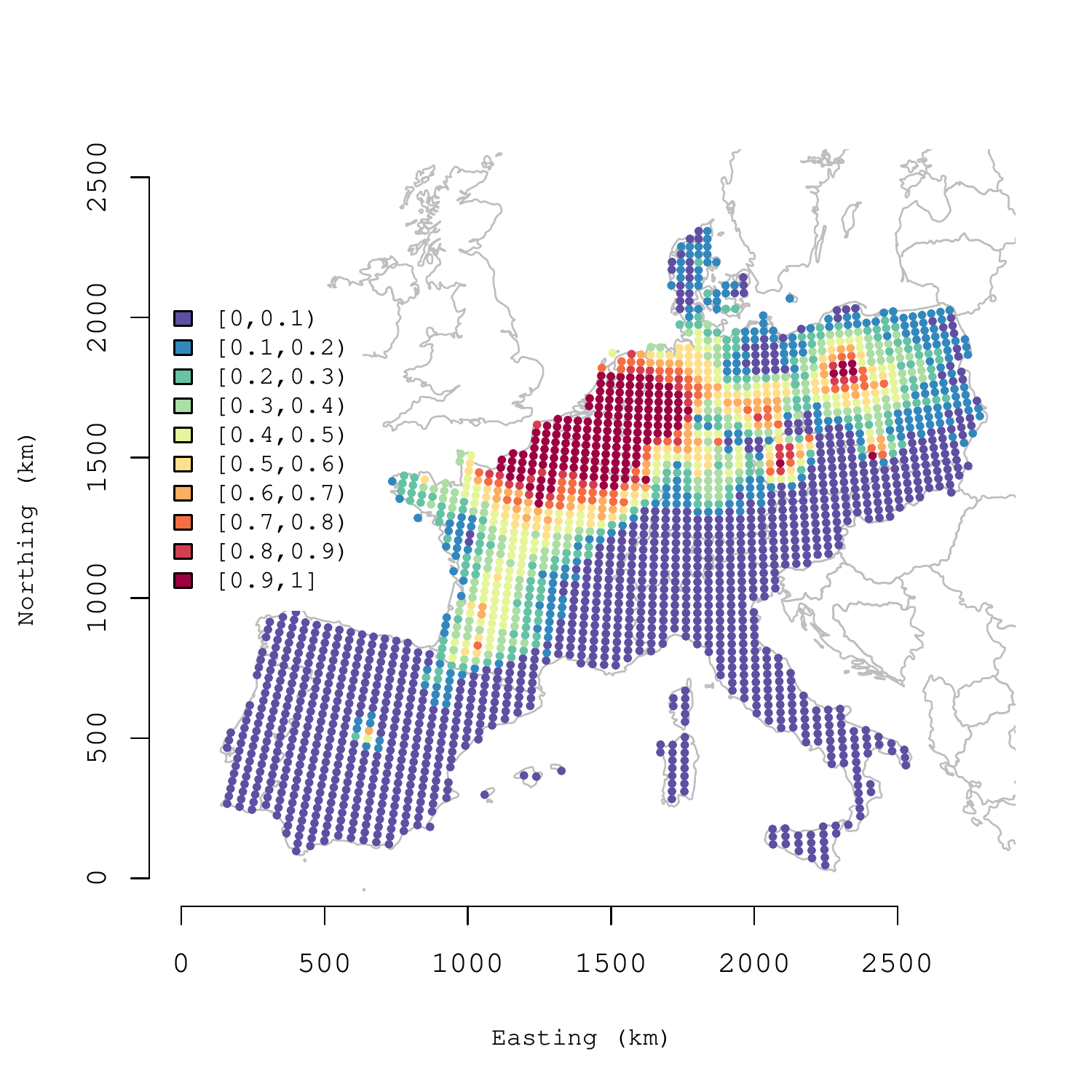}\label{april-3-y-over-50}} 
\\
\subfigure[April 5, 2009 PM$_{10}$]{\includegraphics[trim={0 0.5cm 0 1cm},clip,width=2.in]{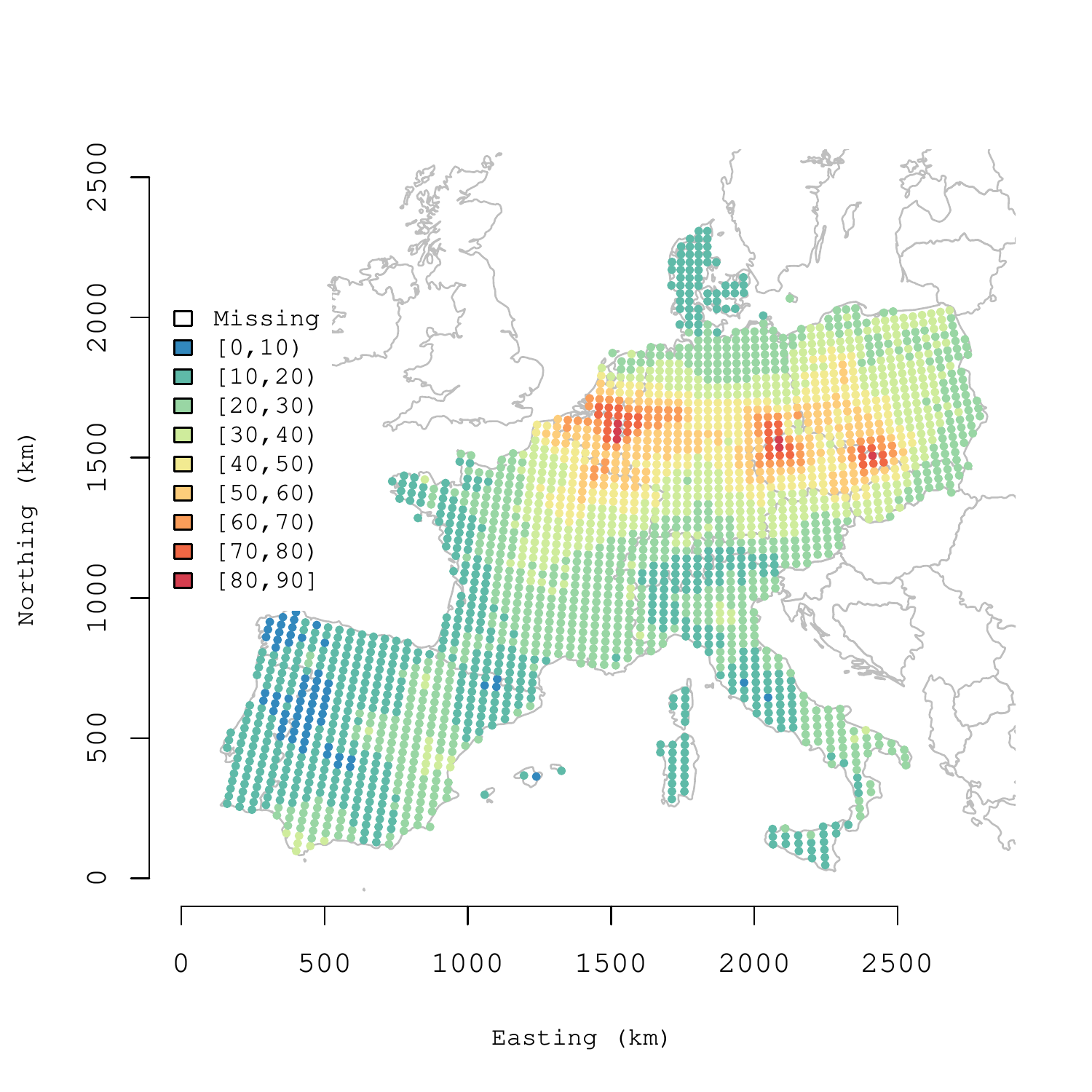}\label{april-5-y-pred}} 
\hfil
\subfigure[April 5, 2009 PM$_{10}$ $>$ 50 $\mu g$ $m^{-3}$]{\includegraphics[trim={0 0.5cm 0 1cm},clip,width=2.in]{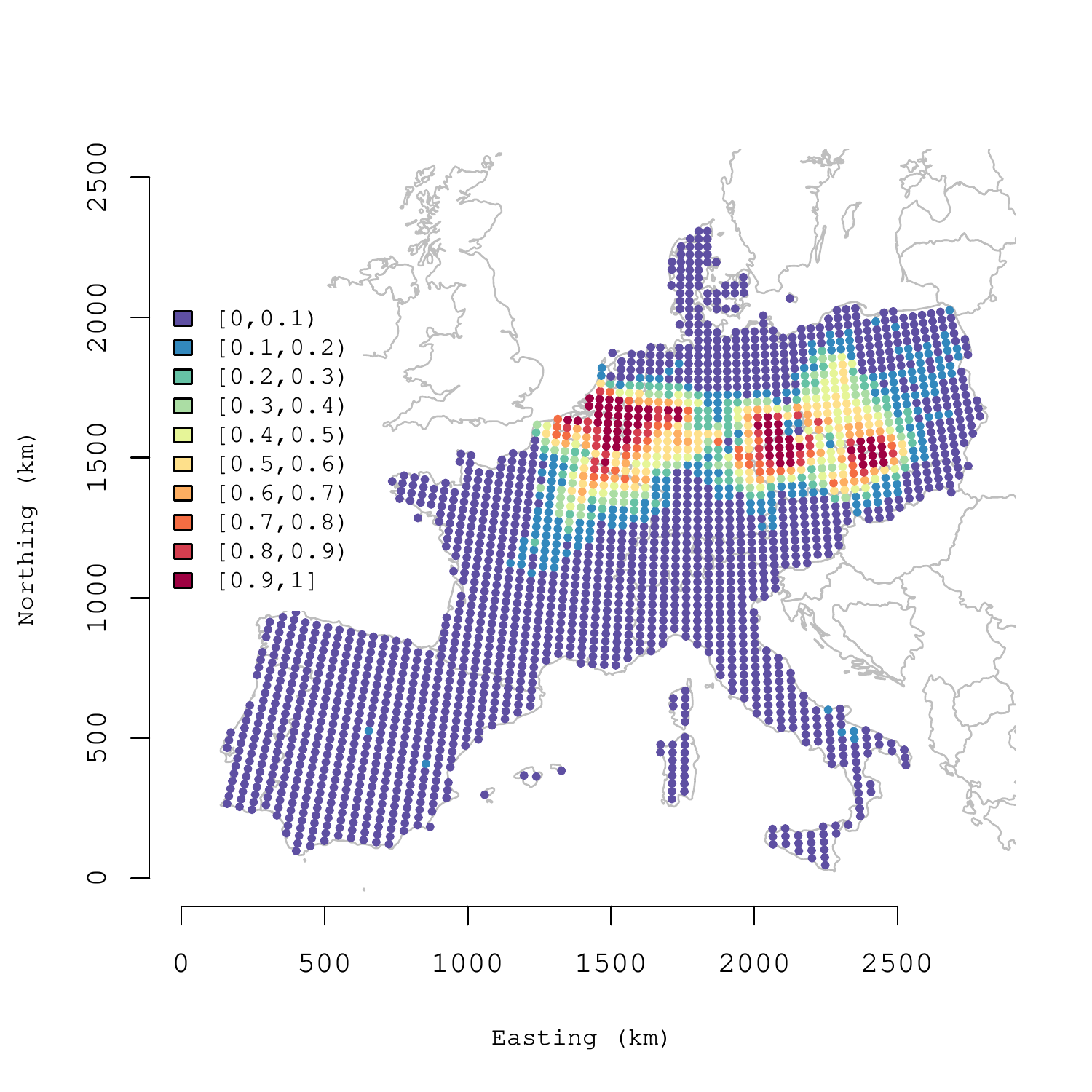}\label{april-5-y-over-50}} 
\caption{Predicted PM$_{10}$ and probability of exceeding 50 $\mu g$ $m^{-3}$ for two example dates.}\label{y-pred}
\end{figure}

In addition to these prediction metrics, maps of posterior predictive summaries at CTM output locations are key inputs to pollution monitoring and mitigation programs. For example, Figure~\ref{y-pred}
 provides maps of the posterior predictive prediction median and the probability of exceeding the 50 $\mu$g m$^{-3}$ regulatory threshold for two example dates. These dates were also examined in \citet[Figure 8]{hamm2015} and the resulting maps are directly comparable. The DNNGP, Figure~\ref{y-pred}, and SVC maps in \citet{hamm2015} show broadly similar patterns, although there are some differences. For example the high pollution over western France and northern Spain on April 3, 2009 is captured more clearly by \citet{hamm2015}. The SVI and SVC models in \citet{hamm2015} did not account for temporal correlation over days---clearly not an accurate assumption. In contrast the DNNGP models smooth over days, which can provide improved predictive performance although the details of highly dynamic events may be less well captured than by the daily specific models used in \citet{hamm2015}.  

The last row in Table~\ref{tab:realParamsDyn} provides the CPU time for delivering 25,000 MCMC iterations. As detailed in Section~\ref{sec:adapt} particular components of the algorithm are easily distributed across multiple CPUs. In particular, partitioning the update of $w(\ell_{i})$'s across multiple CPUs yields substantial computational gains. The DNNGP samplers were implemented in C++ and leveraged OpenMP \citep{openmp} and Intel Math Kernel Library's (MKL) threaded BLAS and LAPACK routines for matrix \citep{mkl}. Running on a single CPU the Adaptive $m$=25 model would require approximately 260 hours. However, when distributed across a 10-core Xeon CPU the total run time was approximately 24 hours.

\section{Conclusion}\label{sec:dconc}
We have addressed the problem of modeling large spatio-temporal datasets, specifically for settings where full inference (with proper accounting for uncertainty) is required at arbitrary resolutions. We presented a new class of dynamic nearest-neighbor Gaussian Process (DNNGP) models over a continuous space-time domain. The DNNGP is a legitimate Gaussian process whose realizations over finite sets enjoy sparse precision matrices, thereby accruing massive computational savings in terms of storage and flops. The DNNGP depends upon the conditional independence of the random effects given its neighbors. We used the strength of a correlation function to construct a parametric distance metric in a spatio-temporal domain. Using monotonicity of covariance functions we showed that it is possible to update neighbor sets using a scalable search algorithm and outlined the steps of a Gibbs' sampler that avoids expensive matrix decompositions and is linear in the number of measurements in terms of storage and flops. 

Analyses combining European CTM outputs and observed data has, to date, focused mainly on spatial analysis per day \citep{DenbyEtAl08a,DenbyEtAl10a,hamm2015}, few studies implement full space-time geostatistical models, e.g., \cite{GraelerEtAl11a}, and none consider such a long time series. The work presented in this paper focuses on DNNGP development to facilitate novel analyses of spatially-indexed time-series data such as PM$_{10}$ concentrations. Here, in addition to improved predictive performance, inference on model covariance parameters provided insight into space-time structures not captured by the LOTOS-EUROS CTM.  Whilst previous analyses of individual days had shown strong residual spatial structure, analysis of this long time-series with explicit time correlation parameters reveals the residual temporal structure dominates.  The temporal range is physically reasonable considering the life-time of PM$_{10}$ is in the order of days and its variability is driven by alternating synoptic meteorological conditions with certain conditions usually lasting for several days to weeks. 

Reproducing the observed variability with a CTM remains challenging, especially for episodic conditions which are associated with particular (stagnant) meteorological conditions or occasional large emissions from, e.g., large wild fires~\citep{RHoniEtAl13a} or dust events~\citep{BirmiliEtAl08a}. A particular issue to be resolved is the lack of detail in the anthropogenic emission variability. This variability is prescribed using static emission profiles for the month of the year, day of the week, and hour of the day. Further detailing through inclusion of meteorological effects may improve the modeling \citep{MuesEtAl14a} and remove the monthly signature found in this analysis. 

The type of analysis that is performed depends on the study objective.  Analysis of individual days is important for the study of individual air pollution events and the associated performance of the CTM \citep{hamm2015}.  The analysis presented in this paper affords a different perspective by identifying long-term space-time structures that offer insight into the performance of the CTM. The DNNGP also yields more accurate predictions than previous studies of these same data. 

Apart from massive scalability, the DNNGP retains the versatility of process-based modeling and can be used as a sparsity-inducing proper prior in any Bayesian hierarchical model designed to deliver full inference at arbitrary spatio-temporal resolutions for massive spatio-temporal datasets. We have developed DNNGP assuming an isotropic non-stationary spatio-temporal covariance structure. However, it can also be potentially extended to certain classes of non-stationary space-time covariances \citep[see Section S3 of the supplemental article][]{dnngpsupp}. Even more generally, the DNNGP can be used for any spatio-temporal random effect in the second stage of specification in hierarchical models for non-Gaussian responses. 
Full posterior distributions for the underlying spatio-temporal process are available at any arbitrary location and time point. Thus, DNNGP can potentially be deployed for statistical downscaling of spatio-temporal datasets obtained at coarser resolutions (e.g. climate downscaling). We also plan to migrate our lower level C++ code into the \texttt{spBayes} R package for wider and friendlier accessibility to DNNGP models. 


\begin{description}


\item[Acknowledgments:] The authors thank the Associate Editor and anonymous reviewers for their suggestions which considerably improved the manuscript. The authors also acknowledge Arjo Segers (TNO) for his support with the LOTOS-EUROS CTM. Andrew Finley was supported by National Science Foundation (NSF) DMS-1513481, EF-1137309, EF-1241874, and EF-1253225, as well as NASA Carbon Monitoring System grants. Sudipto Banerjee was supported by NSF DMS-1513654. 

\end{description}

\begin{supplement} [id=DNNGPSupplement]
\stitle{Supplement to "Non-separable Dynamic Nearest-Neighbor Gaussian Process Models for Large spatio-temporal Data With an Application to Particulate Matter Analysis"}
\slink[doi]{COMPLETED BY THE TYPESETTER}
\sdatatype{.pdf}
\sdescription{File containing supplementary materials including a formal construction of eligible sets, additional simulation experiments and possible extension of DNNGP to model non-stationary covariances.}
\end{supplement}

\bibliographystyle{imsart-nameyear}
\bibliography{DNNGPbib}

\begin{thebibliography}{76}

\bibitem[\protect\citeauthoryear{Allcroft and Glasbey}{2003}]{all03}
\begin{barticle}[author]
\bauthor{\bsnm{Allcroft},~\bfnm{D.~J.}\binits{D.~J.}} \AND
  \bauthor{\bsnm{Glasbey},~\bfnm{C.~A.}\binits{C.~A.}}
(\byear{2003}).
\btitle{A Latent Gaussian Markov Random Field Model for Spatio-temporal
  Rainfall Disaggregation}.
\bjournal{Journal of the Royal Statistical society, Series C}
\bvolume{52}
\bpages{487–-498}.
\end{barticle}
\endbibitem

\bibitem[\protect\citeauthoryear{Bai, Song and Raghunathan}{2012}]{bai12}
\begin{barticle}[author]
\bauthor{\bsnm{Bai},~\bfnm{Y.}\binits{Y.}},
  \bauthor{\bsnm{Song},~\bfnm{P.~X.~K.}\binits{P.~X.~K.}} \AND
  \bauthor{\bsnm{Raghunathan},~\bfnm{T.~E.}\binits{T.~E.}}
(\byear{2012}).
\btitle{Bayesian Dynamic Modeling for Large Space-time Datasets using Gaussian
  Predictive Processes}.
\bjournal{Journal of the Royal Statistical society, Series B}
\bvolume{74}
\bpages{799–-824}.
\end{barticle}
\endbibitem

\bibitem[\protect\citeauthoryear{Banerjee, Carlin and Gelfand}{2014}]{ban14}
\begin{binbook}[author]
\bauthor{\bsnm{Banerjee},~\bfnm{S.}\binits{S.}},
  \bauthor{\bsnm{Carlin},~\bfnm{B.~P.}\binits{B.~P.}} \AND
  \bauthor{\bsnm{Gelfand},~\bfnm{A.~E.}\binits{A.~E.}}
(\byear{2014}).
\btitle{Hierarchical Modeling and Analysis for Spatial Data},
\bedition{second} ed.
\bpublisher{Chapman \& Hall/CRC}, \baddress{Boca Raton, FL}.
\end{binbook}
\endbibitem

\bibitem[\protect\citeauthoryear{Banerjee et~al.}{2008}]{ban08}
\begin{barticle}[author]
\bauthor{\bsnm{Banerjee},~\bfnm{S.}\binits{S.}},
  \bauthor{\bsnm{Gelfand},~\bfnm{A.~E.}\binits{A.~E.}},
  \bauthor{\bsnm{Finley},~\bfnm{A.~O.}\binits{A.~O.}} \AND
  \bauthor{\bsnm{Sang},~\bfnm{H.}\binits{H.}}
(\byear{2008}).
\btitle{Gaussian Predictive Process Models for Large Spatial Datasets}.
\bjournal{Journal of the Royal Statistical society, Series B}
\bvolume{70}
\bpages{825-848}.
\end{barticle}
\endbibitem

\bibitem[\protect\citeauthoryear{Bevilacqua et~al.}{2012}]{bevil12}
\begin{barticle}[author]
\bauthor{\bsnm{Bevilacqua},~\bfnm{Moreno}\binits{M.}},
  \bauthor{\bsnm{Gaetan},~\bfnm{Carlo}\binits{C.}},
  \bauthor{\bsnm{Mateu},~\bfnm{Jorge}\binits{J.}} \AND
  \bauthor{\bsnm{Porcu},~\bfnm{Emilio}\binits{E.}}
(\byear{2012}).
\btitle{Estimating Space and Space-Time Covariance Functions for Large Data
  Sets: A Weighted Composite Likelihood Approach}.
\bjournal{Journal of the American Statistical Association}
\bvolume{107}
\bpages{268-280}.
\bdoi{10.1080/01621459.2011.646928}
\end{barticle}
\endbibitem

\bibitem[\protect\citeauthoryear{Bevilacqua et~al.}{2015}]{bevil15}
\begin{barticle}[author]
\bauthor{\bsnm{Bevilacqua},~\bfnm{M.}\binits{M.}},
  \bauthor{\bsnm{Fass},~\bfnm{{\`o}~A.}\binits{{\`o}.~A.}},
  \bauthor{\bsnm{Gaetan},~\bfnm{C}\binits{C.}} \AND
  \bauthor{\bsnm{Velandia},~\bfnm{D.}\binits{D.}}
(\byear{2015}).
\btitle{Covariance tapering for multivariate Gaussian random fields
  estimation}.
\bjournal{Statistical Methods and Application}.
\bdoi{DOI: 10.1007/s10260-015-0338-3}
\end{barticle}
\endbibitem

\bibitem[\protect\citeauthoryear{Birmili et~al.}{2008}]{BirmiliEtAl08a}
\begin{barticle}[author]
\bauthor{\bsnm{Birmili},~\bfnm{W.}\binits{W.}},
  \bauthor{\bsnm{Schepanski},~\bfnm{K.}\binits{K.}},
  \bauthor{\bsnm{Ansmann},~\bfnm{A.}\binits{A.}},
  \bauthor{\bsnm{Spindler},~\bfnm{G.}\binits{G.}},
  \bauthor{\bsnm{Tegen},~\bfnm{I.}\binits{I.}},
  \bauthor{\bsnm{Wehner},~\bfnm{B.}\binits{B.}},
  \bauthor{\bsnm{Nowak},~\bfnm{A.}\binits{A.}},
  \bauthor{\bsnm{Reimer},~\bfnm{E.}\binits{E.}},
  \bauthor{\bsnm{Mattis},~\bfnm{I.}\binits{I.}},
  \bauthor{\bsnm{Muller},~\bfnm{K.}\binits{K.}},
  \bauthor{\bsnm{Bruggemann},~\bfnm{E.}\binits{E.}},
  \bauthor{\bsnm{Gnauk},~\bfnm{T.}\binits{T.}},
  \bauthor{\bsnm{Herrmann},~\bfnm{H.}\binits{H.}},
  \bauthor{\bsnm{Wiedensohler},~\bfnm{A.}\binits{A.}},
  \bauthor{\bsnm{Althausen},~\bfnm{D.}\binits{D.}},
  \bauthor{\bsnm{Schladitz},~\bfnm{A.}\binits{A.}},
  \bauthor{\bsnm{Tuch},~\bfnm{T.}\binits{T.}} \AND
  \bauthor{\bsnm{Loschau},~\bfnm{G.}\binits{G.}}
(\byear{2008}).
\btitle{A Case of Extreme Particulate Matter Concentrations over Central Europe
  Caused by Dust Emitted over the Southern Ukraine}.
\bjournal{Atmospheric Chemistry and Physics}
\bvolume{8}
\bpages{997-1016}.
\end{barticle}
\endbibitem

\bibitem[\protect\citeauthoryear{Brauer et~al.}{2011}]{BrauerEtAl11a}
\begin{barticle}[author]
\bauthor{\bsnm{Brauer},~\bfnm{Michael}\binits{M.}},
  \bauthor{\bsnm{Amann},~\bfnm{Markus}\binits{M.}},
  \bauthor{\bsnm{Burnett},~\bfnm{Rick~T.}\binits{R.~T.}},
  \bauthor{\bsnm{Cohen},~\bfnm{Aaron}\binits{A.}},
  \bauthor{\bsnm{Dentener},~\bfnm{Frank}\binits{F.}},
  \bauthor{\bsnm{Ezzati},~\bfnm{Majid}\binits{M.}},
  \bauthor{\bsnm{Henderson},~\bfnm{Sarah~B.}\binits{S.~B.}},
  \bauthor{\bsnm{Krzyzanowski},~\bfnm{Michal}\binits{M.}},
  \bauthor{\bsnm{Martin},~\bfnm{Randall~V.}\binits{R.~V.}},
  \bauthor{\bsnm{Van~Dingenen},~\bfnm{Rita}\binits{R.}},
  \bauthor{\bparticle{van} \bsnm{Donkelaar},~\bfnm{Aaron}\binits{A.}} \AND
  \bauthor{\bsnm{Thurston},~\bfnm{George~D.}\binits{G.~D.}}
(\byear{2011}).
\btitle{Exposure Assessment for Estimation of the Global Burden of Disease
  Attributable to Outdoor Air Pollution}.
\bjournal{Environmental Science and Technology}
\bvolume{46}
\bpages{652-660}.
\bdoi{http://dx.doi.org/10.1021/es2025752}
\end{barticle}
\endbibitem

\bibitem[\protect\citeauthoryear{Brunekreef and Holgate}{2002}]{BrunekreefH02a}
\begin{barticle}[author]
\bauthor{\bsnm{Brunekreef},~\bfnm{B.}\binits{B.}} \AND
  \bauthor{\bsnm{Holgate},~\bfnm{S.~T.}\binits{S.~T.}}
(\byear{2002}).
\btitle{Air Pollution and Health}.
\bjournal{The Lancet}
\bvolume{360}
\bpages{1233-1242}.
\end{barticle}
\endbibitem

\bibitem[\protect\citeauthoryear{Candiani et~al.}{2013}]{CandianiEtAl13a}
\begin{barticle}[author]
\bauthor{\bsnm{Candiani},~\bfnm{Gabriele}\binits{G.}},
  \bauthor{\bsnm{Carnevale},~\bfnm{Claudio}\binits{C.}},
  \bauthor{\bsnm{Finzi},~\bfnm{Giovanna}\binits{G.}},
  \bauthor{\bsnm{Pisoni},~\bfnm{Enrico}\binits{E.}} \AND
  \bauthor{\bsnm{Volta},~\bfnm{Marialuisa}\binits{M.}}
(\byear{2013}).
\btitle{A Comparison of Reanalysis Techniques: Applying Optimal Interpolation
  and Ensemble {K}alman Filtering to Improve Air Quality Monitoring at
  Mesoscale}.
\bjournal{Science of the Total Environment}
\bvolume{458-460}
\bpages{7-14}.
\bdoi{http://dx.doi.org/10.1016/j.scitotenv.2013.03.089}
\end{barticle}
\endbibitem

\bibitem[\protect\citeauthoryear{{European Commission}}{2015}]{ECStandards15}
\begin{barticle}[author]
\bauthor{\bsnm{{European Commission}}}
(\byear{2015}).
\btitle{European Union Air Quality Standards}.
\bjournal{http://ec.europa.eu/environment/air/quality/standards.htm}.
\end{barticle}
\endbibitem

\bibitem[\protect\citeauthoryear{Crainiceanu, Diggle and
  Rowlingson}{2008}]{cra08}
\begin{barticle}[author]
\bauthor{\bsnm{Crainiceanu},~\bfnm{C.~M.}\binits{C.~M.}},
  \bauthor{\bsnm{Diggle},~\bfnm{P.~J.}\binits{P.~J.}} \AND
  \bauthor{\bsnm{Rowlingson},~\bfnm{B.}\binits{B.}}
(\byear{2008}).
\btitle{Bivariate Binomial Spatial Modeling of Loa Loa Prevalence in Tropical
  Africa}.
\bjournal{Journal of the American Statistical Association}
\bvolume{103}
\bpages{21-37}.
\end{barticle}
\endbibitem

\bibitem[\protect\citeauthoryear{Cressie and Huang}{1999}]{cres99}
\begin{barticle}[author]
\bauthor{\bsnm{Cressie},~\bfnm{N.}\binits{N.}} \AND
  \bauthor{\bsnm{Huang},~\bfnm{H.~C.}\binits{H.~C.}}
(\byear{1999}).
\btitle{Classes of Nonseparable, Spatio-temporal Stationary Covariance
  Functions}.
\bjournal{Journal of the American Statistical Association}
\bvolume{94}
\bpages{1330–-1340}.
\end{barticle}
\endbibitem

\bibitem[\protect\citeauthoryear{Cressie and Johannesson}{2008}]{cres08}
\begin{barticle}[author]
\bauthor{\bsnm{Cressie},~\bfnm{N.}\binits{N.}} \AND
  \bauthor{\bsnm{Johannesson},~\bfnm{G.}\binits{G.}}
(\byear{2008}).
\btitle{Fixed Rank Kriging for Very Large Data Sets}.
\bjournal{Journal of the Royal Statistical society, Series B}
\bvolume{70}
\bpages{209-226}.
\end{barticle}
\endbibitem

\bibitem[\protect\citeauthoryear{Cressie, Shi and Kang}{2010}]{cres10}
\begin{barticle}[author]
\bauthor{\bsnm{Cressie},~\bfnm{N.}\binits{N.}},
  \bauthor{\bsnm{Shi},~\bfnm{T.}\binits{T.}} \AND
  \bauthor{\bsnm{Kang},~\bfnm{E.~L.}\binits{E.~L.}}
(\byear{2010}).
\btitle{Fixed Rank Filtering for Spatio-temporal Data}.
\bjournal{Journal of Computational and Graphical Statistics}
\bvolume{19}
\bpages{724–-745}.
\end{barticle}
\endbibitem

\bibitem[\protect\citeauthoryear{Cressie and Wikle}{2011}]{creswikle11}
\begin{binbook}[author]
\bauthor{\bsnm{Cressie},~\bfnm{Noel A.~C.}\binits{N.~A.~C.}} \AND
  \bauthor{\bsnm{Wikle},~\bfnm{Christopher~K.}\binits{C.~K.}}
(\byear{2011}).
\btitle{Statistics for Spatio-temporal Data}.
\bseries{Wiley series in probability and statistics}.
\bpublisher{Hoboken, N.J. Wiley}.
\end{binbook}
\endbibitem

\bibitem[\protect\citeauthoryear{Dagum and Menon}{1998}]{openmp}
\begin{barticle}[author]
\bauthor{\bsnm{Dagum},~\bfnm{Leonardo}\binits{L.}} \AND
  \bauthor{\bsnm{Menon},~\bfnm{Ramesh}\binits{R.}}
(\byear{1998}).
\btitle{OpenMP: An Industry Standard API for Shared-memory Programming}.
\bjournal{Computational Science \& Engineering, IEEE}
\bvolume{5}
\bpages{46--55}.
\end{barticle}
\endbibitem

\bibitem[\protect\citeauthoryear{Datta et~al.}{2016a}]{datta14}
\begin{barticle}[author]
\bauthor{\bsnm{Datta},~\bfnm{A.}\binits{A.}},
  \bauthor{\bsnm{Banerjee},~\bfnm{S.}\binits{S.}},
  \bauthor{\bsnm{Finley},~\bfnm{A.~O.}\binits{A.~O.}} \AND
  \bauthor{\bsnm{Gelfand},~\bfnm{A.~E.}\binits{A.~E.}}
(\byear{2016}a).
\btitle{Hierarchical Nearest-Neighbor Gaussian Process Models for Large
  Geostatistical Datasets}.
\bjournal{Journal of the American Statistical Association}
\bpages{In Press}.
\bdoi{DOI:10.1080/01621459.2015.1044091}
\end{barticle}
\endbibitem

\bibitem[\protect\citeauthoryear{Datta et~al.}{2016b}]{dnngpsupp}
\begin{barticle}[author]
\bauthor{\bsnm{Datta},~\bfnm{A.}\binits{A.}},
  \bauthor{\bsnm{Banerjee},~\bfnm{S.}\binits{S.}},
  \bauthor{\bsnm{Finley},~\bfnm{A.~O.}\binits{A.~O.}},
  \bauthor{\bsnm{Hamm},~\bfnm{N.~A.~S.}\binits{N.~A.~S.}} \AND
  \bauthor{\bsnm{Schaap},~\bfnm{M.}\binits{M.}}
(\byear{2016}b).
\btitle{Supplement to "Non-separable Dynamic Nearest-Neighbor Gaussian Process
  Models for Large spatio-temporal Data With an Application to Particulate
  Matter Analysis"}.
\end{barticle}
\endbibitem

\bibitem[\protect\citeauthoryear{Denby et~al.}{2008}]{DenbyEtAl08a}
\begin{barticle}[author]
\bauthor{\bsnm{Denby},~\bfnm{B.}\binits{B.}},
  \bauthor{\bsnm{Schaap},~\bfnm{M.}\binits{M.}},
  \bauthor{\bsnm{Segers},~\bfnm{A.}\binits{A.}},
  \bauthor{\bsnm{Builtjes},~\bfnm{P.}\binits{P.}} \AND
  \bauthor{\bsnm{Horalek},~\bfnm{J.}\binits{J.}}
(\byear{2008}).
\btitle{Comparison of Two Data Assimilation Methods for Assessing {PM10}
  Exceedances on the {E}uropean Scale}.
\bjournal{Atmospheric Environment}
\bvolume{42}
\bpages{7122-7134}.
\bdoi{10.1016/j.atmosenv.2008.05.058}
\end{barticle}
\endbibitem

\bibitem[\protect\citeauthoryear{Denby et~al.}{2010}]{DenbyEtAl10a}
\begin{barticle}[author]
\bauthor{\bsnm{Denby},~\bfnm{B.}\binits{B.}},
  \bauthor{\bsnm{Sundvor},~\bfnm{I.}\binits{I.}},
  \bauthor{\bsnm{Cassiani},~\bfnm{M.}\binits{M.}}, \bauthor{\bparticle{de}
  \bsnm{Smet},~\bfnm{P.}\binits{P.}}, \bauthor{\bparticle{de}
  \bsnm{Leeuw},~\bfnm{F.}\binits{F.}} \AND
  \bauthor{\bsnm{Horalek},~\bfnm{J.}\binits{J.}}
(\byear{2010}).
\btitle{Spatial Mapping of Ozone and SO2 Trends in Europe}.
\bjournal{Science Of The Total Environment}
\bvolume{408}
\bpages{4795-4806}.
\bdoi{10.1016/j.scitotenv.2010.06.021}
\end{barticle}
\endbibitem

\bibitem[\protect\citeauthoryear{Du, Zhang and Mandrekar}{2009}]{du09}
\begin{barticle}[author]
\bauthor{\bsnm{Du},~\bfnm{J.}\binits{J.}},
  \bauthor{\bsnm{Zhang},~\bfnm{H.}\binits{H.}} \AND
  \bauthor{\bsnm{Mandrekar},~\bfnm{V.~S.}\binits{V.~S.}}
(\byear{2009}).
\btitle{Fixed-domain Asymptotic Properties of Tapered Maximum Likelihood
  Estimators}.
\bjournal{Annals of Statistics}
\bvolume{37}
\bpages{3330-3361}.
\end{barticle}
\endbibitem

\bibitem[\protect\citeauthoryear{Eeftens et~al.}{2012}]{EeftensEtAl12a}
\begin{barticle}[author]
\bauthor{\bsnm{Eeftens},~\bfnm{M.}\binits{M.}},
  \bauthor{\bsnm{Tsai},~\bfnm{M.~Y.}\binits{M.~Y.}},
  \bauthor{\bsnm{Ampe},~\bfnm{C.}\binits{C.}},
  \bauthor{\bsnm{Anwander},~\bfnm{B.}\binits{B.}},
  \bauthor{\bsnm{Beelen},~\bfnm{R.}\binits{R.}},
  \bauthor{\bsnm{Bellander},~\bfnm{T.}\binits{T.}},
  \bauthor{\bsnm{Cesaroni},~\bfnm{G.}\binits{G.}},
  \bauthor{\bsnm{Cirach},~\bfnm{M.}\binits{M.}},
  \bauthor{\bsnm{Cyrys},~\bfnm{J.}\binits{J.}}, \bauthor{\bparticle{de}
  \bsnm{Hoogh},~\bfnm{K.}\binits{K.}},
  \bauthor{\bsnm{De~Nazelle},~\bfnm{A.}\binits{A.}}, \bauthor{\bparticle{de}
  \bsnm{Vocht},~\bfnm{F.}\binits{F.}},
  \bauthor{\bsnm{Declercq},~\bfnm{C.}\binits{C.}},
  \bauthor{\bsnm{Dedele},~\bfnm{A.}\binits{A.}},
  \bauthor{\bsnm{Eriksen},~\bfnm{K.}\binits{K.}},
  \bauthor{\bsnm{Galassi},~\bfnm{C.}\binits{C.}},
  \bauthor{\bsnm{Grazuleviciene},~\bfnm{R.}\binits{R.}},
  \bauthor{\bsnm{Grivas},~\bfnm{G.}\binits{G.}},
  \bauthor{\bsnm{Heinrich},~\bfnm{J.}\binits{J.}},
  \bauthor{\bsnm{Hoffmann},~\bfnm{B.}\binits{B.}},
  \bauthor{\bsnm{Iakovides},~\bfnm{M.}\binits{M.}},
  \bauthor{\bsnm{Ineichen},~\bfnm{A.}\binits{A.}},
  \bauthor{\bsnm{Katsouyanni},~\bfnm{K.}\binits{K.}},
  \bauthor{\bsnm{Korek},~\bfnm{M.}\binits{M.}},
  \bauthor{\bsnm{Kramer},~\bfnm{U.}\binits{U.}},
  \bauthor{\bsnm{Kuhlbusch},~\bfnm{T.}\binits{T.}},
  \bauthor{\bsnm{Lanki},~\bfnm{T.}\binits{T.}},
  \bauthor{\bsnm{Madsen},~\bfnm{C.}\binits{C.}},
  \bauthor{\bsnm{Meliefste},~\bfnm{K.}\binits{K.}},
  \bauthor{\bsnm{Molter},~\bfnm{A.}\binits{A.}},
  \bauthor{\bsnm{Mosler},~\bfnm{G.}\binits{G.}},
  \bauthor{\bsnm{Nieuwenhuijsen},~\bfnm{M.}\binits{M.}},
  \bauthor{\bsnm{Oldenwening},~\bfnm{M.}\binits{M.}},
  \bauthor{\bsnm{Pennanen},~\bfnm{A.}\binits{A.}},
  \bauthor{\bsnm{Probst-Hensch},~\bfnm{N.}\binits{N.}},
  \bauthor{\bsnm{Quass},~\bfnm{U.}\binits{U.}},
  \bauthor{\bsnm{Raaschou-Nielsen},~\bfnm{O.}\binits{O.}},
  \bauthor{\bsnm{Ranzi},~\bfnm{A.}\binits{A.}},
  \bauthor{\bsnm{Stephanou},~\bfnm{E.}\binits{E.}},
  \bauthor{\bsnm{Sugiri},~\bfnm{D.}\binits{D.}},
  \bauthor{\bsnm{Udvardy},~\bfnm{O.}\binits{O.}},
  \bauthor{\bsnm{Vaskoevi},~\bfnm{E.}\binits{E.}},
  \bauthor{\bsnm{Weinmayr},~\bfnm{G.}\binits{G.}},
  \bauthor{\bsnm{Brunekreef},~\bfnm{B.}\binits{B.}} \AND
  \bauthor{\bsnm{Hoek},~\bfnm{G.}\binits{G.}}
(\byear{2012}).
\btitle{Spatial Variation of {PM2.5}, {PM10}, {PM2.5} Absorbance and {PM}coarse
  Concentrations Between and Within 20 European Study Areas and the
  Relationship with {NO2} - Results of the ESCAPE Project}.
\bjournal{Atmospheric Environment}
\bvolume{62}
\bpages{303-317}.
\bdoi{10.1016/j.atmosenv.2012.08.038}
\end{barticle}
\endbibitem

\bibitem[\protect\citeauthoryear{Eidsvik et~al.}{2014}]{eidsvik14}
\begin{barticle}[author]
\bauthor{\bsnm{Eidsvik},~\bfnm{J.}\binits{J.}},
  \bauthor{\bsnm{Shaby},~\bfnm{B.~A.}\binits{B.~A.}},
  \bauthor{\bsnm{Reich},~\bfnm{B.~J.}\binits{B.~J.}},
  \bauthor{\bsnm{Wheeler},~\bfnm{M.}\binits{M.}} \AND
  \bauthor{\bsnm{Niemi},~\bfnm{Jared}\binits{J.}}
(\byear{2014}).
\btitle{Estimation and Prediction in Spatial Models with Block Composite
  Likelihoods}.
\bjournal{Journal of Computational and Graphical Statistics}
\bvolume{23}
\bpages{295-315}.
\end{barticle}
\endbibitem

\bibitem[\protect\citeauthoryear{Finley, Banerjee and McRoberts}{2009}]{fin09}
\begin{barticle}[author]
\bauthor{\bsnm{Finley},~\bfnm{Andrew~O.}\binits{A.~O.}},
  \bauthor{\bsnm{Banerjee},~\bfnm{Sudipto}\binits{S.}} \AND
  \bauthor{\bsnm{McRoberts},~\bfnm{Ronald~E.}\binits{R.~E.}}
(\byear{2009}).
\btitle{Hierarchical Spatial Models for Predicting Tree Species Assemblages
  across Large Domains}.
\bjournal{Ann. Appl. Stat.}
\bvolume{3}
\bpages{1052--1079}.
\bdoi{10.1214/09-AOAS250}
\end{barticle}
\endbibitem

\bibitem[\protect\citeauthoryear{Finley, Banerjee and Gelfand}{2012}]{fbg12}
\begin{barticle}[author]
\bauthor{\bsnm{Finley},~\bfnm{A.~O.}\binits{A.~O.}},
  \bauthor{\bsnm{Banerjee},~\bfnm{S.}\binits{S.}} \AND
  \bauthor{\bsnm{Gelfand},~\bfnm{A.~E.}\binits{A.~E.}}
(\byear{2012}).
\btitle{Bayesian Dynamic Modeling for Large Space-time Datasets using Gaussian
  Predictive Processes}.
\bjournal{Journal of Geographical Systems}
\bvolume{14}
\bpages{29–-47}.
\end{barticle}
\endbibitem

\bibitem[\protect\citeauthoryear{Flemming et~al.}{2009}]{FlemmingEtAl09a}
\begin{barticle}[author]
\bauthor{\bsnm{Flemming},~\bfnm{J.}\binits{J.}},
  \bauthor{\bsnm{Inness},~\bfnm{A.}\binits{A.}},
  \bauthor{\bsnm{Flentje},~\bfnm{H.}\binits{H.}},
  \bauthor{\bsnm{Huijnen},~\bfnm{V.}\binits{V.}},
  \bauthor{\bsnm{Moinat},~\bfnm{P.}\binits{P.}},
  \bauthor{\bsnm{Schultz},~\bfnm{M.~G.}\binits{M.~G.}} \AND
  \bauthor{\bsnm{Stein},~\bfnm{O.}\binits{O.}}
(\byear{2009}).
\btitle{Coupling Global Chemistry Transport Models to {ECMWF}'s Integrated
  Forecast System}.
\bjournal{Geoscientific Model Development}
\bvolume{2}
\bpages{253-265}.
\end{barticle}
\endbibitem

\bibitem[\protect\citeauthoryear{Furrer, Genton and Nychka}{2006}]{fur06}
\begin{barticle}[author]
\bauthor{\bsnm{Furrer},~\bfnm{R.}\binits{R.}},
  \bauthor{\bsnm{Genton},~\bfnm{M.~G.}\binits{M.~G.}} \AND
  \bauthor{\bsnm{Nychka},~\bfnm{D.}\binits{D.}}
(\byear{2006}).
\btitle{Covariance Tapering for Interpolation of Large Spatial Datasets}.
\bjournal{Journal of Computational and Graphical Statistics}
\bvolume{15}
\bpages{503-523}.
\end{barticle}
\endbibitem

\bibitem[\protect\citeauthoryear{Gelfand, Banerjee and Gamerman}{2005}]{gelf05}
\begin{barticle}[author]
\bauthor{\bsnm{Gelfand},~\bfnm{A.~E.}\binits{A.~E.}},
  \bauthor{\bsnm{Banerjee},~\bfnm{S.}\binits{S.}} \AND
  \bauthor{\bsnm{Gamerman},~\bfnm{D.}\binits{D.}}
(\byear{2005}).
\btitle{Spatial Process Modelling for Univariate and Multivariate Dynamic
  Spatial Data}.
\bjournal{Environmetrics}
\bvolume{16}
\bpages{465–-479}.
\end{barticle}
\endbibitem

\bibitem[\protect\citeauthoryear{Gelfand and Ghosh}{1998}]{gelf98}
\begin{barticle}[author]
\bauthor{\bsnm{Gelfand},~\bfnm{A.~E.}\binits{A.~E.}} \AND
  \bauthor{\bsnm{Ghosh},~\bfnm{S.~K.}\binits{S.~K.}}
(\byear{1998}).
\btitle{Model Choice: A Minimum Posterior Predictive Loss Approach}.
\bjournal{Biometrika}
\bvolume{85}
\bpages{1--11}.
\end{barticle}
\endbibitem

\bibitem[\protect\citeauthoryear{Gelfand et~al.}{2010}]{geldigfuegut}
\begin{bbook}[author]
\bauthor{\bsnm{Gelfand},~\bfnm{A.~E.}\binits{A.~E.}},
  \bauthor{\bsnm{Diggle},~\bfnm{P.~J.}\binits{P.~J.}},
  \bauthor{\bsnm{Fuentes},~\bfnm{M.}\binits{M.}} \AND
  \bauthor{\bsnm{Guttorp},~\bfnm{P}\binits{P.}}
(\byear{2010}).
\btitle{Handbook of Spatial Statistics}.
\bpublisher{Boca Raton, FL: CRC Press}.
\end{bbook}
\endbibitem

\bibitem[\protect\citeauthoryear{Gneiting}{2002}]{gnei02}
\begin{barticle}[author]
\bauthor{\bsnm{Gneiting},~\bfnm{T}\binits{T.}}
(\byear{2002}).
\btitle{Nonseparable, Stationary Covariance Functions for Space–time Data}.
\bjournal{Journal of the American Statistical Association}
\bvolume{97}
\bpages{590--600}.
\end{barticle}
\endbibitem

\bibitem[\protect\citeauthoryear{Gneiting, Genton and Guttorp}{2007}]{ggg07}
\begin{bincollection}[author]
\bauthor{\bsnm{Gneiting},~\bfnm{T.}\binits{T.}},
  \bauthor{\bsnm{Genton},~\bfnm{M.~G.}\binits{M.~G.}} \AND
  \bauthor{\bsnm{Guttorp},~\bfnm{P.}\binits{P.}}
(\byear{2007}).
\btitle{Geostatistical Space-time Models, Stationarity, Separability and Full
  Symmetry}.
In \bbooktitle{Statistics of SpatioTemporal Systems}
\bpages{151–-175}.
\bpublisher{Chapman and Hall\ CRC Press}
\bnote{(eds Finkenstaedt, B. and Held, L. and Isham, V.)}.
\end{bincollection}
\endbibitem

\bibitem[\protect\citeauthoryear{Gneiting and Guttorp}{2010}]{gnei10}
\begin{barticle}[author]
\bauthor{\bsnm{Gneiting},~\bfnm{T.}\binits{T.}} \AND
  \bauthor{\bsnm{Guttorp},~\bfnm{P.}\binits{P.}}
(\byear{2010}).
\btitle{Continuous-parameter Spatio-temporal Processes}.
\bjournal{Handbook of Spatial Statistics}
\bpages{427-436}.
\bnote{Gelfand, A. E., Diggle, P., Fuentes, M. and Guttorp, P., editors,
  Chapman and Hall/CRC, pp. 427-436}.
\end{barticle}
\endbibitem

\bibitem[\protect\citeauthoryear{Gr{\"a}ler, Gerharz and
  Pebesma}{2011}]{GraelerEtAl11a}
\begin{barticle}[author]
\bauthor{\bsnm{Gr{\"a}ler},~\bfnm{Benedikt}\binits{B.}},
  \bauthor{\bsnm{Gerharz},~\bfnm{Lydia}\binits{L.}} \AND
  \bauthor{\bsnm{Pebesma},~\bfnm{Edzer}\binits{E.}}
(\byear{2011}).
\btitle{Spatio-temporal Analysis and Interpolation of PM10 Measurements in
  Europe}.
\bjournal{ETC/ACM Technical Paper}
\bvolume{10}.
\end{barticle}
\endbibitem

\bibitem[\protect\citeauthoryear{Gramacy and Apley}{2015}]{gram14}
\begin{barticle}[author]
\bauthor{\bsnm{Gramacy},~\bfnm{Robert~B.}\binits{R.~B.}} \AND
  \bauthor{\bsnm{Apley},~\bfnm{Daniel~W.}\binits{D.~W.}}
(\byear{2015}).
\btitle{Local Gaussian Process Approximation for Large Computer Experiments}.
\bjournal{Journal of Computational and Graphical Statistics}
\bvolume{24}
\bpages{561-578}.
\bdoi{10.1080/10618600.2014.914442}
\end{barticle}
\endbibitem

\bibitem[\protect\citeauthoryear{Hamm et~al.}{2015}]{hamm2015}
\begin{barticle}[author]
\bauthor{\bsnm{Hamm},~\bfnm{N.~A.~S.}\binits{N.~A.~S.}},
  \bauthor{\bsnm{Finley},~\bfnm{A.~O.}\binits{A.~O.}},
  \bauthor{\bsnm{Schaap},~\bfnm{M.}\binits{M.}} \AND
  \bauthor{\bsnm{Stein},~\bfnm{A.}\binits{A.}}
(\byear{2015}).
\btitle{A Spatially Varying Coefficient Model for Mapping PM10 Air Quality at
  the European scale}.
\bjournal{Atmospheric Environment}
\bvolume{102}
\bpages{393--405}.
\end{barticle}
\endbibitem

\bibitem[\protect\citeauthoryear{Hendriks et~al.}{2013}]{HendriksEtAl13a}
\begin{barticle}[author]
\bauthor{\bsnm{Hendriks},~\bfnm{C.}\binits{C.}},
  \bauthor{\bsnm{Kranenburg},~\bfnm{R.}\binits{R.}},
  \bauthor{\bsnm{Kuenen},~\bfnm{J.}\binits{J.}}, \bauthor{\bparticle{van}
  \bsnm{Gijlswijk},~\bfnm{R.}\binits{R.}},
  \bauthor{\bsnm{Kruit},~\bfnm{R.~Wichink}\binits{R.~W.}},
  \bauthor{\bsnm{Segers},~\bfnm{A.}\binits{A.}}, \bauthor{\bparticle{van~der}
  \bsnm{Gon},~\bfnm{H.~Denier}\binits{H.~D.}} \AND
  \bauthor{\bsnm{Schaap},~\bfnm{M.}\binits{M.}}
(\byear{2013}).
\btitle{The Origin of Ambient Particulate Matter Concentrations in the
  Netherlands}.
\bjournal{Atmospheric Environment}
\bvolume{69}
\bpages{289-303}.
\end{barticle}
\endbibitem

\bibitem[\protect\citeauthoryear{Higdon}{2001}]{hig01}
\begin{barticle}[author]
\bauthor{\bsnm{Higdon},~\bfnm{D.}\binits{D.}}
(\byear{2001}).
\btitle{Space and Space Time Modeling using Process Convolutions}.
\bjournal{Technical Report, Institute of Statistics and Decision Sciences, Duke
  University, Durham}.
\end{barticle}
\endbibitem

\bibitem[\protect\citeauthoryear{Hoek et~al.}{2013}]{HoekEtAl13a}
\begin{barticle}[author]
\bauthor{\bsnm{Hoek},~\bfnm{G.}\binits{G.}},
  \bauthor{\bsnm{Krishnan},~\bfnm{R.~M.}\binits{R.~M.}},
  \bauthor{\bsnm{Beelen},~\bfnm{R.}\binits{R.}},
  \bauthor{\bsnm{Peters},~\bfnm{A.}\binits{A.}},
  \bauthor{\bsnm{Ostro},~\bfnm{B.}\binits{B.}},
  \bauthor{\bsnm{Brunekreef},~\bfnm{B.}\binits{B.}} \AND
  \bauthor{\bsnm{Kaufman},~\bfnm{J.~D.}\binits{J.~D.}}
(\byear{2013}).
\btitle{Long-term Air Pollution Exposure and Cardio- respiratory Mortality: A
  Review}.
\bjournal{Environmental Health}
\bvolume{12}
\bpages{43}.
\bdoi{10.1186/1476-069x-12-43}
\end{barticle}
\endbibitem

\bibitem[\protect\citeauthoryear{Intel}{2015}]{mkl}
\begin{bmisc}[author]
\bauthor{\bsnm{Intel}}
(\byear{2015}).
\btitle{Math Kernel Library}.
\bhowpublished{http://developer.intel.com/software/products/mkl/}.
\end{bmisc}
\endbibitem

\bibitem[\protect\citeauthoryear{Jones and Zhang}{1997}]{jones97}
\begin{bincollection}[author]
\bauthor{\bsnm{Jones},~\bfnm{R.~H.}\binits{R.~H.}} \AND
  \bauthor{\bsnm{Zhang},~\bfnm{Y.}\binits{Y.}}
(\byear{1997}).
\btitle{Models for Continuous Stationary Space-time Processes}.
In \bbooktitle{Modelling Longitudinal and Spatially Correlated Data}
\bpages{289--298}.
\bpublisher{New York: Springer}
\bnote{(eds T. G. Gregoire, D. R. Brillinger, P. J. Diggle, E. Russek-Cohen, W.
  G. Warren and R. D. Wolfinger)}.
\end{bincollection}
\endbibitem

\bibitem[\protect\citeauthoryear{Kammann and Wand}{2003}]{kam03}
\begin{barticle}[author]
\bauthor{\bsnm{Kammann},~\bfnm{E.~E.}\binits{E.~E.}} \AND
  \bauthor{\bsnm{Wand},~\bfnm{M.~P.}\binits{M.~P.}}
(\byear{2003}).
\btitle{Geoadditive Models}.
\bjournal{Applied Statistics}
\bvolume{52}
\bpages{1-18}.
\end{barticle}
\endbibitem

\bibitem[\protect\citeauthoryear{Katzfuss}{2016}]{katzfussmultires}
\begin{barticle}[author]
\bauthor{\bsnm{Katzfuss},~\bfnm{Matthias}\binits{M.}}
(\byear{2016}).
\btitle{A multi-resolution approximation for massive spatial datasets}.
\bjournal{Journal of the American Statistical Association}
\bvolume{ja}.
\bdoi{10.1080/01621459.2015.1123632}
\end{barticle}
\endbibitem

\bibitem[\protect\citeauthoryear{Katzfuss and Cressie}{2012}]{katz12}
\begin{barticle}[author]
\bauthor{\bsnm{Katzfuss},~\bfnm{M.}\binits{M.}} \AND
  \bauthor{\bsnm{Cressie},~\bfnm{N.}\binits{N.}}
(\byear{2012}).
\btitle{Bayesian Hierarchical Spatio-temporal Smoothing for Very Large
  Datasets}.
\bjournal{Environmetrics}
\bvolume{23}
\bpages{94--107}.
\end{barticle}
\endbibitem

\bibitem[\protect\citeauthoryear{Kaufman, Scheverish and Nychka}{2008}]{kauf08}
\begin{barticle}[author]
\bauthor{\bsnm{Kaufman},~\bfnm{C.~G.}\binits{C.~G.}},
  \bauthor{\bsnm{Scheverish},~\bfnm{M.~J.}\binits{M.~J.}} \AND
  \bauthor{\bsnm{Nychka},~\bfnm{D.~W.}\binits{D.~W.}}
(\byear{2008}).
\btitle{Covariance Tapering for Likelihood-Based Estimation in Large Spatial
  Data Sets}.
\bjournal{Journal of the American Statistical Association}
\bvolume{103}
\bpages{1545-1555}.
\end{barticle}
\endbibitem

\bibitem[\protect\citeauthoryear{Kyriakidis and Journel}{1999}]{kyria99}
\begin{barticle}[author]
\bauthor{\bsnm{Kyriakidis},~\bfnm{P.~C.}\binits{P.~C.}} \AND
  \bauthor{\bsnm{Journel},~\bfnm{A.~G.}\binits{A.~G.}}
(\byear{1999}).
\btitle{Geostatistical Space-time Models: A Review}.
\bjournal{Mathematical Geology}
\bvolume{31}
\bpages{651–-684}.
\end{barticle}
\endbibitem

\bibitem[\protect\citeauthoryear{Lloyd and Atkinson}{2004}]{LloydA04a}
\begin{barticle}[author]
\bauthor{\bsnm{Lloyd},~\bfnm{C.~D.}\binits{C.~D.}} \AND
  \bauthor{\bsnm{Atkinson},~\bfnm{P.~M.}\binits{P.~M.}}
(\byear{2004}).
\btitle{Increased Accuracy of Geostatistical Prediction of Nitrogen Dioxide in
  the {U}nited {K}ingdom with Secondary Data}.
\bjournal{International Journal of Applied Earth Observation and
  Geoinformation}
\bvolume{5}
\bpages{293-305}.
\bdoi{http://dx.doi.org/10.1016/j.jag.2004.07.004}
\end{barticle}
\endbibitem

\bibitem[\protect\citeauthoryear{Loomis et~al.}{2013}]{LoomisEtAl13a}
\begin{barticle}[author]
\bauthor{\bsnm{Loomis},~\bfnm{D}\binits{D.}},
  \bauthor{\bsnm{Grosse},~\bfnm{Y}\binits{Y.}},
  \bauthor{\bsnm{Lauby-Secretan},~\bfnm{B.}\binits{B.}},
  \bauthor{\bsnm{El~Ghissassi},~\bfnm{F.}\binits{F.}},
  \bauthor{\bsnm{Bouvard},~\bfnm{V.}\binits{V.}},
  \bauthor{\bsnm{Benbrahim-Tallaa},~\bfnm{L.}\binits{L.}},
  \bauthor{\bsnm{Guha},~\bfnm{N.}\binits{N.}},
  \bauthor{\bsnm{Baan},~\bfnm{R.}\binits{R.}},
  \bauthor{\bsnm{Mattock},~\bfnm{H.}\binits{H.}} \AND
  \bauthor{\bsnm{Straif},~\bfnm{S.}\binits{S.}}
(\byear{2013}).
\btitle{The Carcinogenicity of Outdoor Air Pollution}.
\bjournal{The Lancet Oncology}
\bvolume{14}
\bpages{1262-1263}.
\end{barticle}
\endbibitem

\bibitem[\protect\citeauthoryear{Manders, Schaap and
  Hoogerbrugge}{2009}]{MandersSH09a}
\begin{barticle}[author]
\bauthor{\bsnm{Manders},~\bfnm{A.~M.~M.}\binits{A.~M.~M.}},
  \bauthor{\bsnm{Schaap},~\bfnm{M.}\binits{M.}} \AND
  \bauthor{\bsnm{Hoogerbrugge},~\bfnm{R.}\binits{R.}}
(\byear{2009}).
\btitle{Testing the Capability of the Chemistry Transport Model LOTOS-EUROS to
  Forecast PM10 Levels in the Netherlands}.
\bjournal{Atmospheric Environment}
\bvolume{43}
\bpages{4050-4059}.
\bdoi{10.1016/j.atmosenv.2009.05.006}
\end{barticle}
\endbibitem

\bibitem[\protect\citeauthoryear{Mues et~al.}{2014}]{MuesEtAl14a}
\begin{barticle}[author]
\bauthor{\bsnm{Mues},~\bfnm{A.}\binits{A.}},
  \bauthor{\bsnm{Kuenen},~\bfnm{J.}\binits{J.}},
  \bauthor{\bsnm{Hendriks},~\bfnm{C.}\binits{C.}},
  \bauthor{\bsnm{Manders},~\bfnm{A.}\binits{A.}},
  \bauthor{\bsnm{Segers},~\bfnm{A.}\binits{A.}},
  \bauthor{\bsnm{Scholz},~\bfnm{Y.}\binits{Y.}},
  \bauthor{\bsnm{Hueglin},~\bfnm{C.}\binits{C.}},
  \bauthor{\bsnm{Builtjes},~\bfnm{P.}\binits{P.}} \AND
  \bauthor{\bsnm{Schaap},~\bfnm{M.}\binits{M.}}
(\byear{2014}).
\btitle{Sensitivity of air pollution simulations with LOTOS-EUROS to the
  temporal distribution of anthropogenic emissions}.
\bjournal{Atmospheric Chemistry and Physics}
\bvolume{14}
\bpages{939-955}.
\bdoi{10.5194/acp-14-939-2014}
\end{barticle}
\endbibitem

\bibitem[\protect\citeauthoryear{Omidi and Mohammadzadeh}{2015}]{omidi15}
\begin{barticle}[author]
\bauthor{\bsnm{Omidi},~\bfnm{Mehdi}\binits{M.}} \AND
  \bauthor{\bsnm{Mohammadzadeh},~\bfnm{Mohsen}\binits{M.}}
(\byear{2015}).
\btitle{A New Method to Build Spatio-temporal Covariance Functions: Analysis of
  Ozone Data}.
\bjournal{Statistical Papers}
\bpages{1-15}.
\bdoi{10.1007/s00362-015-0674-2}
\end{barticle}
\endbibitem

\bibitem[\protect\citeauthoryear{Pfeifer and Deutsch}{1980a}]{pfief80a}
\begin{barticle}[author]
\bauthor{\bsnm{Pfeifer},~\bfnm{P.~E.}\binits{P.~E.}} \AND
  \bauthor{\bsnm{Deutsch},~\bfnm{S.~J.}\binits{S.~J.}}
(\byear{1980}a).
\btitle{Independence and Sphericity Tests for the Residuals of Space–time
  ARMA Models}.
\bjournal{Communications in Statistics - Simulation and Computation}
\bvolume{9}
\bpages{533–-549}.
\end{barticle}
\endbibitem

\bibitem[\protect\citeauthoryear{Pfeifer and Deutsch}{1980b}]{pfief80b}
\begin{barticle}[author]
\bauthor{\bsnm{Pfeifer},~\bfnm{P.~E.}\binits{P.~E.}} \AND
  \bauthor{\bsnm{Deutsch},~\bfnm{S.~J.}\binits{S.~J.}}
(\byear{1980}b).
\btitle{Stationarity and Invertibility Regions for Low Order STARMA Models.}
\bjournal{Communications in Statistics - Simulation and Computation}
\bvolume{9}
\bpages{551–-562}.
\end{barticle}
\endbibitem

\bibitem[\protect\citeauthoryear{Pouliot et~al.}{2012}]{PouliotEtAl12a}
\begin{barticle}[author]
\bauthor{\bsnm{Pouliot},~\bfnm{G.}\binits{G.}},
  \bauthor{\bsnm{Pierce},~\bfnm{T.}\binits{T.}}, \bauthor{\bparticle{van~der}
  \bsnm{Gon},~\bfnm{H.~D.}\binits{H.~D.}},
  \bauthor{\bsnm{Schaap},~\bfnm{M.}\binits{M.}},
  \bauthor{\bsnm{Moran},~\bfnm{M.}\binits{M.}} \AND
  \bauthor{\bsnm{Nopmongcol},~\bfnm{U.}\binits{U.}}
(\byear{2012}).
\btitle{Comparing Emission Inventories and Model-ready Emission Datasets
  between {E}urope and {N}orth {A}merica for the {AQMEII} Project}.
\bjournal{Atmospheric Environment}
\bvolume{53}
\bpages{4-14}.
\end{barticle}
\endbibitem

\bibitem[\protect\citeauthoryear{Rasmussen and Williams}{2005}]{rasm08}
\begin{binbook}[author]
\bauthor{\bsnm{Rasmussen},~\bfnm{C.~E.}\binits{C.~E.}} \AND
  \bauthor{\bsnm{Williams},~\bfnm{C.~K.~I.}\binits{C.~K.~I.}}
(\byear{2005}).
\btitle{Gaussian Processes for Machine Learning},
\bedition{first} ed.
\bpublisher{The MIT Press}, \baddress{Cambridge, MA}.
\end{binbook}
\endbibitem

\bibitem[\protect\citeauthoryear{R'Honi et~al.}{2013}]{RHoniEtAl13a}
\begin{barticle}[author]
\bauthor{\bsnm{R'Honi},~\bfnm{Y.}\binits{Y.}},
  \bauthor{\bsnm{Clarisse},~\bfnm{L.}\binits{L.}},
  \bauthor{\bsnm{Clerbaux},~\bfnm{C.}\binits{C.}},
  \bauthor{\bsnm{Hurtmans},~\bfnm{D.}\binits{D.}},
  \bauthor{\bsnm{Duflot},~\bfnm{V.}\binits{V.}},
  \bauthor{\bsnm{Turquety},~\bfnm{S.}\binits{S.}},
  \bauthor{\bsnm{Ngadi},~\bfnm{Y.}\binits{Y.}} \AND
  \bauthor{\bsnm{Coheur},~\bfnm{P.~F.}\binits{P.~F.}}
(\byear{2013}).
\btitle{Exceptional Emissions of NH3 and HCOOH in the 2010 Russian Wildfires}.
\bjournal{Atmospheric Chemistry and Physics}
\bvolume{13}
\bpages{4171-4181}.
\bdoi{10.5194/acp-13-4171-2013}
\end{barticle}
\endbibitem

\bibitem[\protect\citeauthoryear{Rue and Held}{2005}]{rueheld04}
\begin{binbook}[author]
\bauthor{\bsnm{Rue},~\bfnm{Håvard}\binits{H.}} \AND
  \bauthor{\bsnm{Held},~\bfnm{Leonhard}\binits{L.}}
(\byear{2005}).
\btitle{Gaussian Markov Random Fields : Theory and Applications}.
\bseries{Monographs on statistics and applied probability}.
\bpublisher{Chapman \& Hall/CRC}, \baddress{Boca Raton, FL}.
\end{binbook}
\endbibitem

\bibitem[\protect\citeauthoryear{Sang and Huang}{2012}]{sang12}
\begin{barticle}[author]
\bauthor{\bsnm{Sang},~\bfnm{H.}\binits{H.}} \AND
  \bauthor{\bsnm{Huang},~\bfnm{J.~Z.}\binits{J.~Z.}}
(\byear{2012}).
\btitle{A Full Scale Approximation of Covariance Functions for Large Spatial
  Data Sets}.
\bjournal{Journal of the Royal Statistical society, Series B}
\bvolume{74}
\bpages{111-132}.
\end{barticle}
\endbibitem

\bibitem[\protect\citeauthoryear{Schaap et~al.}{2008}]{SchaapEtAl08a}
\begin{barticle}[author]
\bauthor{\bsnm{Schaap},~\bfnm{Martijn}\binits{M.}},
  \bauthor{\bsnm{Timmermans},~\bfnm{Renske M.~A.}\binits{R.~M.~A.}},
  \bauthor{\bsnm{Roemer},~\bfnm{Michiel}\binits{M.}},
  \bauthor{\bsnm{Boersen},~\bfnm{G.~A.~C.}\binits{G.~A.~C.}},
  \bauthor{\bsnm{Builtjes},~\bfnm{Peter}\binits{P.}},
  \bauthor{\bsnm{Sauter},~\bfnm{Ferd}\binits{F.}},
  \bauthor{\bsnm{Velders},~\bfnm{Guus}\binits{G.}} \AND
  \bauthor{\bsnm{Beck},~\bfnm{Jeanette}\binits{J.}}
(\byear{2008}).
\btitle{The {LOTOS-EUROS} Model: Description, Validation and Latest
  Developments}.
\bjournal{International Journal of Environment and Pollution}
\bvolume{32}
\bpages{270-290}.
\end{barticle}
\endbibitem

\bibitem[\protect\citeauthoryear{Shaby and Ruppert}{2012}]{shabytaper}
\begin{barticle}[author]
\bauthor{\bsnm{Shaby},~\bfnm{B.~A.}\binits{B.~A.}} \AND
  \bauthor{\bsnm{Ruppert},~\bfnm{D.}\binits{D.}}
(\byear{2012}).
\btitle{Tapered Covariance: Bayesian Estimation and Asymptotics}.
\bjournal{Journal of Computational and Graphical Statistics}
\bvolume{21}
\bpages{433--452}.
\end{barticle}
\endbibitem

\bibitem[\protect\citeauthoryear{Spiegelhalter et~al.}{2002}]{spieg02}
\begin{barticle}[author]
\bauthor{\bsnm{Spiegelhalter},~\bfnm{David~J.}\binits{D.~J.}},
  \bauthor{\bsnm{Best},~\bfnm{Nicola~G.}\binits{N.~G.}},
  \bauthor{\bsnm{Carlin},~\bfnm{Bradley~P.}\binits{B.~P.}} \AND
  \bauthor{\bparticle{van~der} \bsnm{Linde},~\bfnm{Angelika}\binits{A.}}
(\byear{2002}).
\btitle{Bayesian Measures of Model Complexity and Fit}.
\bjournal{Journal of the Royal Statistical Society B}
\bvolume{64}
\bpages{583--639}.
\end{barticle}
\endbibitem

\bibitem[\protect\citeauthoryear{Stein}{2005}]{stein05b}
\begin{barticle}[author]
\bauthor{\bsnm{Stein},~\bfnm{M.~L.}\binits{M.~L.}}
(\byear{2005}).
\btitle{Space–time Covariance functions}.
\bjournal{Journal of the American Statistical Association}
\bvolume{100}
\bpages{310–-321}.
\end{barticle}
\endbibitem

\bibitem[\protect\citeauthoryear{Stein}{2007}]{stein07}
\begin{barticle}[author]
\bauthor{\bsnm{Stein},~\bfnm{M.~L.}\binits{M.~L.}}
(\byear{2007}).
\btitle{Spatial Variation of Total Column Ozone on a Global Scale}.
\bjournal{Annals of Applied Statistics}
\bvolume{1}
\bpages{191-210}.
\end{barticle}
\endbibitem

\bibitem[\protect\citeauthoryear{Stein}{2008}]{stein08}
\begin{barticle}[author]
\bauthor{\bsnm{Stein},~\bfnm{M.~L.}\binits{M.~L.}}
(\byear{2008}).
\btitle{A Modeling Approach for Large Spatial Datasets}.
\bjournal{Journal of the Korean Statistical Society}
\bvolume{37}
\bpages{3-10}.
\end{barticle}
\endbibitem

\bibitem[\protect\citeauthoryear{Stein}{2013}]{stein2013}
\begin{barticle}[author]
\bauthor{\bsnm{Stein},~\bfnm{Michael~L.}\binits{M.~L.}}
(\byear{2013}).
\btitle{On a Class of Space–time Intrinsic Random Functions}.
\bjournal{Bernoulli}
\bvolume{19}
\bpages{387--408}.
\bdoi{10.3150/11-BEJ405}
\end{barticle}
\endbibitem

\bibitem[\protect\citeauthoryear{Stein}{2014}]{Stein13}
\begin{barticle}[author]
\bauthor{\bsnm{Stein},~\bfnm{Michael~L.}\binits{M.~L.}}
(\byear{2014}).
\btitle{Limitations on Low Rank Approximations for Covariance Matrices of
  Spatial Data}.
\bjournal{Spatial Statistics}
\bvolume{8}
\bpages{1--19}.
\end{barticle}
\endbibitem

\bibitem[\protect\citeauthoryear{Stein, Chi and Welty}{2004}]{stein04}
\begin{barticle}[author]
\bauthor{\bsnm{Stein},~\bfnm{M.~L.}\binits{M.~L.}},
  \bauthor{\bsnm{Chi},~\bfnm{Z.}\binits{Z.}} \AND
  \bauthor{\bsnm{Welty},~\bfnm{L.~J.}\binits{L.~J.}}
(\byear{2004}).
\btitle{Approximating Likelihoods for Large Spatial Data Sets}.
\bjournal{Journal of the Royal Statistical society, Series B}
\bvolume{66}
\bpages{275-296}.
\end{barticle}
\endbibitem

\bibitem[\protect\citeauthoryear{Stern et~al.}{2008}]{SternEtAl08a}
\begin{barticle}[author]
\bauthor{\bsnm{Stern},~\bfnm{R.}\binits{R.}},
  \bauthor{\bsnm{Builtjes},~\bfnm{P.}\binits{P.}},
  \bauthor{\bsnm{Schaap},~\bfnm{M.}\binits{M.}},
  \bauthor{\bsnm{Timmermans},~\bfnm{R.}\binits{R.}},
  \bauthor{\bsnm{Vautard},~\bfnm{R.}\binits{R.}},
  \bauthor{\bsnm{Hodzic},~\bfnm{A.}\binits{A.}},
  \bauthor{\bsnm{Memmesheimer},~\bfnm{M.}\binits{M.}},
  \bauthor{\bsnm{Feldmann},~\bfnm{H.}\binits{H.}},
  \bauthor{\bsnm{Renner},~\bfnm{E.}\binits{E.}},
  \bauthor{\bsnm{Wolke},~\bfnm{R.}\binits{R.}} \AND
  \bauthor{\bsnm{Kerschbaumer},~\bfnm{A.}\binits{A.}}
(\byear{2008}).
\btitle{A {Model} Inter-comparison Study Focussing on Episodes with Elevated
  {PM}10 Concentrations}.
\bjournal{Atmospheric Environment}
\bvolume{42}
\bpages{4567-4588}.
\end{barticle}
\endbibitem

\bibitem[\protect\citeauthoryear{Stoffer}{1986}]{stoff86}
\begin{barticle}[author]
\bauthor{\bsnm{Stoffer},~\bfnm{D.~S.}\binits{D.~S.}}
(\byear{1986}).
\btitle{Estimation and Identification of Space–time ARMAX Models in the
  Presence of Missing Data}.
\bjournal{Journal of the American Statistical Association}
\bvolume{81}
\bpages{762–-772}.
\end{barticle}
\endbibitem

\bibitem[\protect\citeauthoryear{Stroud, M¨uller and Sans´o}{2001}]{stroud01}
\begin{barticle}[author]
\bauthor{\bsnm{Stroud},~\bfnm{J.~R.}\binits{J.~R.}},
  \bauthor{\bsnm{M¨uller},~\bfnm{P.}\binits{P.}} \AND
  \bauthor{\bsnm{Sans´o},~\bfnm{B.}\binits{B.}}
(\byear{2001}).
\btitle{Dynamic Models for Spatiotemporal Data}.
\bjournal{Journal of the Royal Statistical society, Series B}
\bvolume{63}
\bpages{673–-689}.
\end{barticle}
\endbibitem

\bibitem[\protect\citeauthoryear{van~de Kassteele and
  Stein}{2006}]{vdKassteeleS06a}
\begin{barticle}[author]
\bauthor{\bparticle{van~de} \bsnm{Kassteele},~\bfnm{J.}\binits{J.}} \AND
  \bauthor{\bsnm{Stein},~\bfnm{A.}\binits{A.}}
(\byear{2006}).
\btitle{A Model for External Drift Kriging with Uncertain Covariates applied to
  Air Quality Measurements and Dispersion Model Output}.
\bjournal{Environmetrics}
\bvolume{17}
\bpages{309-322}.
\bdoi{10.1002/env.771}
\end{barticle}
\endbibitem

\bibitem[\protect\citeauthoryear{Vecchia}{1988}]{ve88}
\begin{barticle}[author]
\bauthor{\bsnm{Vecchia},~\bfnm{A.~V.}\binits{A.~V.}}
(\byear{1988}).
\btitle{Estimation and Model Identification for Continuous Spatial Processes}.
\bjournal{Journal of the Royal Statistical society, Series B}
\bvolume{50}
\bpages{297-312}.
\end{barticle}
\endbibitem

\bibitem[\protect\citeauthoryear{Vecchia}{1992}]{ve92}
\begin{barticle}[author]
\bauthor{\bsnm{Vecchia},~\bfnm{A.~V.}\binits{A.~V.}}
(\byear{1992}).
\btitle{A New Method of Prediction for Spatial Regression Models with
  Correlated Errors}.
\bjournal{Journal of the Royal Statistical society, Series B}
\bvolume{54}
\bpages{813-830}.
\end{barticle}
\endbibitem

\bibitem[\protect\citeauthoryear{Xu, Liang and Genton}{2014}]{gang14}
\begin{barticle}[author]
\bauthor{\bsnm{Xu},~\bfnm{G.}\binits{G.}},
  \bauthor{\bsnm{Liang},~\bfnm{F.}\binits{F.}} \AND
  \bauthor{\bsnm{Genton},~\bfnm{M.~G.}\binits{M.~G.}}
(\byear{2014}).
\btitle{A Bayesian Spatio-Temporal Geostatistical Model with an Auxiliary
  Lattice for Large Datasets}.
\bjournal{Statistica Sinica}
\bpages{In press}.
\end{barticle}
\endbibitem

\bibitem[\protect\citeauthoryear{Yeniay and Goktas}{2002}]{rmspe02}
\begin{barticle}[author]
\bauthor{\bsnm{Yeniay},~\bfnm{O.}\binits{O.}} \AND
  \bauthor{\bsnm{Goktas},~\bfnm{A.}\binits{A.}}
(\byear{2002}).
\btitle{A Comparison of Partial Least Squares Regression with Other Prediction
  Methods}.
\bjournal{Hacettepe Journal of Mathematics and Statistics}
\bvolume{31}
\bpages{99-111}.
\end{barticle}
\endbibitem

\end{thebibliography}
\label{lastpage}



\section*{Supplementary material}

# SUPPLEMENT TO "NON-SEPARABLE DYNAMIC NEAREST-NEIGHBOR GAUSSIAN PROCESS MODELS FOR LARGE SPATIO-TEMPORAL DATA WITH AN APPLICATION TO PARTICULATE MATTER ANALYSIS"

By Abhirup Datta, Sudipto Banerjee, Andrew O. Finley, Nicholas A.S. Hamm and Martijn Schaap

**S1. Eligible Sets.** We provide formal definitions of the eligible sets. Recall from Section 4.1 that, for any location $s$, $A(\mathbf{s}, V, m)$ is the set of $m$-nearest neighbors of $\mathbf{s}$ in $V$. So $\mathbf{s} \in V$ implies that $\mathbf{s} \in A(\mathbf{s}, V, m)$ for all $m \geq 1$. For each $(\mathbf{s}_i, t_j)$ in $\mathcal{R}$, we define the *eligible set*

(S1)
$$E(\mathbf{s}_i, t_j) = \bigcup_{k=1}^{m} \{(\mathbf{s}, t_{j-k}) \,|\, \mathbf{s} \in A(\mathbf{s}_i, S, [m/k])\} \bigcup \{(\mathbf{s}, t_j) \,|\, \mathbf{s} \in A(\mathbf{s}_i, S_i, m)\}$$

where for any positive number $x$, $[x]$ denotes the greatest integer not exceeding $x$. So the eligible set for a space-time point consists of $m$-nearest neighbors from the time levels $j$ and $j-1$, $[m/2]$ nearest neighbors from time level $j-2$ and so on upto $[m/m] = 1$ nearest neighbor from time level $j-m$. This is also illustrated in Figure 3(c). So the size of $E(\mathbf{s}_i, t_j)$ does not exceed $m + \sum_{k=1}^{m} [m/k]$. As $m$ is typically chosen to be around 20, this sum is approximately $4m$.

For any point $t$ outside $T$, let $t[k]$ denote the $k^{th}$ nearest time point of $t$ in $T$. Then, we define the eligible set for any $(\mathbf{s}, t)$ outside $\mathcal{R}$ as

(S2)
$$E(\mathbf{s}, t) = \bigcup_{k=1}^{m} \{(\mathbf{s}, t[k]) \,|\, \mathbf{s} \in A(\mathbf{s}, S, [m/k])\}$$

The eligible sets do not depend on the covariance parameters $\boldsymbol{\theta}$. We now show that for any point $(\mathbf{s}, t)$ in $\mathcal{L}$, the eligible set $E(\mathbf{s}, t)$ defined by Equations S1 and S2 contains $m$-nearest neighbors of $(\mathbf{s}, t)$ for all values of $\boldsymbol{\theta}$ as long as the underlying covariance function $C(h, u \,|\, \boldsymbol{\theta})$ possess natural monotonicity.

PROPOSITION S1. *If $C(h, u \,|\, \boldsymbol{\theta})$ satisfies natural monotonicity defined in Section 4.1 for every value of $\boldsymbol{\theta}$, then, for every $(\boldsymbol{s}, t)$, the eligible set $E(\boldsymbol{s}, t)$ defined in Equations S1 and S2 contains $N_\theta(\boldsymbol{s}, t)$ for all $\boldsymbol{\theta}$*

PROOF. We only prove for $(\mathbf{s}, t) = (\mathbf{s}_i, t_j) \in \mathcal{R}$. The proof for $(\mathbf{s}, t) \notin \mathcal{R}$ is similar. We assume that $(\mathbf{s}_u, t_{j-k}) \in N_{\boldsymbol{\theta}}(\mathbf{s}_i, t_j)$ for some $\boldsymbol{\theta}$, $u \le N$ and $k \ge 1$. Also let $\mathbf{s}_i[l]$ denote the $l^{th}$ nearest neighbor of $\mathbf{s}_i$ among $\{\mathbf{s}_1, \mathbf{s}_2, \ldots, \mathbf{s}_N\}$. So, $\mathbf{s}_u = \mathbf{s}_i[l]$ for some $l \ge 1$. Therefore, by natural monotonicity of $C$, we have $C((\mathbf{s}_i, t_j), (\mathbf{s}_i[a], t_{j-k}) \,|\, \boldsymbol{\theta}) \ge C((\mathbf{s}_i, t_j), (\mathbf{s}_i[l], t_{j-k}) \,|\, \boldsymbol{\theta})$ for all $1 \le a \le l$. One more application of natural monotonicity implies that $C((\mathbf{s}_i, t_j), (\mathbf{s}_i[a], t_{j-b}) \,|\, \boldsymbol{\theta}) > C((\mathbf{s}_i, t_j), (\mathbf{s}_i[a], t_{j-k}) \,|\, \boldsymbol{\theta})$ for all $1 \le b \le k$. As $(\mathbf{s}_u, t_{j-k}) \in N_{\boldsymbol{\theta}}(\mathbf{s}_i, t_j)$, then so does $(\mathbf{s}_i[a], t_{j-b})$ for all $a \le l$ and $b \le k$. Therefore, $lk \le m$ i.e. $l \le [m/k]$. □

## S2. Comparisons with Local Approximation GP.

Gramacy and Apley (2015) proposed a Local Approximation Gaussian Process (LAGP) to predict a function $y(\mathbf{s}_0)$ at a location $s_0$ given the observations $\{y(\mathbf{s}_1), y(\mathbf{s}_2), \ldots, y(\mathbf{s}_n)\}$ for $n$ locations. For predicting at each new location $s_0$, a small neighbor set of size $m$ is selected from $\{\mathbf{s}_1, \mathbf{s}_2, \ldots, \mathbf{s}_n\}$. However, unlike Nearest Neighbor Gaussian Processes (Datta et al., 2016), the neighbor sets doesn't merely consist of $m$ nearest neighbors of $s_0$. Instead, locations are augmented to an initial set of $m_0 < m$ nearest neighbors based on minimization of the predictive Mean Square Error for $y(\mathbf{s}_0)$. Simulation studies detailed in Gramacy and Apley (2015) suggests that LAGP produces improved prediction over the fully nearest neighbor based approach while the computational costs for the two methods are comparable.

Although promising in terms of predictive performance and computational scalability, LAGP is not a proper Gaussian Process. It essentially provides a non-stationary local approximation to a Gaussian Process at every predictive location. LAGP thereby can only be used for predicting the response and lacks the versatility of a full GP based approach.

For the European $PM_{10}$ data, we used a semiparametric regression model with the raw CTM output as a covariate and a spatio-temporally varying random intercept (Equation 5.3) which was modeled using a DNNGP prior. LAGP, in its current form, cannot be used for hierarchical models like spatially varying intercept or spatially varying coefficient models where a GP is used to model latent random effects instead of the response.

Moreover, LAGP uses an isotropic squared exponential covariance function for specifying the GP. In a spatio-temporal setup, this amounts to the assumption that the variation along the spatial and temporal directions are on the same scale. These assumptions are violated when we are modeling an anisotropic or non-separable space-time function. LAGP, unlike Dynamic NNGP, cannot accommodate such space-time non-separable functions, and as seen in simulation experiments detailed in Table S1, performs poorly when

Table S1

*Average RMSPE numbers for LAGP and Dynamic NNGP*

|  | LAGP | Simple DNNGP | Adaptive DNNGP |
|---|---|---|---|
| Case 1 | 0.94 | 0.92 | 0.90 |
| Case 2 | 0.61 | 0.55 | 0.56 |
| Case 3 | 0.89 | 0.83 | 0.83 |

the space and time variations differ significantly.

Finally, one of our main objectives for the spatio-temporal analysis of European $PM_{10}$ dataset was to understand the underlying spatial and temporal dependence of $PM_{10}$. Indeed, the covariance parameter estimates in Table 2 reveals that the effective spatial range is around 45 to 60 km and the temporal range is around 30 to 33 days (Figure 5). Understanding the spatio-temporal structure after adjusting for the CTM output may help identify the physical processes missing in the CTM and subsequently improve the CTM itself. LAGP is not a proper Gaussian Process and does not have any global space and time decay parameters and is hence unsuitable for such spatio-temporal analysis.

Nevertheless, as recommended by the reviewers, we compared the predictive performance of Dynamic NNGP and LAGP via simulation experiments. We generated data from the model

$$(S1) \qquad\qquad y(\mathbf{s}, t) = w(\mathbf{s}, t) + \epsilon(\mathbf{s}, t)$$

where $\epsilon \overset{\text{iid}}{\sim} N(0, \tau^2)$ and $w(\mathbf{s}, t)$ are the realizations of a GP with a non-separable covariance structure specified in (6.1). We used the same three sets of parameters as detailed in Section 6 which corresponds to 1) short spatial range and long temporal range, 2) long spatial and temporal range, and; 3) long spatial range and short temporal range. For each case we generated the data on a $15 \times 15 \times 15$ space-time grid and computed the RMSPE based on 500 holdout locations. The process was repeated 50 times for each of the three scenarios and the average RMSPE numbers are reported in Table S1. We observe that for Case 3 where the temporal range dominates, we see LAGP tends to perform worse. This is expected as LAGP assumes an isotropic covariance function ignoring the disparity in space and time variation scales. Overall, Dynamic NNGP performs better for all the three scenarios.

**S3. Non-stationary covariances.** The Gneiting covariance function (6.1) used to construct the Dynamic NNGP is a stationary covariance function. However, we can also use Dynamic NNGP to approximate certain class of non-stationary covariance functions. Gelfand et al. (2004) proposed

a Spatially Varying Linear Model of Coregionalization (SVLMC) model to create non-stationary spatial processes as spatially weighted sum of stationary processes. For univariate responses, this reduces to $w(\mathbf{s}, t) = \sigma^2(\mathbf{s}, t)v(\mathbf{s}, t)$ where $v(\mathbf{s}, t) \sim GP$ with mean 0 and covariance function defined in (6.1) with $\sigma^2 = 1$. Subsequently, there are several ways of modeling $\sigma^2(\mathbf{s}, t)$ If a covariate $x(\mathbf{s}, t)$ is observed, $\sigma^2(\mathbf{s}, t)$ is often modeled as $x(\mathbf{s}, t)^\phi \sigma^2$ for some unknown $\phi \geq 0$. This introduces covariate dependent non-stationarity. In such case, due to non-stationarity, physical nearest neighbors of a location may not correspond to locations with highest correlations for the process $w(\mathbf{s}, t)$. Hence, the simple or the adaptive neighbor selection methods described in Section 4 which relies on the above mentioned principle, may lead to conditioning sets with very few points with high correlation with a given point. Hence, we suggest modeling the stationary process $v(\mathbf{s}, t)$ as a DNNGP instead of $w(\mathbf{s}, t)$. Letting $\mathbf{v}$ the vector formed by stacking up $v(\mathbf{s}, t)$ over the set of observed locations, the joint likelihood will be specified as

$$\prod_{(\mathbf{s},t)} N(y(\mathbf{s}, t) \,|\, x(\mathbf{s}, t)\beta + \sigma x(\mathbf{s}, t)^{\phi/2}v(\mathbf{s}, t), \tau^2) \times N(\mathbf{v} \,|\, \mathbf{0}, \tilde{\mathbf{C}}(\boldsymbol{\theta}))$$

(S1) $$\times \, p(\boldsymbol{\theta}) \times p(\sigma) \times p(\phi) \times IG(\tau^2 \,|\, a_\tau, b_\tau) \times N(\boldsymbol{\beta} \,|\, \boldsymbol{\mu}_\beta, \mathbf{V}_\beta) \,,$$

where $\tilde{\mathbf{C}}(\boldsymbol{\theta})$ is the DNNGP covariance matrix created from the original non-separable covariance matrix for the process $v(\mathbf{s}, t)$. The Gibbs' sampler steps are similar to those detailed in Sections 5.1 and 5.2. The spatio-temporal random effects $v(\mathbf{s}, t)$ are updated sequentially and have conjugate normal distributions similar to those defined in (5.4). The covariance parameters $\boldsymbol{\theta}$ are updated using the random walk Metropolis step (5.5). The additional parameters $\sigma$ and $\phi$ are also easily updated using the Metropolis random walk step.

Alternatively, the non-stationary variance $\sigma^2(\mathbf{s}, t)$ can be modeled as a smooth spatial process for e.g. $\log \sigma^2(\mathbf{s}, t) \sim GP$. In such a case, we can use DNNGPs to approximate the stationary covariances of both $v(\mathbf{s}, t)$ and $u(\mathbf{s}, t) = \log \sigma^2(\mathbf{s}, t)$. However, unlike $v(\mathbf{s}, t)$ the full conditional for $u(\mathbf{s}, t)$ is not Gaussian. Instead, $p(u(\mathbf{s}, t) \,|\, \cdot)$ is proportional to

$$N(y(\mathbf{s}, t) \,|\, x(\mathbf{s}, t)\beta + \exp(u(\mathbf{s}, t)/2) \, v(\mathbf{s}, t), \tau^2) \times N(\mathbf{u} \,|\, \mathbf{0}, \tilde{\mathbf{C}}_u(\boldsymbol{\theta}_u)) \,,$$

where $\tilde{\mathbf{C}}_u(\boldsymbol{\theta}_u)$ is the DNNGP covariance matrix for $\mathbf{u}$ – the vector of random effects $u(\mathbf{s}, t)$, and $\boldsymbol{\theta}_u$ are the associated covariance parameters. Evaluating this conditional likelihood for all $(\mathbf{s}, t)$ for Metropolis updates will still require $O(n)$ flops and scalability will be retained.

While, in theory, DNNGP can be easily extended to model some non-stationary covariances in a scalable manner, extensive simulation studies

needs to be conducted to actually determine the accuracy of parameter estimation and kriging in a non-stationary setup. MCMC convergence behavior for the spatio-temporal random effects in absence of conjugate Gibbs' updates also needs to be investigated. Furthermore, more general non-stationary structures as specified in Paciorek and Schervish (2006) also needs to be considered. These explorations, however, do not fall within the scope of the current paper and we identify them as areas of future research.

## References.

DATTA, A., BANERJEE, S., FINLEY, A. O. and GELFAND, A. E. (2016). Hierarchical Nearest-Neighbor Gaussian Process Models for Large Geostatistical Datasets. *Journal of the American Statistical Association* In Press.

GELFAND, A., SCHMIDT, A., BANERJEE, S. and C., S. (2004). Nonstationary multivariate process modeling through spatially varying coregionalization. *TEST: An Official Journal of the Spanish Society of Statistics and Operations Research* **13** 263-312.

GRAMACY, R. B. and APLEY, D. W. (2015). Local Gaussian Process Approximation for Large Computer Experiments. *Journal of Computational and Graphical Statistics* **24** 561-578.

PACIOREK, C. J. and SCHERVISH, M. J. (2006). Spatial modelling using a new class of nonstationary covariance functions. *Environmetrics* 483–506.



\end{document}